\documentclass[journal ]{new-aiaa}
\usepackage[utf8]{inputenc}
\usepackage{textcomp}

\usepackage{graphicx}
\usepackage{amsmath}
\usepackage[version=4]{mhchem}
\usepackage{siunitx}
\usepackage{longtable,tabularx}
\usepackage{epstopdf}
\usepackage{overpic}
\usepackage{xcolor}

\usepackage{caption}
\usepackage{subcaption}
\graphicspath{{./Graphics/}}

\usepackage{xcolor}

\title{Transonic buffet characteristics\\ under conditions of free and forced transition}

\author{Pradeep Moise\footnote{Research Fellow, pradeep890@gmail.com.}, Markus Zauner\footnote{Research Fellow, m.zauner@soton.ac.uk} and Neil D. Sandham\footnote{Professor, n.sandham@soton.ac.uk.}}
\affil{University of Southampton, Southampton, SO17 1BJ, United Kingdom}
\author{Sebastian Timme\footnote{Senior Lecturer, sebastian.timme@liverpool.ac.uk} and Wei He\footnote{Research Associate, present address: City, University of London, wei.he.2@city.ac.uk}}
\affil{University of Liverpool, Liverpool, L69 3GH, United Kingdom}
\begin{document}

\maketitle

\begin{abstract}
\setstretch{1.5}
Transonic buffet is commonly associated with self-sustained flow unsteadiness involving shock-wave/boundary-layer interaction over aerofoils and wings. The phenomenon has been classified as either laminar or turbulent based on the state of the boundary layer immediately upstream of the shock foot and distinct mechanisms for the two types have been suggested. The turbulent case is known to be associated with a global linear instability. Herein, large-eddy simulations are used for the first time to make direct comparisons of the two types by examining free- and forced-transition conditions. Corresponding simulations based on the Reynolds-averaged Navier--Stokes equations for the forced-transition case are also performed for comparison with the scale-resolving approach and for linking the findings with existing literature. Coherent flow features are scrutinised using both data-based spectral proper orthogonal decomposition of the time-marched results and operator-based global linear stability and resolvent analyses within the Reynolds-averaged Navier--Stokes framework. It is demonstrated that the essential dynamic features remain the same for the two buffet types (and for the two levels of the aerodynamic modelling hierarchy), suggesting that both types arise due to the same fundamental mechanism. 
\end{abstract}

\newpage
\section*{Nomenclature}

{\renewcommand\arraystretch{1.0}
\noindent\begin{longtable*}{@{}l @{\quad=\quad} l@{}}
$C_p$ & pressure coefficient \\
$C_f$ & skin-friction coefficient \\
$C_L$ & lift coefficient \\
$C_D$ & drag coefficient \\
$(C_D)_p$ & pressure drag coefficient \\
$(C_D)_f$ & skin friction drag coefficient \\
$t$ & dimensionless time \\
$t_0$ & dimensionless time corresponding to when flow is approximately past transients\\
$p$ & dimensionless pressure\\
$(x,y,z)$ & dimensionless Cartesian coordinates representing streamwise direction, direction normal\\ & to streamwise and spanwise directions, and spanwise direction, respectively \\
$(x',y')$ & dimensionless Cartesian coordinates representing chordwise direction and direction normal to \\ & chordwise and spanwise directions, respectively\\
$(\xi, \eta)$ & dimensionless circumferential and radial curvilinear coordinate based on grid, respectively \\
$\alpha$ & angle of attack \\
$M$ & Freestream Mach number \\
$Re$ & Freestream Reynolds number based on chord and freestream velocity \\
$St$ & Strouhal number/dimensionless frequency based on freestream velocity and chord\\
$St_b$ & dimensionless buffet frequency\\
$\boldsymbol{C}$ & cross-spectral density tensor used for spectral proper orthogonal decomposition\\
$\boldsymbol{W}$ & weight associated with the quadrature on the grid \\
$\langle .,. \rangle_{\boldsymbol{W}}$ & a weighted inner product using $\boldsymbol{W}$ \\
$\lambda_i$ & $i-$th eigenvalue of $\boldsymbol{C}$ (ranked in a descending order based on energy) \\
$\boldsymbol{\psi_i}$ & $i-$th eigenfunction of $\boldsymbol{C}$ \\
$(\Delta \xi^+$, $\Delta \eta^+$, $\Delta z^+)$ & grid spacing at the wall in wall units\\
$\phi$ & phase within a buffet cycle \\
$\rho u_\eta\left|_{\eta=0}\right.$ & dimensionless wall-normal component of momentum on aerofoil surface of synthetic jet\\
$A,\:\sigma$ & dimensionless amplitude and standard deviation of Gaussian of the synthetic jet\\
$x_t, x_{ts}, x_{tp}$ & trip location on aerofoil, with subscripts used when specifying suction or pressure surface \\
$k,\omega,\Phi$ & dimensionless spanwise wavenumber, angular frequency and temporal phase of the synthetic jet\\
$x_\mathrm{shock}$ & streamwise location of shock wave\\
$\boldsymbol{S}$, $\boldsymbol{\varOmega}$ & dimensionless symmetric and anti-symmetric components of velocity gradient tensor\\
$Q_\mathrm{inc}$ & incompressible version of second invariant of the dimensionless velocity gradient tensor\\
$||.||_F$& Frobenius norm\\
%\multicolumn{2}{@{}l}{Subscripts}\\
%~ & ~ \\
\multicolumn{2}{@{}l}{Superscripts}\\
$\tilde{()}$ & reconstructed flow field based on the buffet mode from SPOD\\
$\overline{()}$ & time and span-averaged (mean) quantity\\
$()'$ & fluctuating component of a quantity\\
\end{longtable*}}

%%%%%%%%%%%%%%%%%%%%%%%%%%%%%%%%%%%%%%%%%
%\section*{2do list:}
%%%%%%%%%%%%%%%%%%%%%%%%%%%%%%%%%%%%%%%%%
    
%Final checklist:    
%\begin{enumerate}
%    \item Abstract is less than 200 words (current: 175)
%    \item All symbols are introduced only in nomenclature
%    \item All figure captions are less than 25 words as far as possible
%    \item confirm that 22 is Garbaruk et al.
%    \item Remove/comment out this section
%\end{enumerate}    

%%%%%%%%%%%%%%%%%%%%%%%%%%%%%%%%%%%%%%%%
\section{Introduction}
\label{secIntro}
%%%%%%%%%%%%%%%%%%%%%%%%%%%%%%%%%%%%%%%%

%% Background/Intro to intro
Self-sustained unsteadiness that occurs in transonic flows over wings under certain flow conditions can be associated with a phenomenon commonly referred to as transonic buffet~\citep{Helmut1974}. Transonic buffet, or simply `buffet', can lead to significant variations in lift, possibly due to large-scale streamwise motion of shock waves and the quasi-periodic separation and reattachment of the boundary layer. This variation in lift, in turn, can lead to passenger discomfort, challenges with aircraft control, structural fatigue and even failure~\citep{Lee2001}. Thus, transonic buffet is detrimental to aircraft performance and manoeuvrability and, for these reasons, it has been extensively studied with a focus on understanding and controlling its occurrence~\citep{Giannelis2017}.

%% 2D vs 3D, and Buffet/wake modes
%Transonic buffet has been shown to be essentially a two-dimensional phenomenon \citep{Dandois2016,Timme2020}. Thus, 
Although differences between two-dimensional aerofoil and three-dimensional finite-wing buffet scenarios have been reported~\citep{Dandois2016}, numerical studies on infinite wings~\citep{Crouch2019,paladini_2019_PFR,he_timme_2021_jfm,Plante2020} have described key characteristics observed on both finite wings~\citep{Timme2020} and aerofoils. Thus, a large number of studies have examined infinite-straight wings based on common aerofoil profiles (\textit{e.g.}, \citep{Xiao2006a, Crouch2007, Grossi2014, Hartmann2012}). Simulations of such flows using the Reynolds-Averaged Navier--Stokes (RANS) equations have been especially fruitful in gaining insights into the mechanisms associated with transonic buffet. \citet{Crouch2007}, using a global linear stability analysis (GLSA) of steady-state RANS results for the flow over a NACA0012 aerofoil, showed that transonic buffet can be associated with a globally unstable mode. Besides confirming these earlier stability results for a supercritical OAT15A aerofoil, \citet{Sartor2015} performed a resolvent analysis. Through this, they identified an additional optimally amplified mode that resembles a von K\'arm\'an vortex street in the aerofoil wake and occurs at a relatively higher frequency as compared to aerofoil buffet. Unsteady RANS simulations do not seem to capture this wake mode, although they capture buffet mode features (\textit{e.g.}, see Fig.~20 of~\citet{Szubert2016}). By contrast, high-fidelity approaches such as variants of detached-eddy simulations \citep{Deck2005,Grossi2014}, large-eddy simulations (LES)~\citep{Fukushima2018,Zauner2020a,Moise2022} and direct numerical simulations (DNS)~\citep{Zauner2019} capture the occurrence of both.

%% fully turbulent vs laminar buffet  part 1
% While most such simulations examine buffet features at high Reynolds numbers ($Re$, based on freestream conditions and aerofoil chord such that $Re \geq 10^6$), the latter DNS study employed a lower $Re = 5\times 10^5$. Interestingly, in this DNS study, multiple shock waves are observed in the transitional flow field. This should be contrasted with the single shock wave seen at higher $Re$ in the vast majority of studies on buffet, assuming a fully developed turbulent boundary layer.
%This aspect was further explored using LES by \citet{Moise2022}, who showed by simulating conditions below the onset of buffet that these shock waves occur even in the absence of buffet. Thus, the authors suggested that multiple shock waves are related to the laminar/transitional boundary layer evolution which is supported by previous and recent experimental studies \citep{Ackeret1947,ZaunerUKFluids2021}.
%% fully turbulent vs laminar buffet  part 2 (please combine the 2 parts)

Based on the boundary layer characteristics in the region of shock-wave/boundary-layer interaction, transonic buffet can be classified as either laminar or turbulent \citep{Brion2020}. For the former, the boundary layer remains laminar all the way to the shock foot, whereas in the turbulent case the boundary layer fully transitions to turbulence well upstream of the shock wave. Possibly due to the challenges involved in accurately modelling flows with free transition, all studies that employ RANS simulations have focused on turbulent buffet. As noted above, it is only this type which has been shown to occur due to a global instability~\citep{Crouch2007}. Furthermore, all studies on turbulent buffet report only a single shock wave terminating the supersonic region, but laminar buffet with multiple shock waves present has been reported in experiments\footnote{https://doi.org/10.13140/RG.2.2.30817.58725 and  https://www.youtube.com/watch?v=llF2U03SShw}, which requires further scrutiny.

%% tripped buffet and missing link
While computational turbulent buffet studies typically assume fully-turbulent conditions (\textit{i.e.}~a turbulent boundary layer from the leading edge), recently \citet{Garbaruk2021} have examined the effect of forcing transition at different locations on the aerofoil upstream of the shock wave using steady RANS simulations and GLSA. They have shown that the essential characteristics of turbulent buffet remain the same irrespective of the location of transition, which is in agreement with the experimental findings of~\citet{Dor1989}. By contrast, for laminar buffet, it has been proposed by \citet{Dandois2018} that the mechanism sustaining it is different and associated with a bubble-breathing phenomenon. However, it was later demonstrated by \citet{Moise2022} using LES and a modal reconstruction that laminar buffet has characteristics that are essentially similar to those reported in previous literature for turbulent buffet. The simulations in~\citep{Moise2022} were carried out only for laminar buffet, while the steady RANS simulations and GLSA in~\citep{Garbaruk2021} were performed only for turbulent buffet and gaps remain in linking the two studies.

%% outlook on the paper
Direct comparisons of (i) the laminar and turbulent buffet types, (ii) the occurrence of multiple shock waves or a single shock wave, and (iii) the simulation approaches of LES and RANS can help bridge the gaps in our understanding of the relations between them. Motivated by these aspects, we perform LES of both laminar and turbulent buffet under similar flow conditions and compare the findings for the latter type with corresponding steady and unsteady RANS (URANS) results. Furthermore, we extract modal features from the LES and URANS data using spectral proper orthogonal decomposition (SPOD) and compare them with results obtained using GLSA and resolvent analysis of the steady RANS solutions. We continue with an overview of the simulation approaches and analysis tools in Sec.~\ref{secMethod}, with details of flow conditions provided in Sec.~\ref{secTestCase}. The simulation results are presented in Sec.~\ref{secResults}, following which the results from SPOD, GLSA and resolvent analysis are discussed in Sec.~\ref{secDecomNRecon}, with Sec. \ref{secConc} concluding the study. 

%%%%%%%%%%%%%%%%%%%%%%%%%%%%%%%%%%%%%%%%
\section{Methodology}
\label{secMethod}
%%%%%%%%%%%%%%%%%%%%%%%%%%%%%%%%%%%%%%%%
%%%%%%%%%%%%%%%%%%%%%%%%%%%%%%%%%%%%%%%%
\subsection{Large-eddy simulations}
\label{subsecMethodLES}
%%%%%%%%%%%%%%%%%%%%%%%%%%%%%%%%%%%%%%%%
As noted in the previous section, the LES performed in this study can be categorised into two types; the first being the free-transition case, where the boundary layer transitions naturally, and forced-transition (or tripped) cases, where a trip is introduced on either the pressure or suction side of the aerofoil (or both). With the exception of the introduction of the transition forcing, the methodology adopted here for the LES is the same as that detailed in Ref.~\citep{Moise2022}. Specifically, the in-house flow solver SBLI is used to perform the simulations on transonic buffet. It has been used in a number of previous studies to examine laminar buffet \citep{Zauner2019, Zauner2020a, Zauner2020, Moise2022}. The compressible Navier--Stokes equations are solved in a dimensionless form with the required scales being the aerofoil chord, the freestream density, velocity and temperature. A spectral-error based implicit LES approach is adopted \citep{Jacobs2018,Zauner2020Conference}. This has previously been validated against DNS for flows where buffet is observed \citep{Zauner2020}. For spatial discretisation, a fourth-order finite difference scheme is employed, together with a total variation diminishing scheme to capture shock wave features. A third-order Runge-Kutta scheme is used for time discretisation. The dimensionless time step was chosen as $3.2\times 10^{-5}$ and $1.6\times 10^{-5}$ for the free and forced-transition cases, respectively (implying approximately $10^5$ iterations in a buffet cycle). 

For the present aerofoil (straight infinite wing) simulations, the spanwise direction is chosen as periodic. Zonal characteristic boundary conditions (associated with a buffer zone) are applied at the outflow boundary, while integral characteristic boundary conditions are applied on other outer boundaries (see \citet{Moise2022}, figure 1\textit{a}). The aerofoil is treated as an isothermal and no-slip wall, except at the location of the trip. %To trip the boundary layer, a synthetic jet which induces boundary layer transition is employed. Note that turbulent buffet can occur when the Reynolds number is sufficiently high and other conditions are conducive for natural transition to occur close to the leading edge (\textit{e.g.}, \citep{Lee1989}) or when the boundary layer is forced to transition to turbulence in experiments (\textit{e.g.}, \citep{Brion2020}) or numerical studies (\textit{e.g.}, \citep{Xiao2006a, Garbaruk2021}). Here, Due to the moderate values of Reynolds numbers studied here, 
At this location, we perturb the boundary layer using a synthetic jet that exits the aerofoil surface along the wall-normal direction. The wall-normal momentum component, $\rho u_\eta$, of the jet is given by
\begin{equation}
\rho u_\eta\left|_{\eta=0}\right. = \sum\limits_{i=1}^3 A \exp{\left(-\frac{(x-x_t)^2}{2\sigma^2}\right)\sin(k_i^2z)\sin(\omega_i t+\Phi_i)}.
\end{equation}
The variation in the streamwise direction is that of a Gaussian function with parameters $A = 0.05$ and $\sigma = 0.00833$, centred about the chosen transition location $x_t$. The variation in time and spanwise directions are sinusoidal with three modes superposed ($k_i =\{120\pi,160\pi,160\pi\}$, $\omega_i = \{26,88,200\}$, $\Phi_i = \{0,\pi,-\pi/2\}$). %The spanwise wavenumbers and amplitude were chosen for the jet on the suction surface based on results from the free-transition case such that the former is approximately ten times the boundary layer thickness at $x_t$ and the latter is a third of the velocity at the edge of the boundary layer on the suction surface. Other parameters were found by trial-and-error. The same values were retained for the pressure-side forcing as they were found to be adequate to induce transition there as well. 

%The pressure-side trip position, $x_{tp}$, is chosen such that in the region downstream, the natural pressure gradient aids in amplifying the perturbations and triggering a transition. The tripping on the pressure surface was used only to examine the effect of tripping on various sides and to make comparisons with the RANS results. 
%The spanwise direction is denoted by $z$, whereas the streamwise and the third orthogonal Cartesian direction are labelled as $x$ and $y$, respectively. The curvilinear circumferential and radial directions are $\xi$ and $\eta$. The chordwise direction is referred to as $x'$. 

%%%%%%%%%%%%%%%%%%%%%%%%%%%%%%%%%%%%%%%%
%\subsubsection{Boundary-layer tripping}
%\label{secTripping}
%%%%%%%%%%%%%%%%%%%%%%%%%%%%%%%%%%%%%%%%
%{\color{red}The simulations of turbulent buffet are performed by forcing boundary layer transition for flow conditions similar to those at which laminar buffet is otherwise observed. To achieve transition in the LES, we use a trip located on the surface of the aerofoil and modelled by an unsteady, wall-normal jet that perturbs the flow.} 

%%%%%%%%%%%%%%%%%%%%%%%%%%%%%%%%%%%%%%%%
\subsection{RANS simulation}
\label{subsecMethodRANS}
%%%%%%%%%%%%%%%%%%%%%%%%%%%%%%%%%%%%%%%%
% Recent studies \cite{Garbaruk2021, tabatabaei2021} have examined the effect of tripped transition in RANS simulations with different turbulence models and numerical methods. 

At the RANS level of our study, we follow the numerical approach outlined in detail in our previous work~\citep{Timme2020,he_timme_2021_jfm,houtman2022}. Specifically, the nonlinear compressible RANS equations coupled with the negative Spalart--Allmaras turbulence model including the mixing-layer compressibility correction~\cite{SpalartAllmaras,spalart-SACC} are solved. To force transition in the RANS simulations, more precisely to prevent the early development of a fully-turbulent boundary layer, the source term of the turbulence model is deactivated upstream of the transition location. Hence, neither a more sophisticated transition model nor the so-called trip term in the original Spalart--Allmaras turbulence model are employed in the simulations. 
The equations are solved using the DLR-TAU code, which has a second-order, cell-vertex, finite-volume formulation capable of dealing with complex geometries~\citep{Schwamborntau}. 
The inviscid fluxes of the flow model are discretised using a central scheme with matrix artificial dissipation. Gradients of flow variables, where needed, are computed using the Green--Gauss approach. The two-dimensional aerofoil simulations have a no-slip adiabatic boundary condition at the viscous aerofoil wall, while the far field is described as free-stream flow realised by a characteristic boundary condition consistent with the discretisation of the interior fluxes. For accelerated steady-state convergence, the chosen time stepper uses the implicit backward Euler method with lower–upper symmetric Gauss–Seidel iterations, local time stepping (setting the Courant–Friedrichs–Lewy number for the implicit scheme to 50 throughout) and an agglomeration geometric multigrid approach on three grid levels. We typically aim for a ten orders of magnitude reduction in the norm of the density residual. For unsteady time-marching simulations, the temporal derivative is discretised with a second-order backward difference formula and the resulting residual in dual time is converged with the same steady-state time stepper. The dimensionless time-step size (with respect to free-stream velocity and chord length) is set at 0.012 which gives approximately 700 time steps for the buffet mode per cycle. % (and 70 steps for the wake mode which does not feature in URANS simulations though). 
Cauchy convergence control on the drag coefficient is specified at a level of $10^{-8}$ with a minimum of 50 iterations in dual time per real time step. Approximately 190 solution updates per time step are required on average over the entire time history of 55,000 time steps. 

%%%%%%%%%%%%%%%%%%%%%%%%%%%%%%%%%%%%%%%%%%%%%%%%%%%%%%%%%%%%%%%%%%%%%%%%%%%%%%%%
\subsection{Spectral proper orthogonal decomposition}
\label{subSecSPOD}
%%%%%%%%%%%%%%%%%%%%%%%%%%%%%%%%%%%%%%%%%%%%%%%%%%%%%%%%%%%%%%%%%%%%%%%%%%%%%%%%
Spatio-temporal coherent structures in the flow field can be extracted using the SPOD approach~\citep{Lumley1970, Glauser1987, Towne2018}. Herein, we use the streaming algorithm and the numerical code provided by \citet{SCHMIDT201998}. On a domain $\Omega$, this approach solves the eigenvalue problem
\begin{equation}
\int_\Omega \boldsymbol{C}(\mathbf{x}_1,\mathbf{x}_2,St_0)  \,\boldsymbol{W}(\mathbf{x}_2) \,\boldsymbol{{\psi}}_i(\mathbf{x}_2,St_0)\, \text{d}\mathbf{x}_2 =  \lambda_i\,\boldsymbol{\psi}_i(\mathbf{x}_1,St_0)
\label{eqnSPOD}
\end{equation}
at a given Strouhal number, $St_0$. Here, $\boldsymbol{C}$ is computed using Welch's method based on spatio-temporal data collected on the two-dimensional plane, $z = 0$, with $\mathbf{x}_1$ and $\mathbf{x}_2$ being any two points on this plane (with $\mathbf{x}=(x,y)$), while $\boldsymbol{W}$ contains the approximate cell volume associated with the grid points. The eigenvalues are indexed such that $\lambda_1 > \lambda_2 > \lambda_3 > \dots$, implying that the most-energetic SPOD mode is $\boldsymbol{\psi}_1$ with higher indices representing modes of lower energy. These eigenmodes are orthonormal for the weighted inner product based on $\boldsymbol{W}$. It was found that the dominant coherent motion present in the flow field at a given frequency can be represented by the leading mode only, whereas the energy content of higher modes was negligible (not shown herein, but cf.~\citep{Moise2022}). Thus, these higher modes were not further considered in the analysis. The reconstructed flow field, $\mathbf{\tilde{q}}(\mathbf{x},t)$, based on the leading mode only is then obtained by combining the spatio-temporal SPOD modal features with the mean flow field, $\mathbf{\overline{q}}(\mathbf{x})$ and is given by
\begin{equation}
    \mathbf{\tilde{q}}(\mathbf{x},t) =  \mathbf{\overline{q}}(\mathbf{x}) + \mathrm{Re}\left\{\sqrt{\lambda_1}\,\, \boldsymbol{\psi}_1(\mathbf{x},St_b)\,\, \exp{(\mathrm{i}2\pi St_b\, t)}\right\} \label{eqnReconstr}
\end{equation}
See \citep{Moise2022} for a more detailed description of the overall approach. Note that equivalent expressions can also be used for the modes obtained from the operator-based stability and resolvent approaches, discussed below. The temporal data at the grid point locations was sampled at a dimensionless frequency of 12.5 (based on chord length and free-stream velocity), which is about two and one orders of magnitude higher than observed buffet frequencies and the frequencies associated with wake modes reported in Sec.~\ref{secSPOD}, respectively. The signals were collected for a duration of 70 dimensionless time units (approximately 10 buffet cycles) and these were then divided into blocks of 44 time units using a $50\%$ overlap for Welch's method. A similar set-up is chosen for both LES and time-marched URANS data. 
%SPOD was also performed for the results obtained from URANS simulation. 

%%%%%%%%%%%%%%%%%%%%%%%%%%%%%%%%%%%%%%%%%%%%%%%%%%%%%%%%%%%%%%%%%%%%%%%%%%%%%%%%
\subsection{Global linear stability and resolvent analyses}
\label{subSecGLSA}
%%%%%%%%%%%%%%%%%%%%%%%%%%%%%%%%%%%%%%%%%%%%%%%%%%%%%%%%%%%%%%%%%%%%%%%%%%%%%%%%
In contrast to the data-based post-processing using SPOD analysis, the operator-based approach in the RANS study is directly integrated into the linear harmonic incarnation of the chosen flow solver, where the first-discretise-then-linearise Jacobian operator is obtained from a graph-coloured finite-difference approach and stored explicitly. The computation of the Jacobian matrix has been scrutinised in~\cite{he_timme_2021_jfm}. The two-dimensional global stability and resolvent analyses both rely on inner-outer iterative solution schemes. The extraction of global modes uses the implicitly restarted Arnoldi method with shift-invert spectral transformation in the outer iterative level. Specifically, the ARPACK library is adopted for this purpose~\cite{Sorensen_1992_siam}. %The underlying Arnoldi algorithm in the outer iterative level has been detailed in the literature many times and the interested reader is referred to elsewhere. 
The shift-invert transformation results in large linear algebraic problems in the inner iterative level, which are solved using the preconditioned generalised conjugate residual method with inner orthogonalisation and deflated restarting, where zero-level incomplete lower-upper factorisation is the preconditioner~\cite{XU2016385}. The linear system is not normally required to be converged to machine-epsilon levels and an approximate solution is sufficient~\cite{Timme2020}. Based on insight gained in previous work, we set the convergence tolerance to $10^{-8}$. The extraction of resolvent modes makes use of the iterative formulation first exercised in \cite{weiAviation} and later detailed in \cite{houtman2022}. The algorithm follows the time-stepping procedure presented in~\cite{gomez2016} except that periodic states are solved directly in the frequency domain, using the same inner-level iterative solver as for the stability tool, instead of relying on long time integration. Firstly, a random forcing vector is passed through the resolvent operator, which is the inverse of the frequency-shifted Jacobian operator, to find the most amplified response. Secondly, the normalised response is passed through the adjoint resolvent operator to obtain a better approximation to the optimal forcing. Continued iteration will converge to the optimal solution with the associated energy gain obtained from normalising the forcing/response vector with respect to a suitable inner product. Herein we chose the volume-weighted inner product for convenience in comparing with the SPOD modes. Suboptimal modes can be found with a type of deflation. Essentially the same iterative algorithm, referred to as `sketching', was presented in \cite{house2022}, derived from a randomised resolvent approach \citep{ribeiro2020}. The authors stated a rate of convergence which is dependent on the ratio of optimal to first suboptimal singular values. If strong modal behaviour is present in the flow, such that a low rank (or even rank-1) assumption is reasonably adequate~\cite{gomez2016}, the algorithm will converge rapidly. Finally, the convergence level of the inner linear systems is adapted to the current outer-level convergence for reasons of computational efficiency, with an absolute minimum tolerance set at $10^{-8}$. 

\begin{table}[t]
    \centering
    \caption{Cases simulated using LES and URANS, corresponding flow parameters and main results. Parameters $x_{ts} \rightarrow \infty$ and $x_{tp} \rightarrow \infty$ imply that transition is not enforced.}
    \label{tab:mainResults}
    \begin{tabular}{cccccccccccc}
    \hline\hline
Label    & Method &  $x_{ts}$ &  $x_{tp}$         & $\alpha$ & $M$ &  $\overline{C}_L$ & $(\overline{C}_D)_p$ & $(\overline{C}_D)_f$ &     $\overline{C}_D$ &  $St_b$ & PSD$(C_L',St_b)$ \\
    \hline
A5M7F    &  LES          &  $\infty$ &  $\infty$  & $5^\circ$& 0.7 & 0.82 & 0.061 & 0.0034 & 0.065 & 0.11 & 0.34  \\
A5M7S    &  LES          &  0.2 &  $\infty$  & $5^\circ$& 0.7 & 0.83 & 0.043 & 0.0059 & 0.049 & --    & --  \\
A5M735F  &  LES          &  $\infty$ &  $\infty$  & $5^\circ$& 0.735 & 0.69 & 0.076 & 0.0033 & 0.079 & 0.15  & 0.19  \\
A5M735S  &  LES          &  0.2 &  $\infty$  & $5^\circ$& 0.735 & 0.68 & 0.056 & 0.0052 & 0.061 & 0.13  & 0.0028  \\
A5M735P  &  LES          &  $\infty$ &  0.5  & $5^\circ$& 0.735 & 0.75 & 0.082 & 0.0050 & 0.086 & 0.16  & 0.13  \\
A5M735SP &  LES          &  0.2 &  0.5  & $5^\circ$& 0.735 & 0.73 & 0.056 & 0.0069 & 0.063 & 0.16  & 0.0020  \\
A5M735SP &  URANS        &  0.2 &  0.5  & $5^\circ$& 0.735 & 0.77 & 0.042 & 0.0053 & 0.048 & 0.12  & 0.096  \\
A7M7F    &  LES          &  $\infty$ &  $\infty$  & $7^\circ$& 0.7 & 0.95 & 0.108 & 0.0030 & 0.112 & 0.13  & 0.31  \\
A7M7S    &  LES          &  0.2 &  $\infty$  & $7^\circ$& 0.7 & 0.98 & 0.078 & 0.0046 & 0.082 & 0.12  & 0.024  \\
A0M8F    &  LES          &  $\infty$ &  $\infty$  & $0^\circ$& 0.8 &-0.027& 0.067 & 0.0033 & 0.071 & 0.13  & 0.28  \\
A0M8S    &  LES         &  0.2 &  $\infty$  & $0^\circ$& 0.8  & 0.070& 0.058 & 0.0055 & 0.063 & 0.12  & 0.25  \\
    \hline\hline
    \end{tabular}
\end{table}

\section{Configuration}
\label{secTestCase}

This study focuses on simulating buffet features at $Re = 5\times 10^5$ using Dassault Aviation's laminar-flow, supercritical V2C profile with a blunt trailing edge of thickness 0.5\% chord. The exceptions to this are the $Re = 3\times10^6$ cases considered in Sec.~\ref{subSecGLSAResolvent} for the V2C and ONERA's OAT15A aerofoils (Fig.~\ref{figGLSAHighReOAT15a}) and the validation, using the latter, of the RANS simulations (Appendix~\ref{subSecRANSValidation}). For all cases, the fluid is assumed to be a perfect gas with a specific heat ratio of 1.4 and satisfying Fourier's law of heat conduction  with a Prandtl number of 0.72 (and a turbulent Prandtl number of 0.9 for RANS simulations). It is also assumed to be Newtonian and satisfying Sutherland's law, with the Sutherland coefficient as 110.4 at a reference temperature 268.67~K. The incidence angle, freestream Mach number and the transition type are varied in the LES. The main focus will be on a `reference incidence' of $\alpha = 5^\circ$. When the boundary layer tripping is active, the trip positions are $x_{ts} = 0.2$ on the suction surface and/or $x_{tp} = 0.5$ on the pressure surface. With the exception of Appendix~\ref{subSecRANSValidation}, the RANS-level simulations will examine only the case of $\alpha = 5^\circ$ and $M = 0.735$ with transition enforced on both the suction and pressure sides. The main cases studied herein are summarised in Table~\ref{tab:mainResults}. 

%\subsection{Large-eddy simulations}

\begin{figure}[t]
	\centering
	\includegraphics[width=0.495\columnwidth]{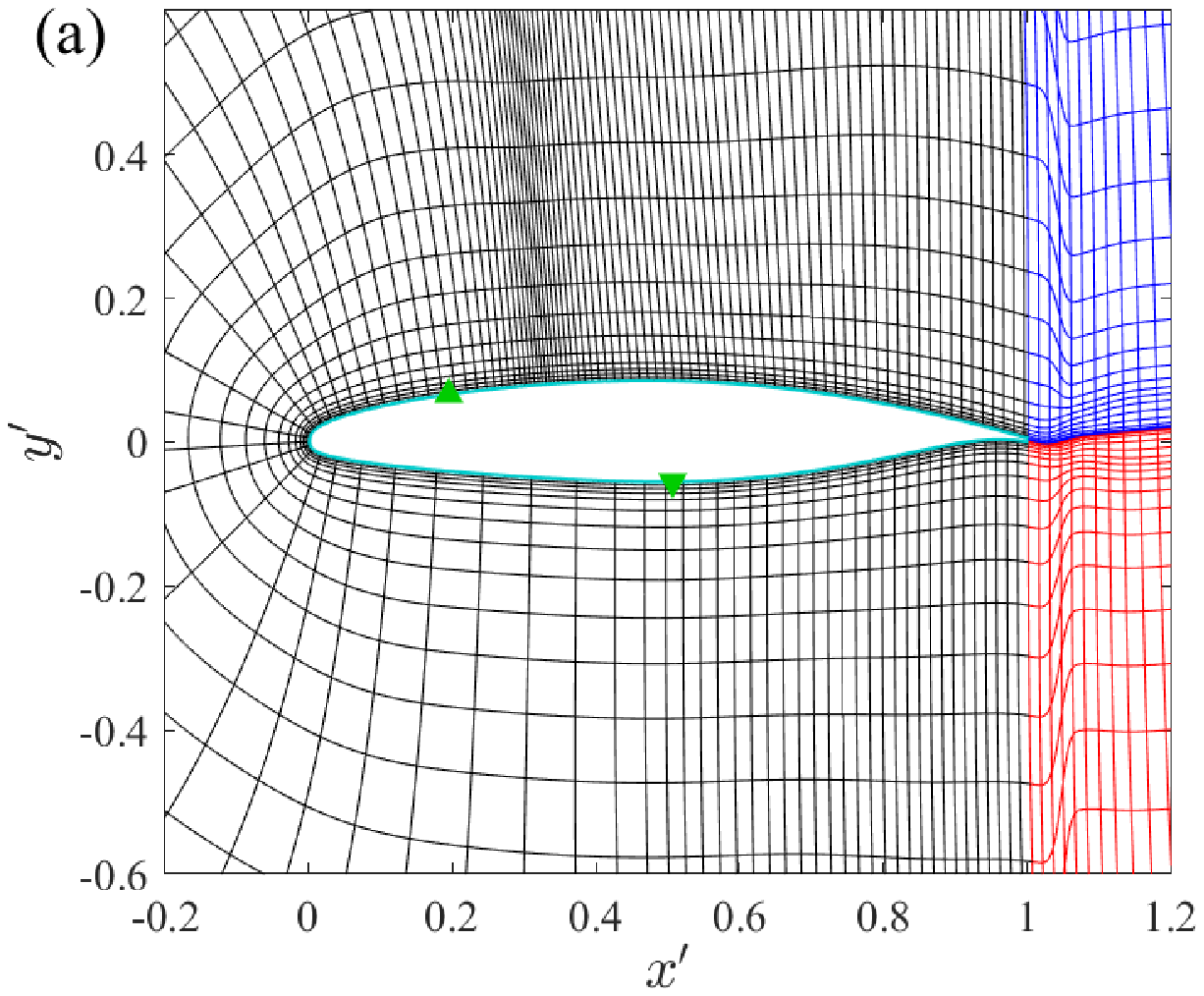}
	\includegraphics[width=0.495\columnwidth]{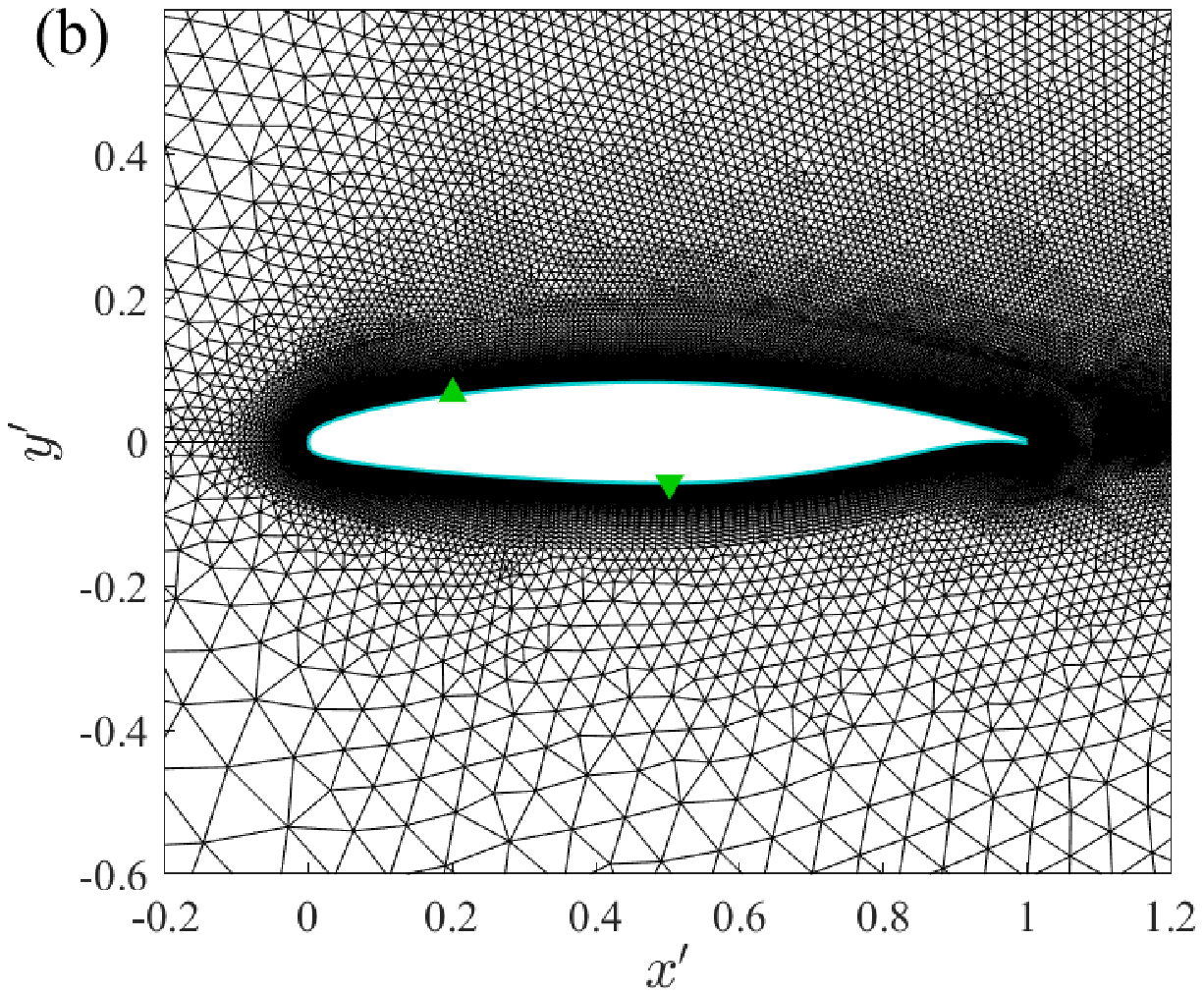}
	\caption{Near-field features of V2C aerofoil grids used for (a) LES (every 15$^\mathrm{th}$ point shown) and (b) RANS-level simulations. %The green triangles highlight tripping locations.
	}
	\label{figGrid}
\end{figure}

Using an in-house open-source code \citep{Zauner2018}, multi-block structured grids (C-H type) were generated for each angle of attack to ensure that the wake flow is properly resolved in the LES. The C-shaped block is of radius 7.5 chord lengths while the downstream blocks have a streamwise extent of 4.5 chord lengths. Typical LES grid features in the vicinity of the aerofoil are shown in Fig.~\ref{figGrid}\textit{a} using every 15$^\mathrm{th}$ point. %Note that although the grid is shown using the chordwise coordinate system ($x'$ and $y'$), the LES was performed using the streamwise coordinate system ($x$ and $y$), with the aerofoil present at an angle of incidence (e.g., see Fig. \ref{figLES_RhoContours_M735}). Thus, a different grid was generated for each incidence studied. In contrast to the approach of changing the freestream velocity vector's orientation to change incidence, this approach has the advantage of having a better resolution in the approximate direction along which the flow in the aerofoil's wake is expected to be oriented in. 
The colours indicate different blocks and the aerofoil itself, while the green symbols indicate the tripping locations. The grid refinement downstream of the trip was determined such that the turbulent boundary layers are appropriately captured \textit{a posteriori}, with $\Delta \xi^+ < 15$ and $\Delta \eta^+ < 1$. The aerofoil is extruded in the spanwise direction with a uniform grid spacing of $5\times10^{-4}$ and $1\times 10^{-3}$ for the cases of forced and free transition, respectively, with $\Delta z^+ \leq 10$. The span is chosen for both cases as $5\%$ of the aerofoil chord based on previous studies \citep{Zauner2020}. For the forced-transition case, the grid contains approximately 200 million points (one block of $1835\times 550\times 100$ points and two blocks of $799\times 564\times 100$). Grid-convergence studies were carried out to confirm that mean-flow and buffet characteristics are not sensitive to the grid spacing used, as demonstrated in Appendix~\ref{AppLESGridConv}. We also note that for laminar buffet, a validation of the LES with DNS and a check on the adequacy of the domain extent have been reported previously (see \citep{Zauner2020a} and \S2.2.1 in \citep{Moise2022}).

%\subsection{RANS simulations}

The two-dimensional domain for the RANS-level simulations extends to a circular far field boundary at a distance of 100 chord lengths discretised with approximately 65,000 dual-cells. The hybrid primary mesh contains approximately 26,000 quadrilateral elements in the quasi-structured near-wall region and 76,000 triangular elements elsewhere, as indicated in Fig.~\ref{figGrid}\textit{b}. The first wall-normal spacing gives a value of less than one in wall units. In the shock region a typical streamwise spacing of $0.5\%$ chord was used. The wake region was especially refined to ensure clear resolution of the wake mode.

%%%%%%%%%%%%%%%%%%%%%%%%%%%%%%%%%%%%%%%%
\section{Simulation results} % alternative title: Conventional unsteady analysis (I don't prefer this)
\label{secResults}
%%%%%%%%%%%%%%%%%%%%%%%%%%%%%%%%%%%%%%%%
The results from the simulations are presented in this section, while those from SPOD, GLSA and resolvent analysis are discussed later in Sec.~\ref{secDecomNRecon}. Flow features observed in the LES are discussed first before comparing with those from steady RANS and URANS simulations. A summary of all the main results from unsteady simulations of the various cases considered are highlighted in Table \ref{tab:mainResults}. 

%%%%%%%%%%%%%%%%%%%%%%%%%%%%%%%%%%%%%%%%
\subsection{LES results}
\label{subSecLESResults}
%%%%%%%%%%%%%%%%%%%%%%%%%%%%%%%%%%%%%%%%

%%%%%%%%%%%%%%%%%%%%%%%%%%%%%%%%%%%%%%%%
\subsubsection{Forcing boundary layer transition}
\label{subSecQCriterion}
%%%%%%%%%%%%%%%%%%%%%%%%%%%%%%%%%%%%%%%%

\begin{figure}[t]
	\centering
	\includegraphics[trim={4cm 5cm 4cm 6cm},clip,width=0.8\columnwidth]{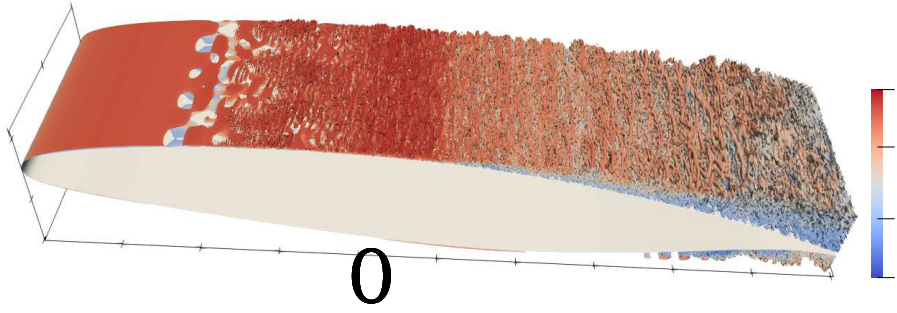}
	\caption{Isosurfaces based on $\boldsymbol{Q}-$criterion and coloured by streamwise velocity at an arbitrary time instant for case A5M735S (LES). The spanwise width has been enlarged for visualisation purposes.}
	\label{figLES_QCriterion}
\end{figure}

The effect of forcing transition on the suction surface in the LES is shown in Fig.~\ref{figLES_QCriterion} using isosurfaces of $Q_\mathrm{inc} = ||\boldsymbol{\varOmega}||_F^2 - ||\boldsymbol{S}||_F^2 = 100$ for a representative case (A5M735S, see Table \ref{tab:mainResults}), highlighting regions where rotation dominates over shear strain. This version of the $Q$-criterion \citep{Hunt1988} explicitly requires $||\boldsymbol{\varOmega}||_F^2 > ||\boldsymbol{S}||_F^2$. Note that the spanwise width is scaled by a factor of four for visualisation purposes. The presence of vortices of a wide range of length scales downstream of the trip location ($x_{ts} = 0.2$) suggests that by $x \approx 0.3$ the boundary layer is fully turbulent. The isosurfaces in the figure are coloured by the local streamwise velocity. The shock position can be inferred from the sudden reduction in velocity at $x \approx 0.5$ (also shown later, \textit{e.g.}, in Fig.~\ref{figLES_RhoContours_M735}) implying that the shock wave is located well downstream of transition. %separation (and the associated shock wave structure) is well downstream of transition.
%boundary layer transition is triggered well upstream of the shock  foot.

%%%%%%%%%%%%%%%%%%%%%%%%%%%%%%%%%%%%%%%%
\subsubsection{Flow features at reference angle of attack}
\label{subSecRefCaseLES}
%%%%%%%%%%%%%%%%%%%%%%%%%%%%%%%%%%%%%%%%

\begin{figure}[t]
	\centering
	\includegraphics[width=.45\textwidth]{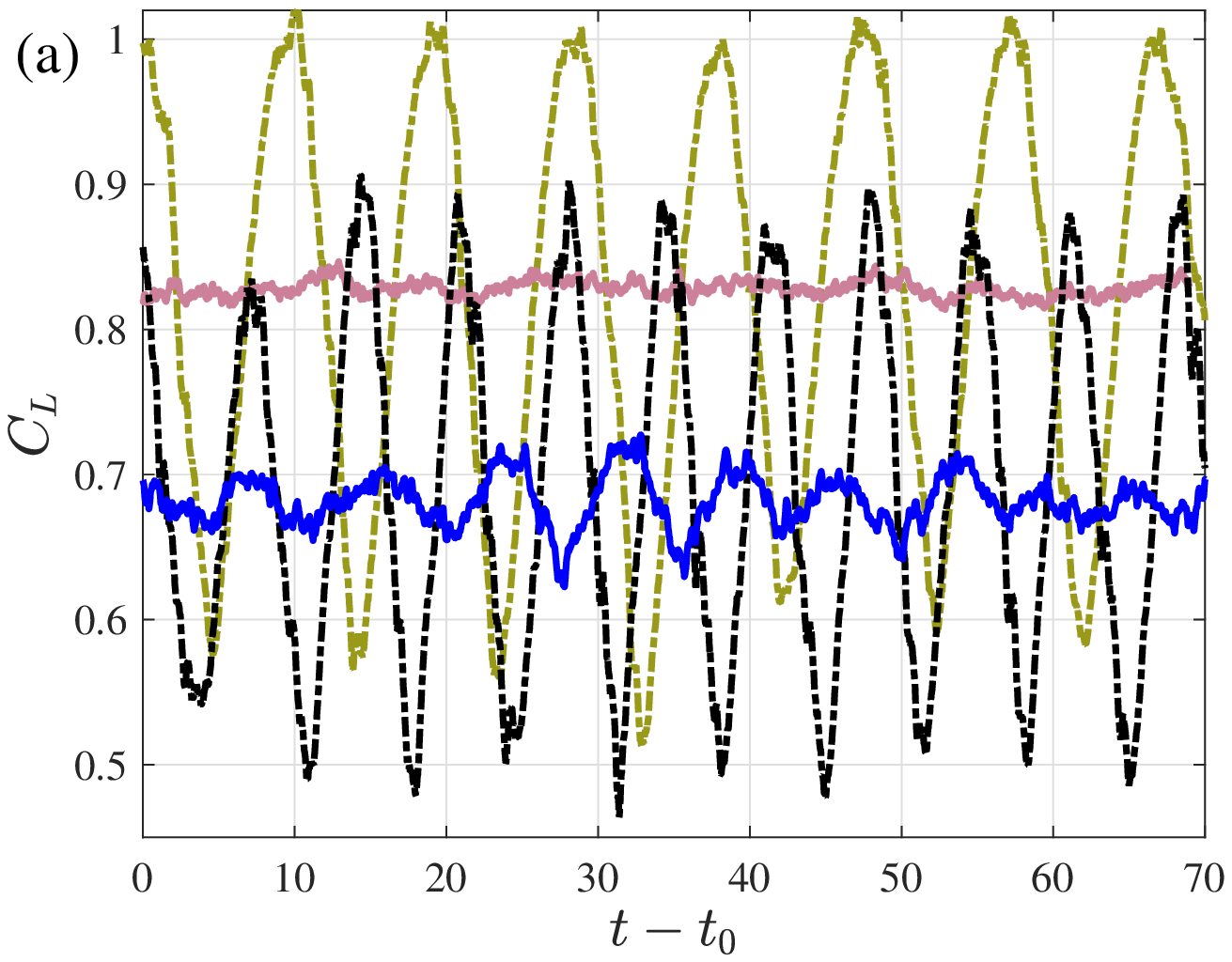}
	\includegraphics[width=.45\textwidth]{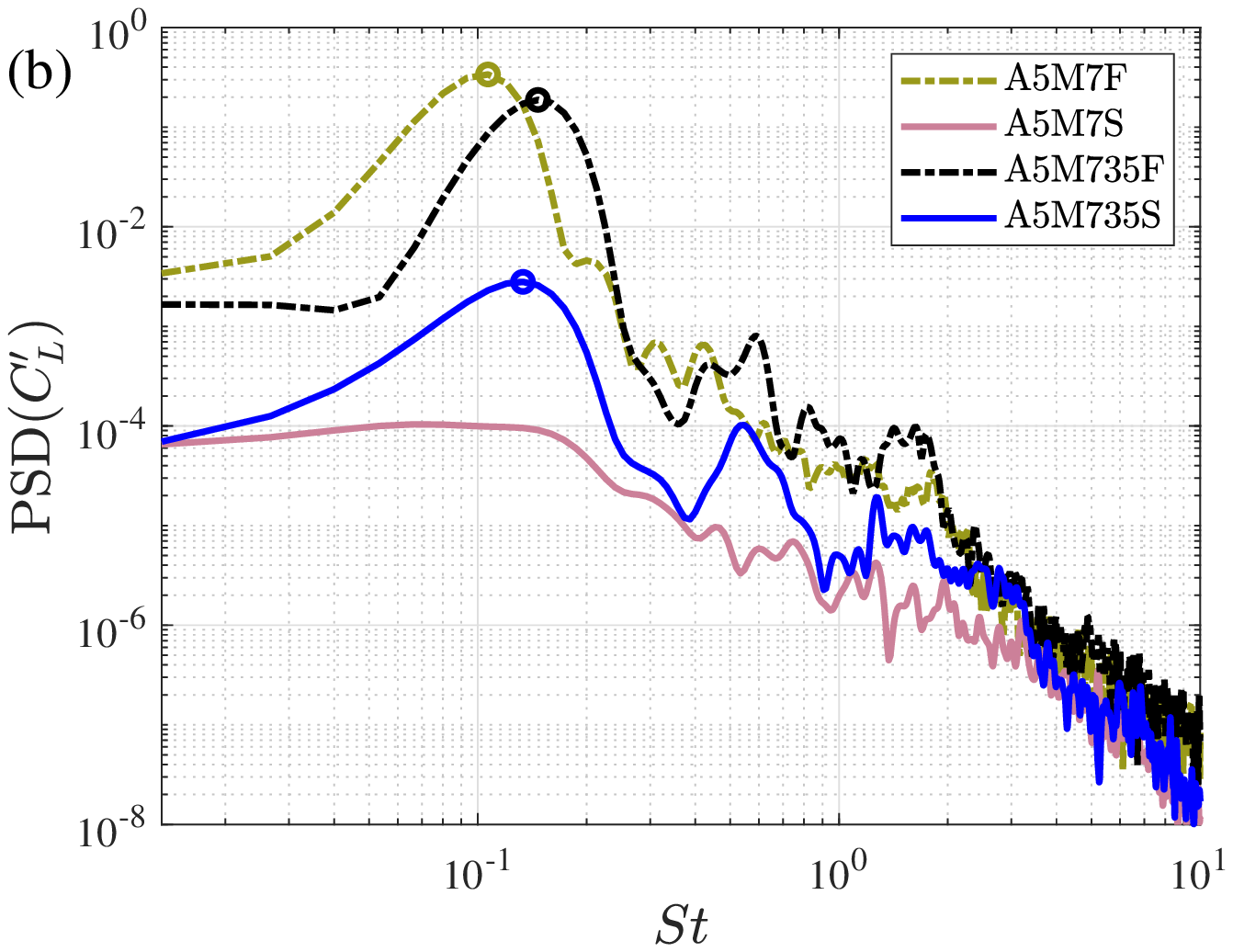}
	\caption{(a) Temporal variation of the lift coefficient past transients and (b) power spectral density of its fluctuating component for different $\boldsymbol{M}$ for $\boldsymbol{\alpha = 5^\circ}$ and $\boldsymbol{Re = 5\times10^5}$ (LES).}
	\label{figLES_ClVsT_PSD_M735}
\end{figure}

\begin{figure}[t]
	\centering
	\includegraphics[trim={0cm 1cm 0cm 2cm},clip,width=.45\textwidth]{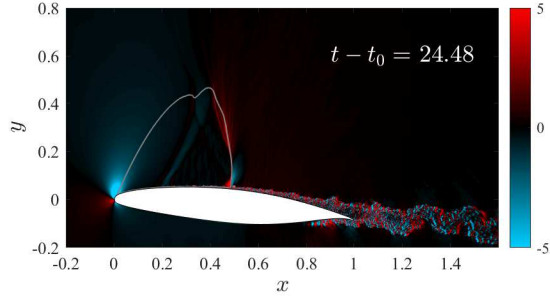}
	\caption{Streamwise density-gradient contours on the $\boldsymbol{z=0}$ plane at an arbitrary time instant for A5M7S (LES). The sonic line is shown using the gray curve.}
	\label{figLES_RhoContours_M7}
\end{figure}

Temporal variations of the lift coefficient, for times past transients ($t> t_0$) at $\alpha = 5^\circ$, are compared for two choices of $M$ and two different transition scenarios in Fig.~\ref{figLES_ClVsT_PSD_M735}\textit{a}. Case A5M7F has been reported in \citep{Moise2022}. Note that $t_0$ is chosen such that a local maximum of $C_L(t)$ occurs at $t = t_0$. The power spectral density (PSD) of the fluctuating component of $C_L$ is compared in Fig.~\ref{figLES_ClVsT_PSD_M735}\textit{b}. Strong lift oscillations, characteristic of buffet, can be seen for the free-transition cases with a peak observed in the PSD at $St_b = 0.11$ and $0.15$ for cases A5M7F and A5M735F, respectively (highlighted by circles). The oscillations are significantly subdued when transition is forced on the suction surface. For A5M7S, no clear lift oscillations or a peak in the PSD are discernible. With an increase in $M$, for case A5M735S, the PSD of the lift fluctuations shows a peak associated with buffet indicating that there is an emerging turbulent buffet at these conditions and that the onset $M$ is in the range (0.7,0.735). Comparing cases A5M735S and A5M735F, we see that the lift oscillations are more irregular. Furthermore, the PSD at the buffet frequency drops by almost two orders of magnitude for the former (\textit{i.e.}, turbulent buffet). Thus, in the parameter range studied, it appears that forcing boundary layer transition leads to a suppression or reduction in strength of buffet when compared to free transition. This suggests two possibilities for forced-transition cases: (i) buffet amplitudes at any flow condition (\textit{i.e.}, $M$, $\alpha$ and $Re$) might be reduced or (ii) buffet onset might be shifted to higher $\alpha$. The latter implies that at higher $\alpha$, there might be relatively higher buffet amplitudes for the forced-transition cases. This is explored later in Sec.~\ref{subSecDeepBuffet}. An important aspect seen here is that the turbulent buffet frequency ($St_b = 0.13$) remains approximately the same as that of laminar buffet ($St_b = 0.15$) at these conditions.
%, which corroborates with the proposition by \citet{Moise2022} that laminar and turbulent buffet have similar mechanisms. This should be contrasted with the bubble-breathing phenomenon reported by \citet{Dandois2018} which occurs at a frequency an order of magnitude higher ($St\approx 1$). 
It can also be inferred from the temporal variations in Fig.~\ref{figLES_ClVsT_PSD_M735}\textit{a} that the time-averaged~$C_L$ is approximately the same at a given~$M$ for both free- and forced-transition cases (see Table \ref{tab:mainResults} as well). Other interesting features are the bumps in the PSD at an intermediate frequency of $St \approx 0.5$ and at a higher frequency $St \approx 1$ for both cases at $M = 0.735$. These features will be discussed later alongside SPOD results.

%latter will be shown in Sec.~\ref{secSPOD} to occur due to wake modes reported by \citet{Moise2022}. 

\begin{figure}[!t]
	\centering
    \includegraphics[trim={0cm 1cm 0cm 2cm},clip,width=.45\textwidth]{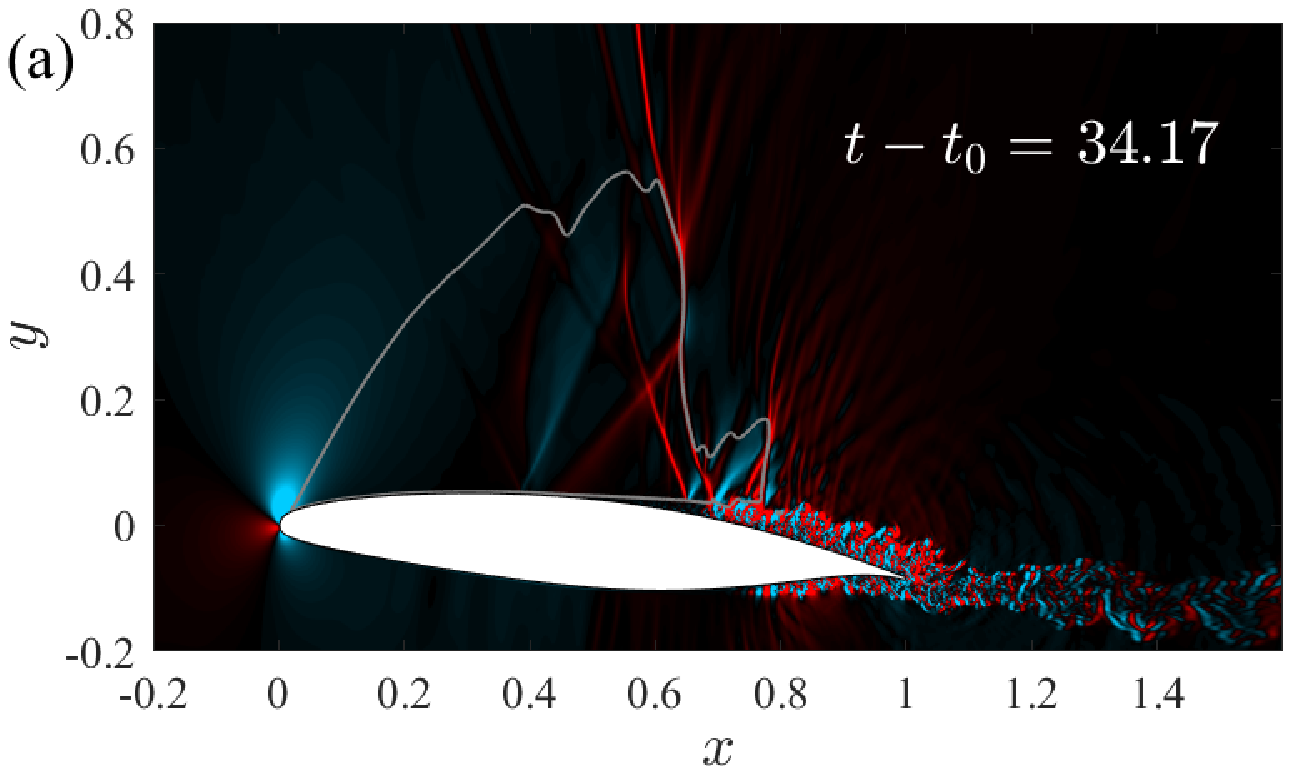}
	\includegraphics[trim={0cm 1cm 0cm 2cm},clip,width=.45\textwidth]{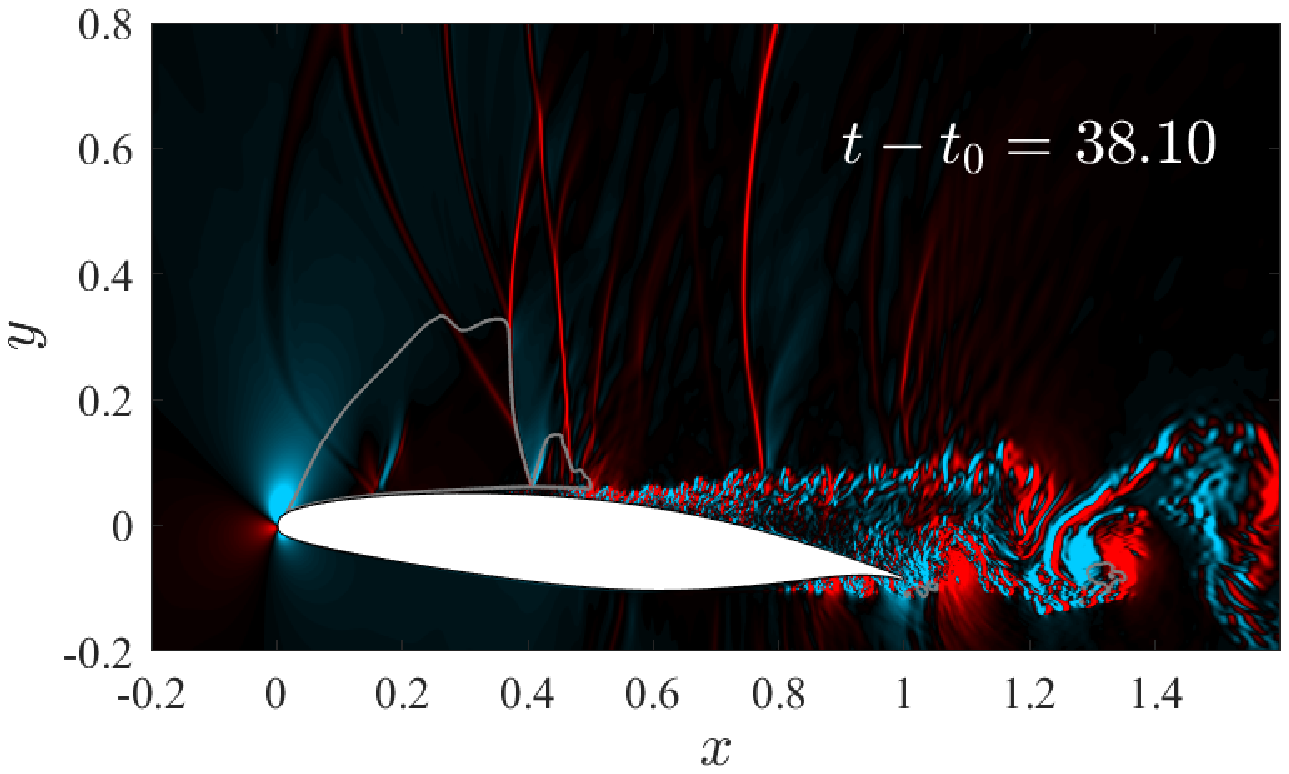}
	\includegraphics[trim={0cm 1cm 0cm 2cm},clip,width=.45\textwidth]{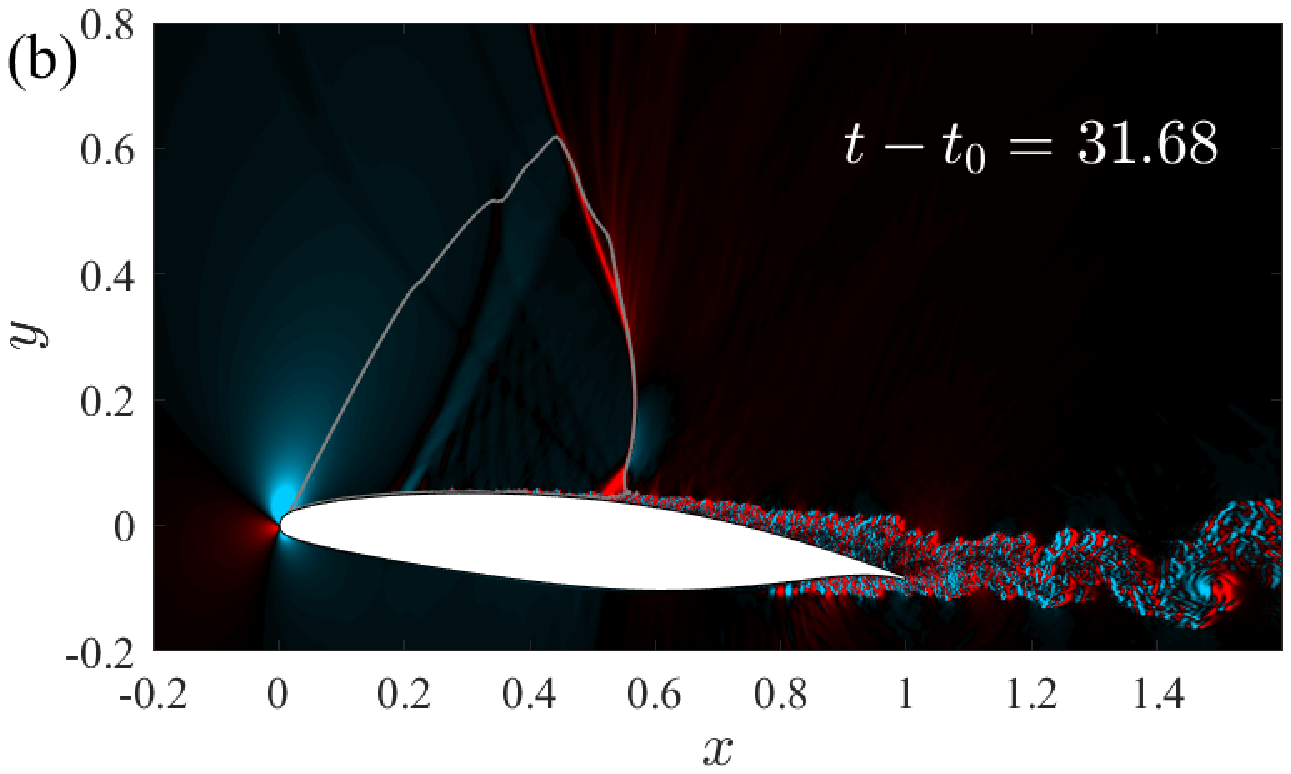}
	\includegraphics[trim={0cm 1cm 0cm 2cm},clip,width=.45\textwidth]{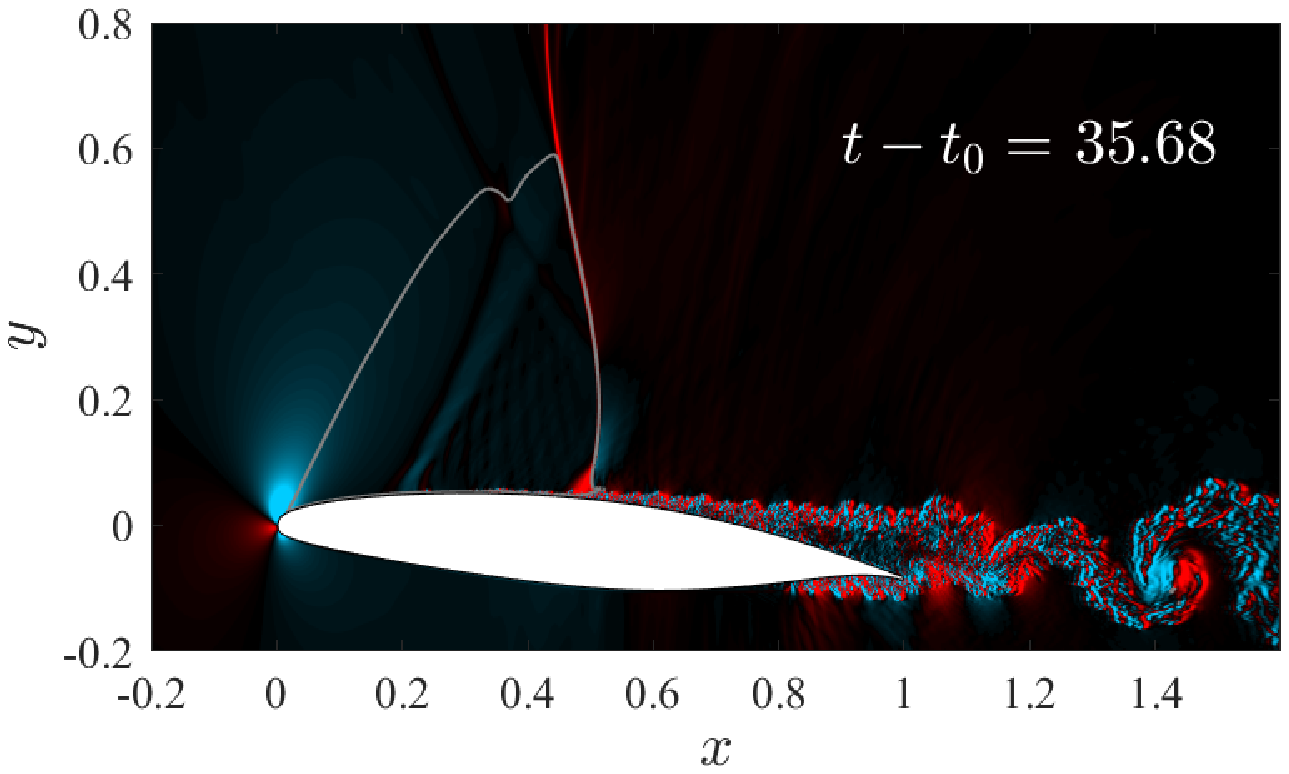}
	\includegraphics[trim={0cm 1cm 0cm 2cm},clip,width=.45\textwidth]{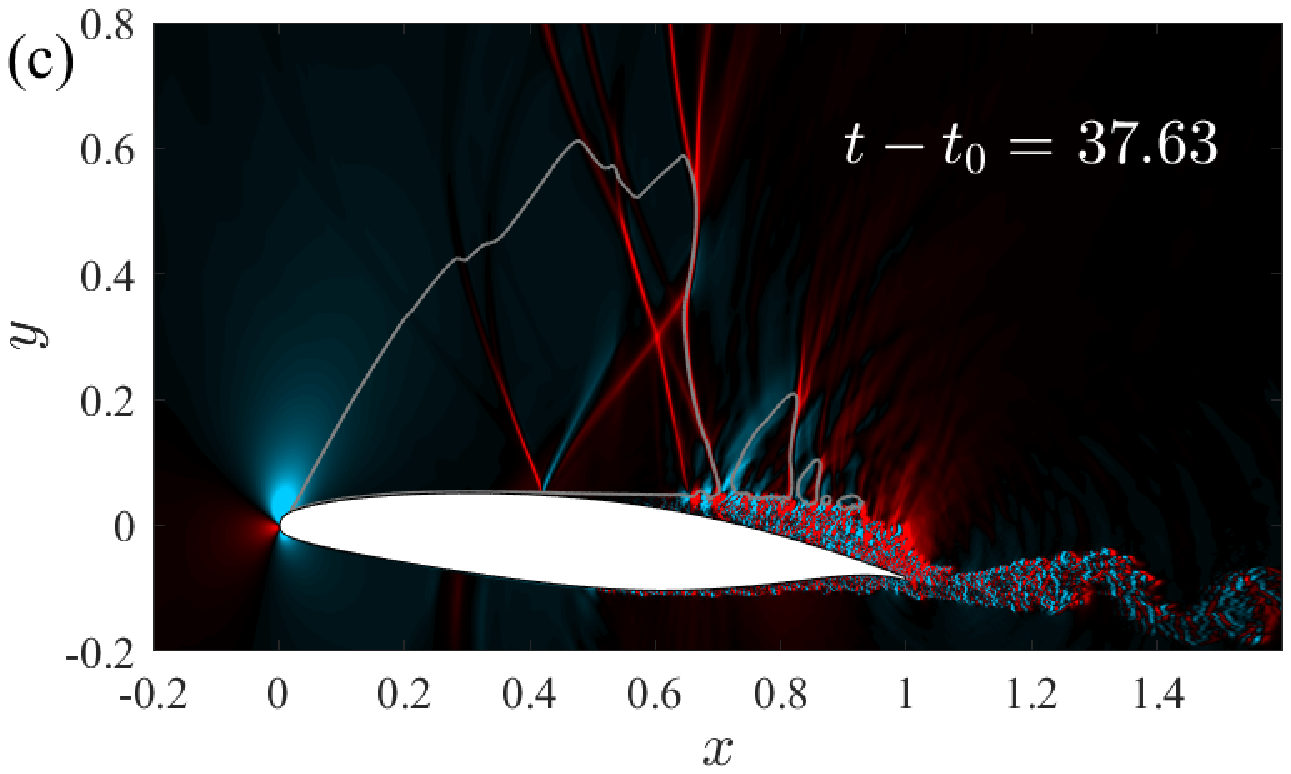}
	\includegraphics[trim={0cm 1cm 0cm 2cm},clip,width=.45\textwidth]{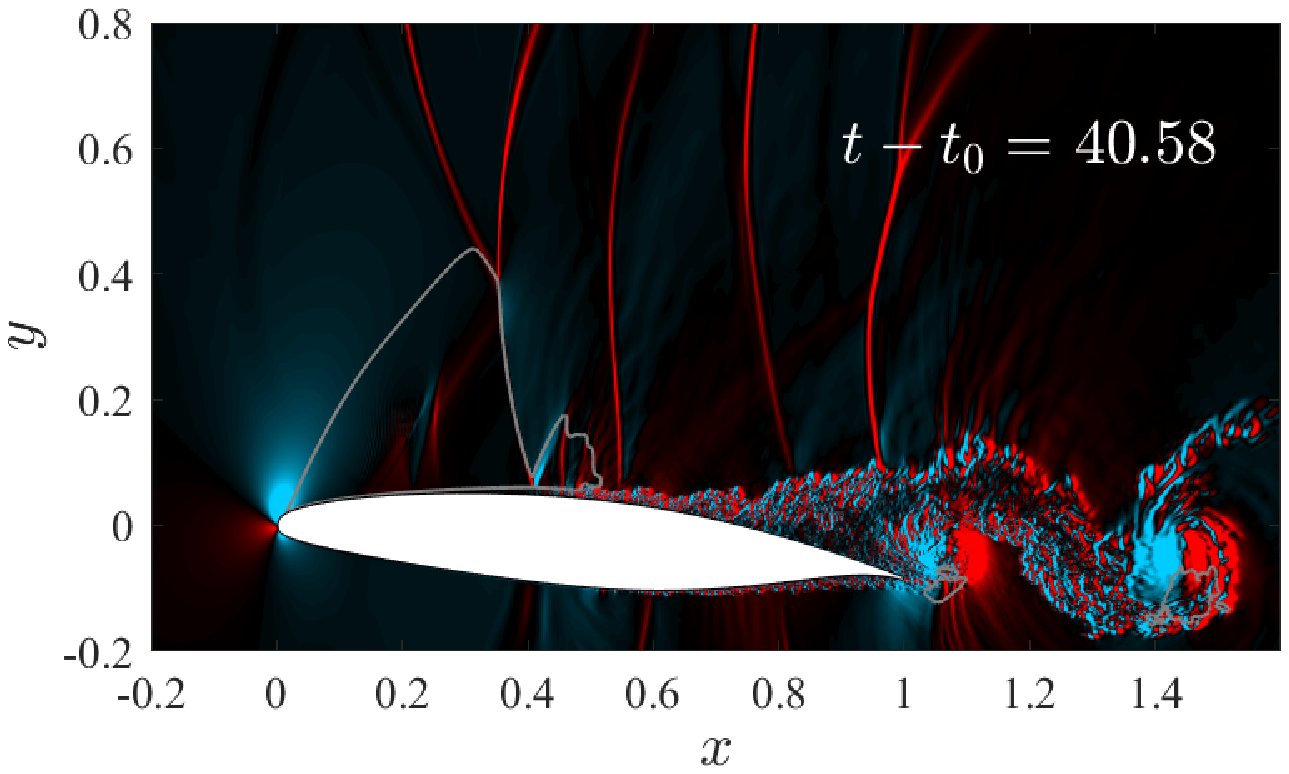}
	\includegraphics[trim={0cm 1cm 0cm 2cm},clip,width=.45\textwidth]{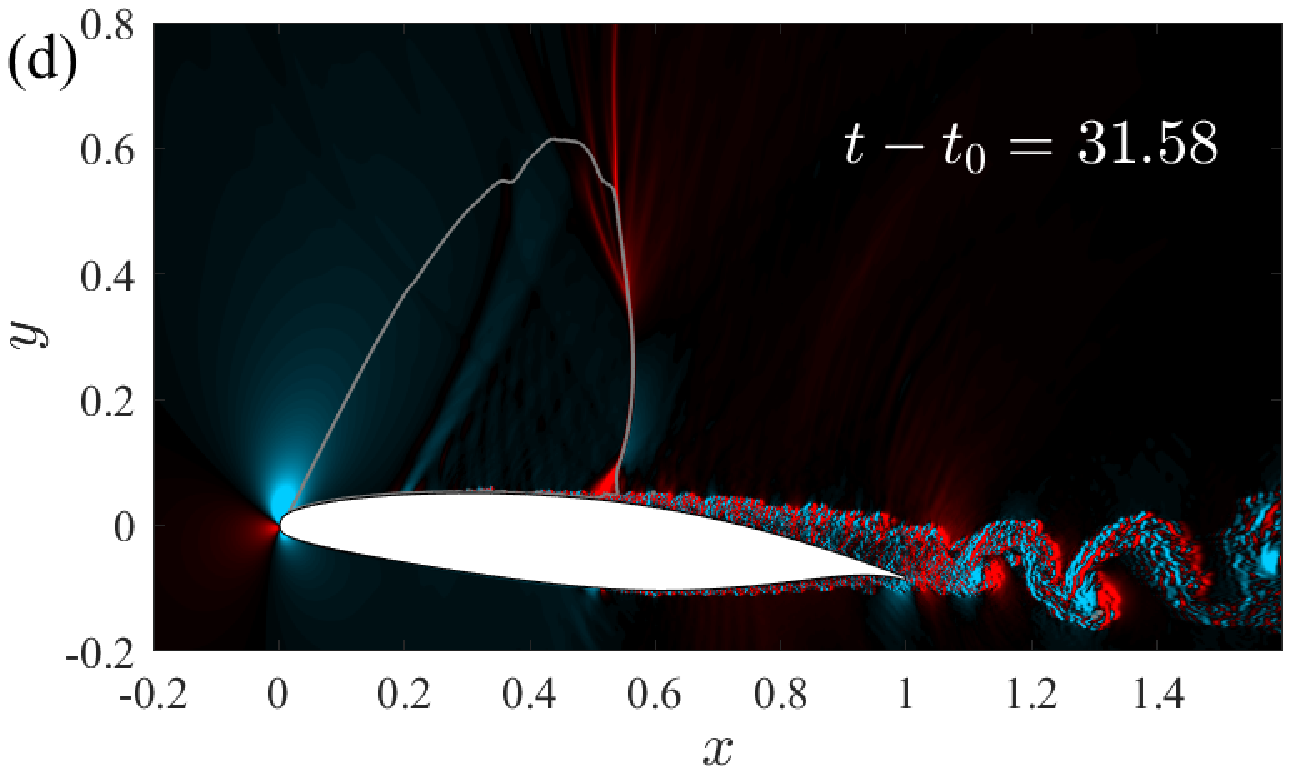}
	\includegraphics[trim={0cm 1cm 0cm 2cm},clip,width=.45\textwidth]{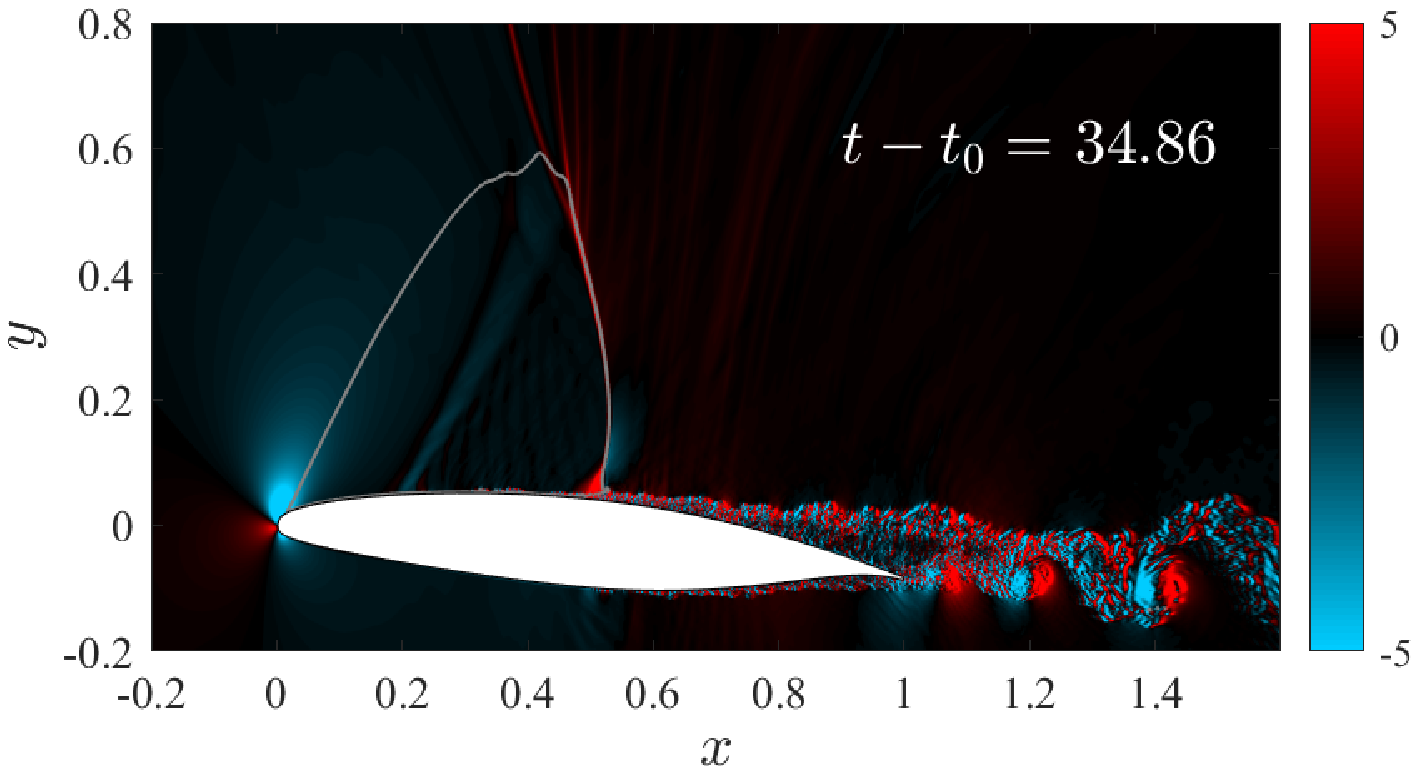}
	\caption{Streamwise density-gradient contours on the $\boldsymbol{z=0}$ plane at high- (left) and low- (right) lift phases for cases (a) A5M735F, (b) A5M735S, (c) A5M735P and (d) A5M735SP cases (LES).}
	\label{figLES_RhoContours_M735}
\end{figure}

Instantaneous contours of the streamwise density gradient on the $z = 0$ plane are plotted in Figs.~\ref{figLES_RhoContours_M7} and~\ref{figLES_RhoContours_M735}. The former shows the flow features at an arbitrary instant for the buffet-free case A5M7S. The latter shows the features of cases A5M735F and A5M735S at approximate times when the lift achieves a local maximum (left) or minimum (right). Animations showing the dynamics for a few buffet cycles are also provided as Supplementary Material. For case A5M735F, the envelope of the supersonic region, delineated by the sonic line (gray curve), varies significantly with time and multiple shock waves are observed, characteristic of laminar buffet at moderate $Re$ (see, \textit{e.g.}, older experiments in \citep{Ackeret1947}, more recent experiments at ONERA\footnote{https://doi.org/10.13140/RG.2.2.30817.58725} and LES in \citep{Caleb2022}). Interestingly, multiple shock waves were also observed in steady inviscid calculations by \cite{Caleb2022}. By contrast, only a single shock wave is present for case A5M735S and the streamwise excursions of the shock are relatively low. This form of buffet, with a single oscillating shock wave, is commonly reported in the literature, \textit{e.g.}, \citep{Crouch2009}. %The difference in shock wave structure between the free- and forced-transition cases might be due to the effect of boundary layer curvature on the potential flow, with the strong streamwise variation of the laminar boundary layer thickness possibly leading to multiple shock waves. % In inviscid simulations , the aerofoil curvature might instead induce the same. 
The reduced buffet amplitude at moderate $Re$ when transition is forced indicates that buffet is sensitive to the boundary layer characteristics upstream of the shock wave, particularly when switching from a laminar to a turbulent state. These findings align with the GLSA results in \citep{Garbaruk2021}, where a buffet mode becomes unstable at lower $\alpha$ when the boundary layer is tripped farther downstream. This suggests that for a constant $\alpha$, buffet intensity would increase with an increase in streamwise extent of the laminar boundary layer. %Finally, comparing the aerofoil wake for both free- and forced-transition cases at the high- and low-lift phases, a stronger vortical organisation resembling the von K\'arm\'an vortex street is present in the latter phase when the separation point is further upstream and the wake is relatively thicker.

%%%%%%%%%%%%%%%%%%%%%%%%%%%%%%%%%%%%%%%%
\subsubsection{Effect of tripping boundary layers on different surfaces of the aerofoil}
\label{secLESTripDiffSides}
%%%%%%%%%%%%%%%%%%%%%%%%%%%%%%%%%%%%%%%%

\begin{figure}[t]
	\centering
	\includegraphics[width=.45\textwidth]{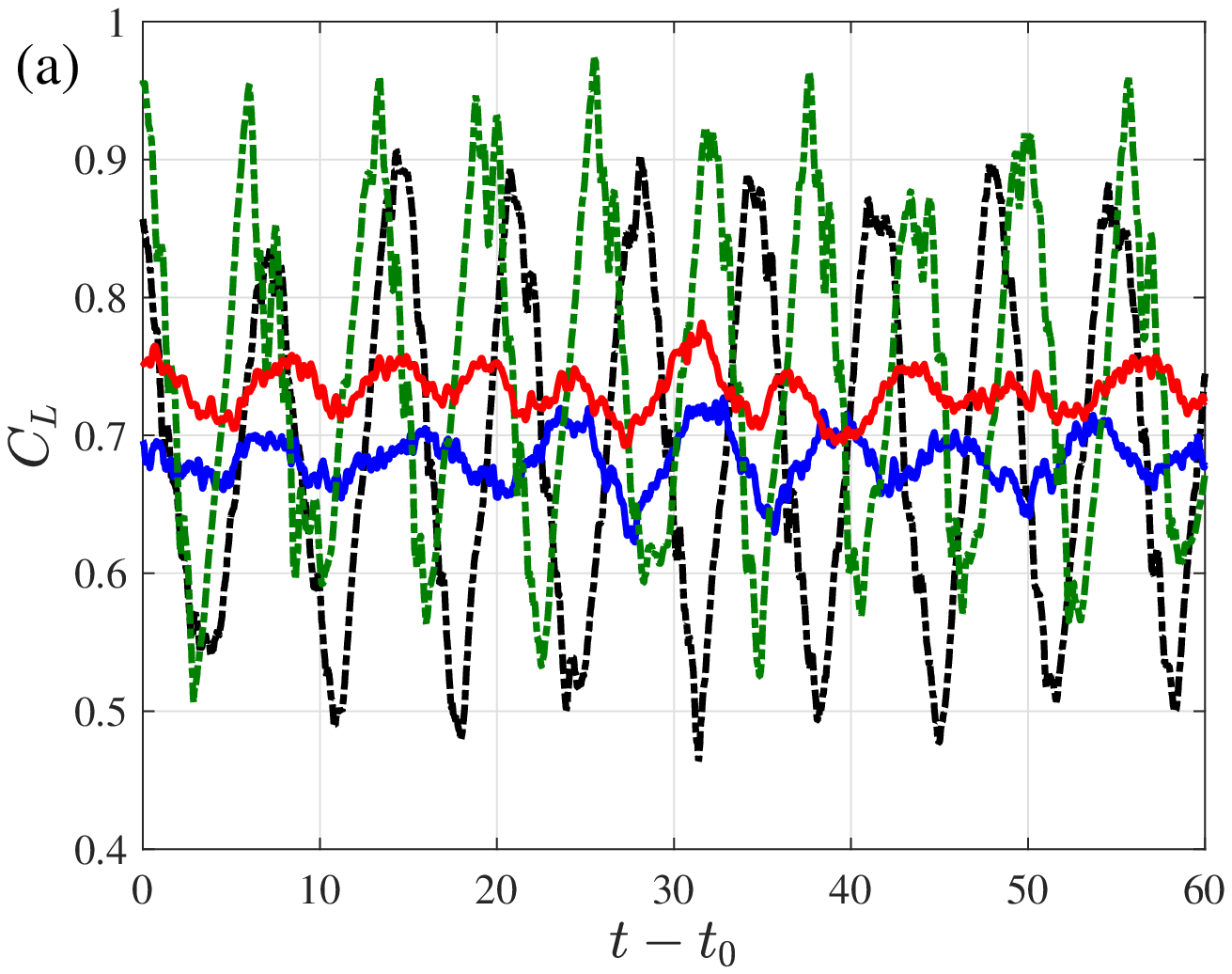}
	\includegraphics[width=.45\textwidth]{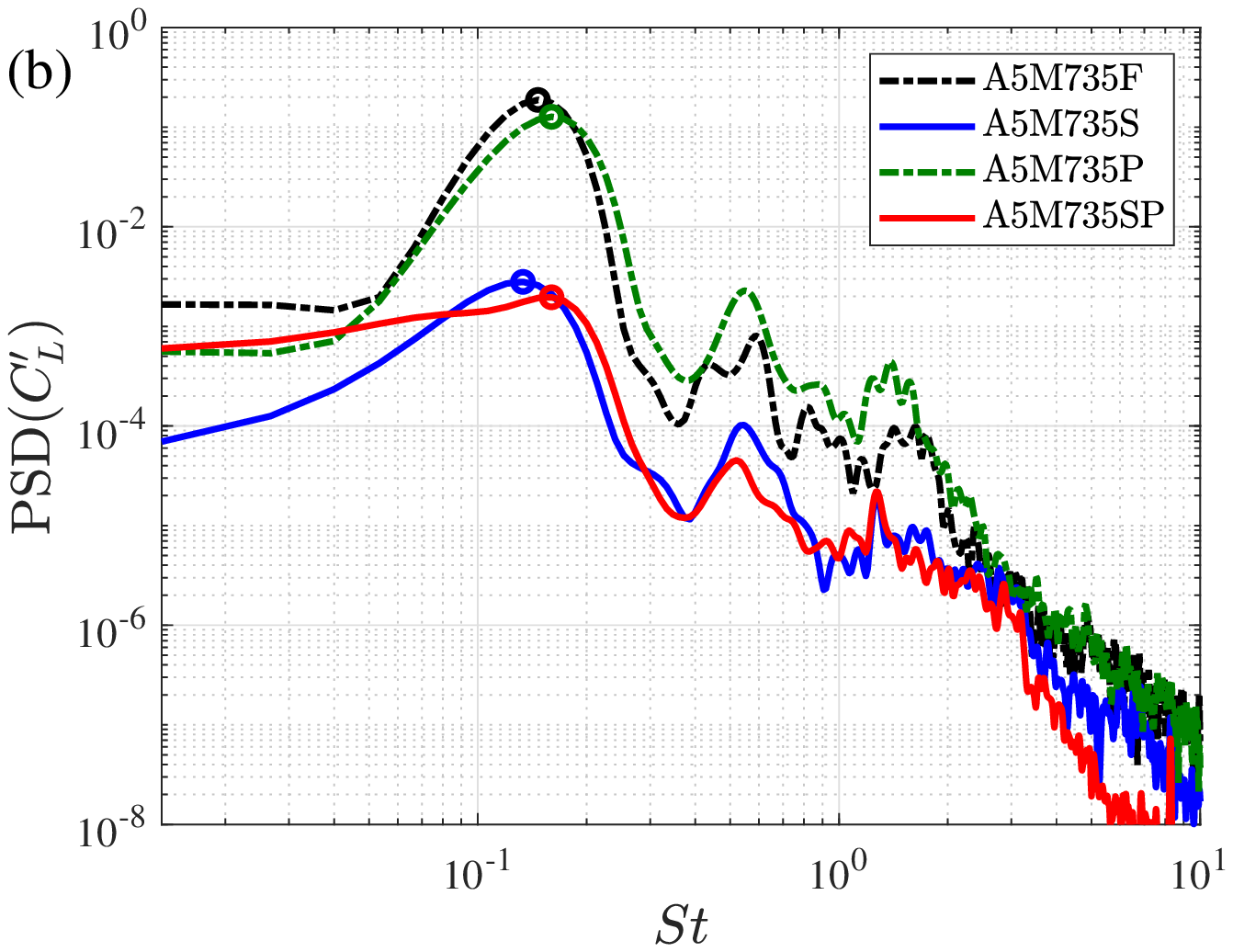}
	\caption{(a) Temporal variation of the lift coefficient past transients and (b) power spectral density of its fluctuating component for different tripping conditions (LES).}
	\label{figLESTripDiffSidesCL}
\end{figure}

The effect of tripping the boundary layer developing on the pressure and suction surfaces is considered here. Dynamical features of four possible cases are compared for parameter settings of $M = 0.735$, $\alpha = 5^\circ$ and $Re = 5\times 10^5$ (\textit{i.e.}, A5M735F, A5M735S, A5M735P and A5M735SP). In addition to the free- and forced-transition cases (suction-side trip) discussed in the previous section, we consider two other cases of tripping the boundary layer only on the pressure side or on both surfaces. The variation of the lift coefficient past transients and the PSD of its fluctuating component are compared for all cases in Fig.~\ref{figLESTripDiffSidesCL}. The buffet frequency is consistent for all cases (see Table~\ref{tab:mainResults}). The value of the PSD of the lift fluctuation at the buffet frequency is similar for cases A5M735F and A5M735P. Cases A5M735S and A5M735SP have lower values that are comparable to each other. Similarities between the cases can also be observed by comparing the high- and low-lift features shown using contours of streamwise density gradient in Fig.~\ref{figLES_RhoContours_M735}.

\begin{figure}[t]
	\centering
	\includegraphics[width=.495\columnwidth]{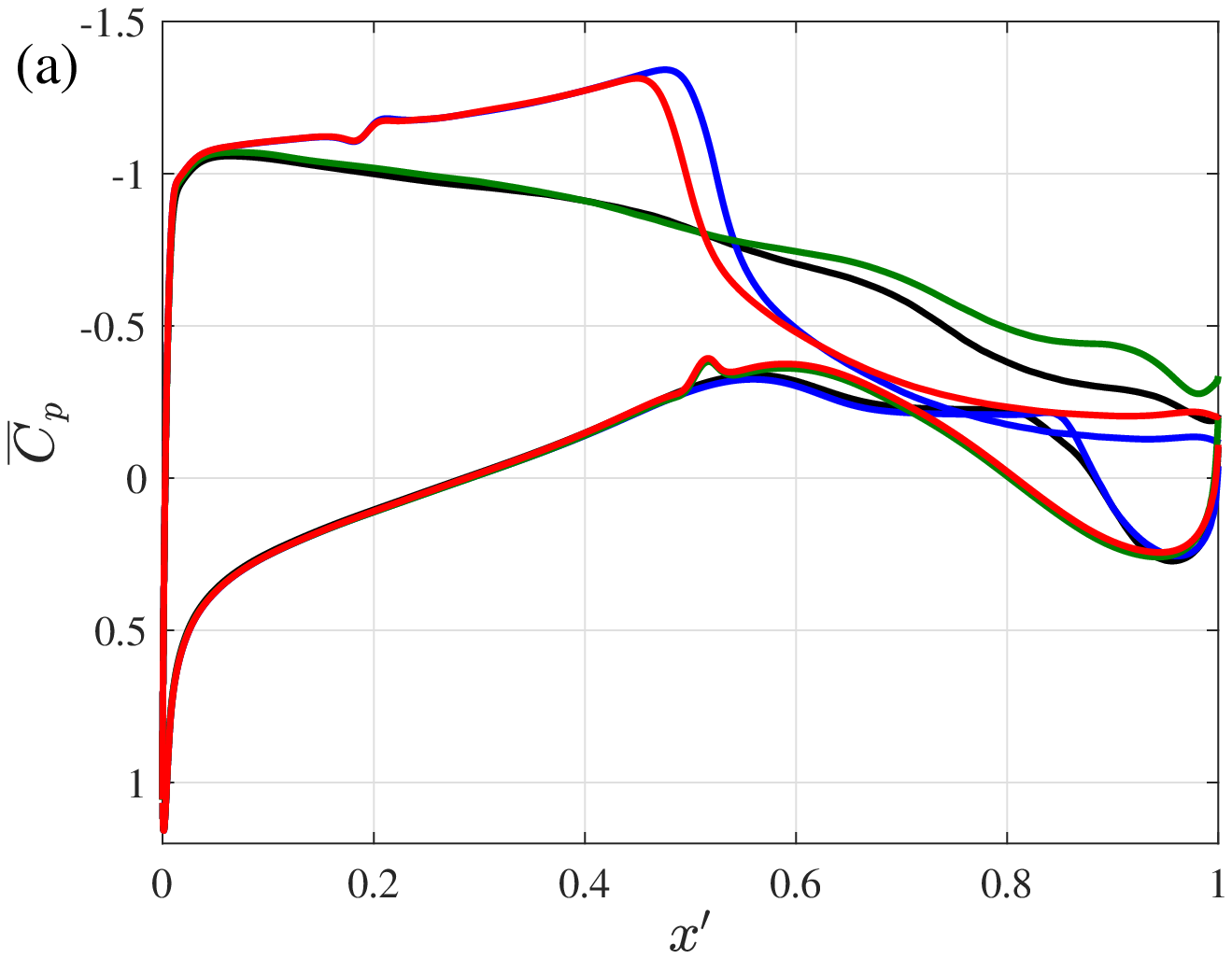}
	\includegraphics[width=.495\columnwidth]{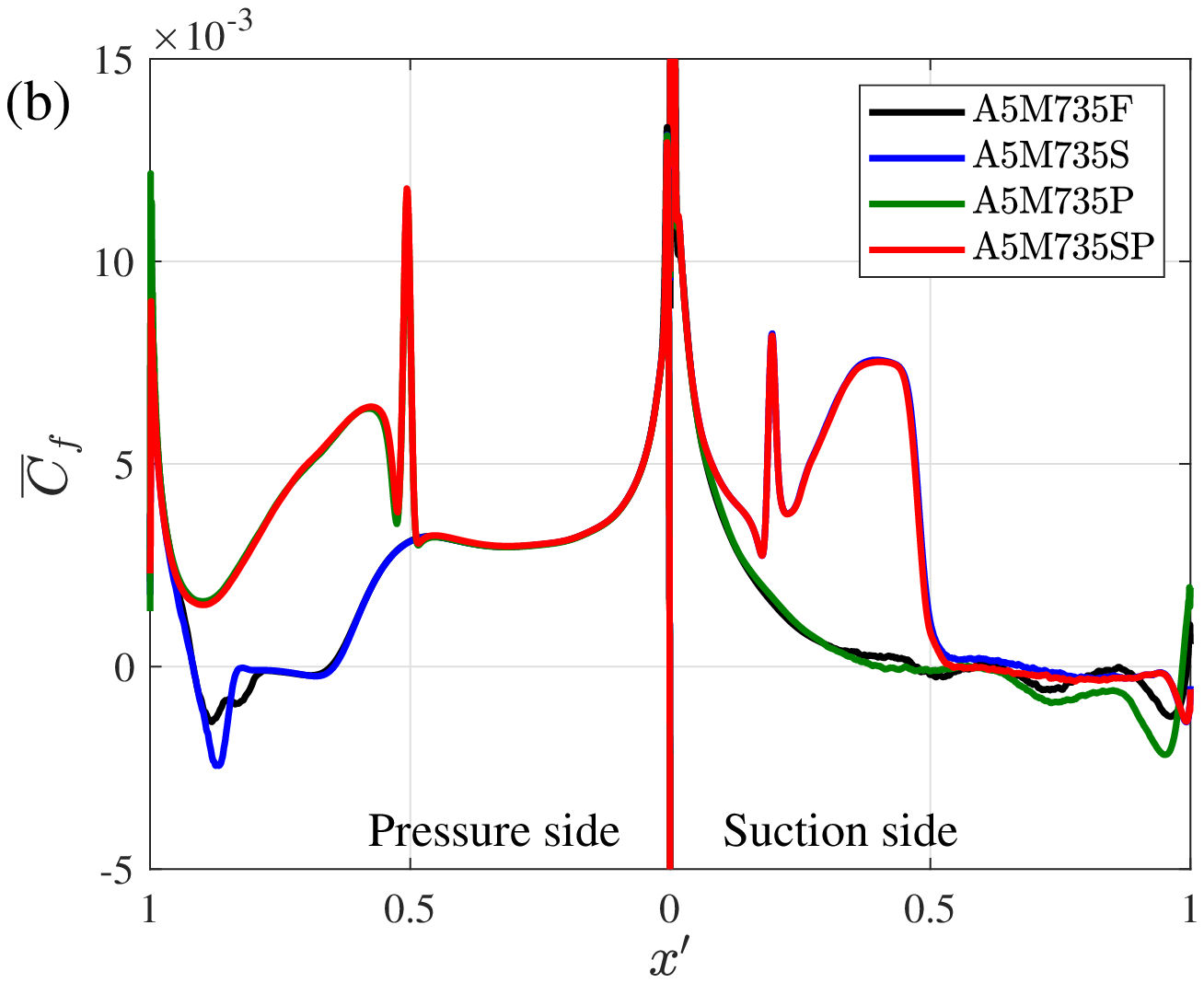}
	\caption{Comparison of (a) pressure coefficient and (b) skin-friction coefficient for different tripping conditions (LES).}
	\label{figLESTripDiffSidesCp}
\end{figure}

The chordwise variation of the mean pressure coefficient $\overline{C}_p$ is compared for the four cases in Fig.~\ref{figLESTripDiffSidesCp}a. On the suction side, cases A5M735F and A5M735P show similar variations, while case A5M735S resembles A5M735SP. On the pressure side, the relations are interchanged. These similarities are clearly due to the effect of tripping. Some differences are observed, including the mean shock position for cases A5M735S and A5M735SP and the reduced pressure near the trailing edge on the suction side whenever the pressure side is tripped (compare case A5M735P with A5M735F, and case A5M735S with A5M735SP). These results indicate that the tripping of the pressure side contributes to global variations in the mean flow field. However, as noted above in the discussions regarding the PSD, there are no significant differences in buffet amplitude or frequency due to this. 
The chordwise variation of $\overline{C}_f$ is compared in Fig.~\ref{figLESTripDiffSidesCp}b. 
The tripping on the suction side causes a relatively strong increase in skin friction in the region $0.2 \leq x' \leq 0.4$, characteristic of a transition to turbulence. % as compared to when a trip is not applied. 
A similar feature can also be observed when the pressure side is tripped (see $0.5 \leq x' \leq 0.5$). From these results, it is concluded that buffet at these incidence angles is not significantly affected by the transition characteristics of the boundary layer developing on the pressure side. This was further confirmed by comparing the SPOD modes at the buffet frequency (see Sec.~\ref{secSPOD}) and also aligns with the results reported in \citet{Garbaruk2021} for turbulent buffet for various trip locations.  

%Thus, the focus of the following section will only be on cases where the trip is employed only on the suction surface.

%%%%%%%%%%%%%%%%%%%%%%%%%%%%%%%%%%%%%%%%
\subsubsection{Flow features at other incidence angles}
\label{subSecDeepBuffet}
%%%%%%%%%%%%%%%%%%%%%%%%%%%%%%%%%%%%%%%%

Laminar and turbulent buffet for other values of the key flow parameters are reported in this section. To examine the influence of incidence angle and Mach number, cases A7M7F, A7M7S, A0M8F and A0M8S were chosen. Case A7M7S was examined in comparison to case A5M735 to explore whether forcing transition always leads to subdued buffet or whether it increases onset values. Case A0M8S was examined because the buffet that occurs for free-transition conditions (\textit{i.e.}, A0M8F) has been shown by \citet{Moise2022} to include shock waves exhibiting periodic fore-aft motion on both sides of the aerofoil, as would be seen for symmetric airfoils at zero angle of attack and classified as ``Type I" buffet in \citep{Giannelis2017}. Note that other than cases A0M8F and A0M8S, all cases discussed in this study exhibit shock wave motion only on the suction side and can be classified as ``Type II" buffet.

\begin{figure}[t]
	\centering
	\includegraphics[width=.45\textwidth]{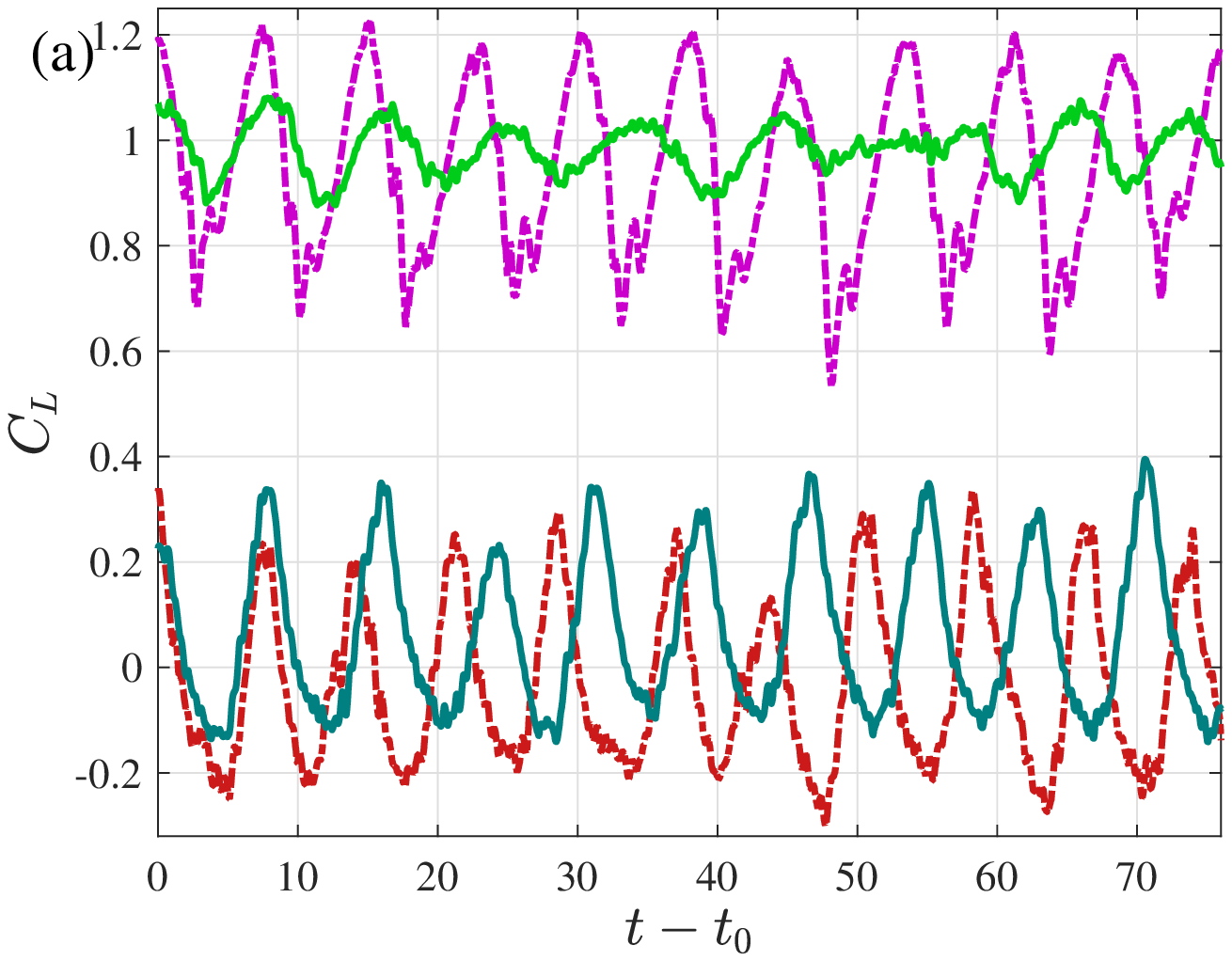}
	\includegraphics[width=.45\textwidth]{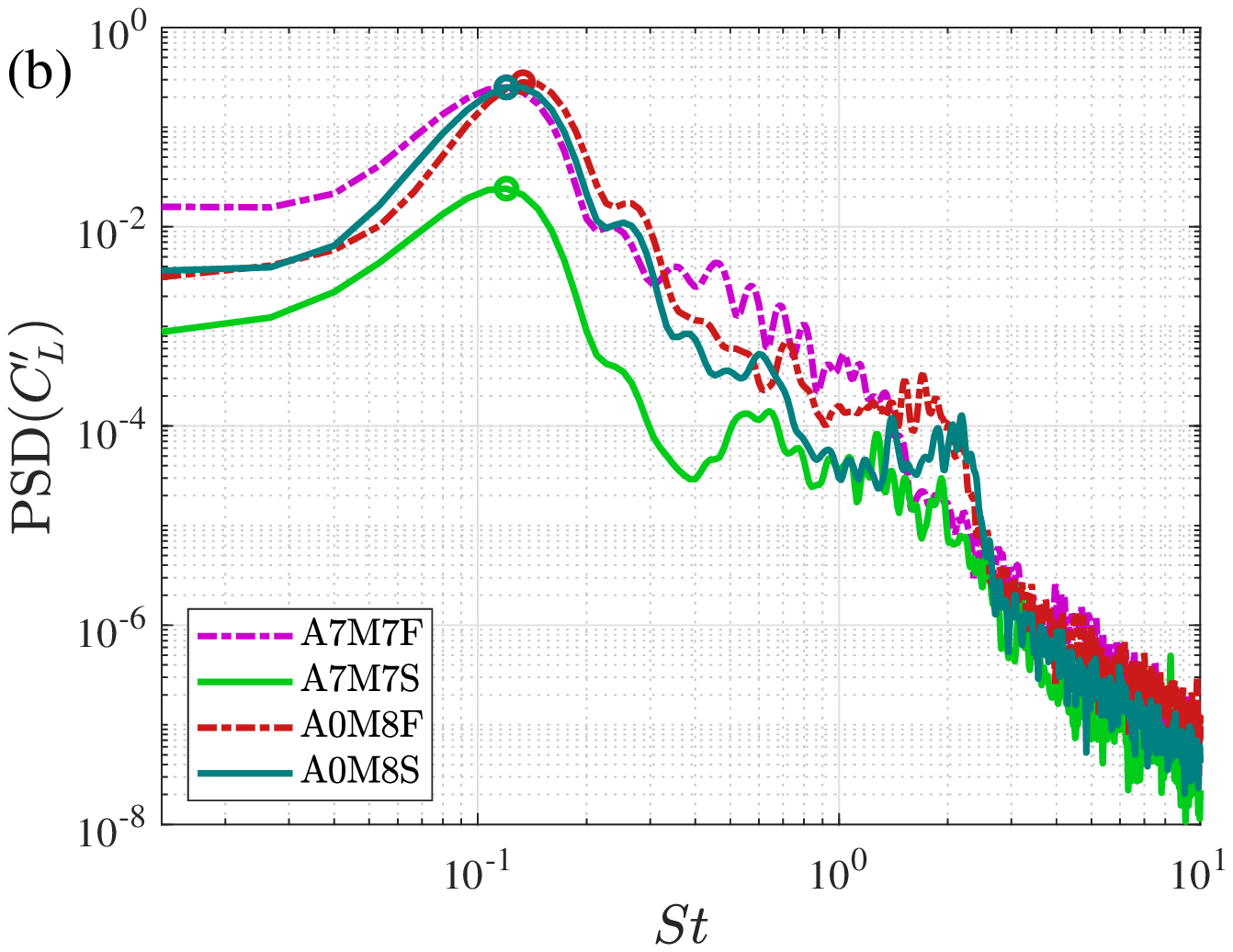}
	\caption{(a) Temporal variation of the lift coefficient past transients and (b) power spectral density of its fluctuating component (LES).}
	\label{figLES_ClVsT_PSD_StrongBuff}
\end{figure}

Temporal variation of the lift coefficient is compared for the cases mentioned above in Fig.~\ref{figLES_ClVsT_PSD_StrongBuff}\textit{a} while the power spectra are compared in Fig.~\ref{figLES_ClVsT_PSD_StrongBuff}\textit{b}. It is evident that all cases exhibit buffet. There are no significant differences between the free- and forced-transition cases of A0M8F and A0M8S. For $\alpha = 7^\circ$ and $M = 0.7$, tripping the boundary layer leads to more irregular oscillations of weaker amplitude. The latter can also be inferred from the PSD of lift fluctuations shown in Fig.~\ref{figLES_ClVsT_PSD_StrongBuff}\textit{b}. Note that such differences are also observed for the case for A5M735 (see Fig.~\ref{figLES_ClVsT_PSD_M735}), suggesting that boundary layer tripping leads to reduced buffet amplitude, although frequencies remain similar. To further develop this discussion point, simulations were carried out at an ever higher incidence of $\alpha = 8^\circ$ at $M = 0.7$ (not shown). It was found that the flow stalls for free-transition conditions and buffet-like features are absent, whereas when the transition on the suction side is forced, there is large-amplitude motion of the shock wave. This would support an alternative conclusion that the buffet envelope is merely shifted by forcing transition. However, we caution that for free-transition conditions, the current grids are not well suited to capture strong leading edge flow separation, while for the forced-transition case, the shock wave was found to almost reach $x_{ts}$ during parts of the buffet cycle, so the interaction becomes partly transitional. 

\begin{figure}[t]
	\centering
	\includegraphics[trim={0cm 1cm 0cm 2cm},clip,width=.45\textwidth]{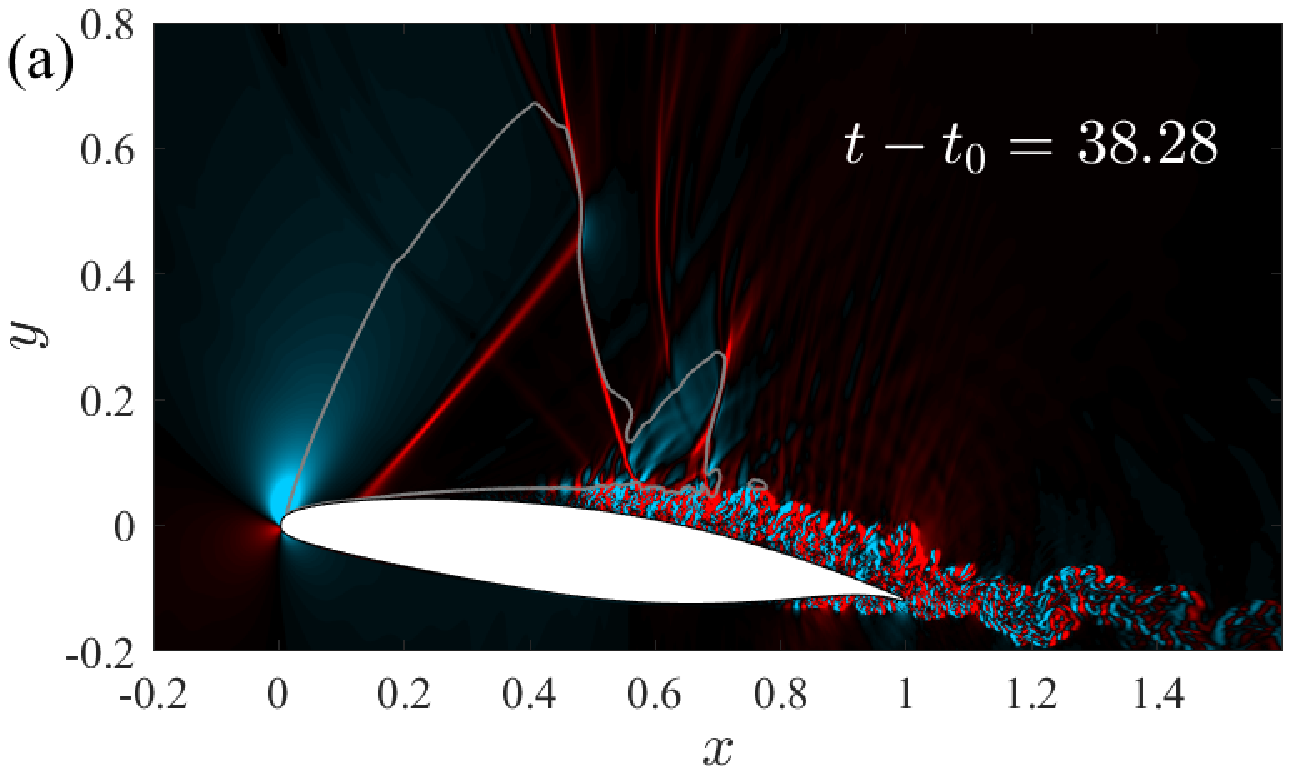}
	\includegraphics[trim={0cm 1cm 0cm 2cm},clip,width=.45\textwidth]{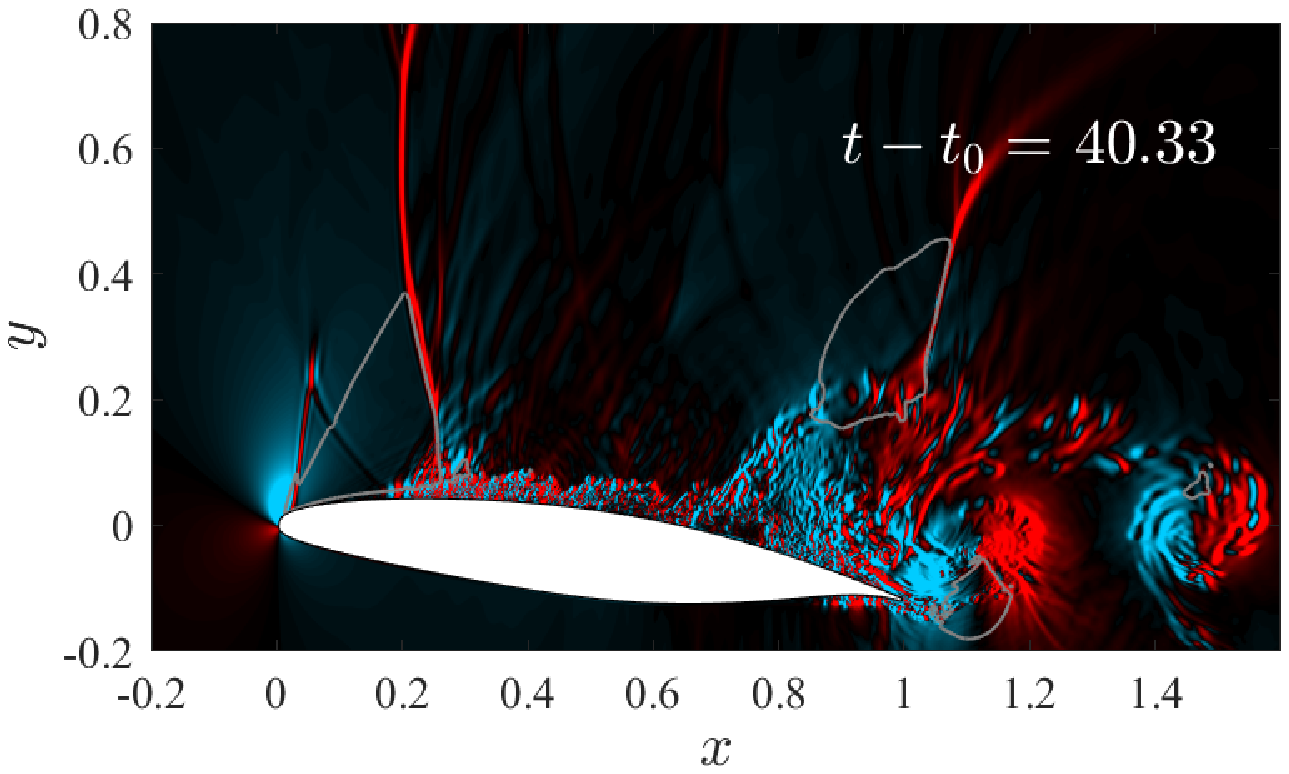}
	\includegraphics[trim={0cm 1cm 0cm 2cm},clip,width=.45\textwidth]{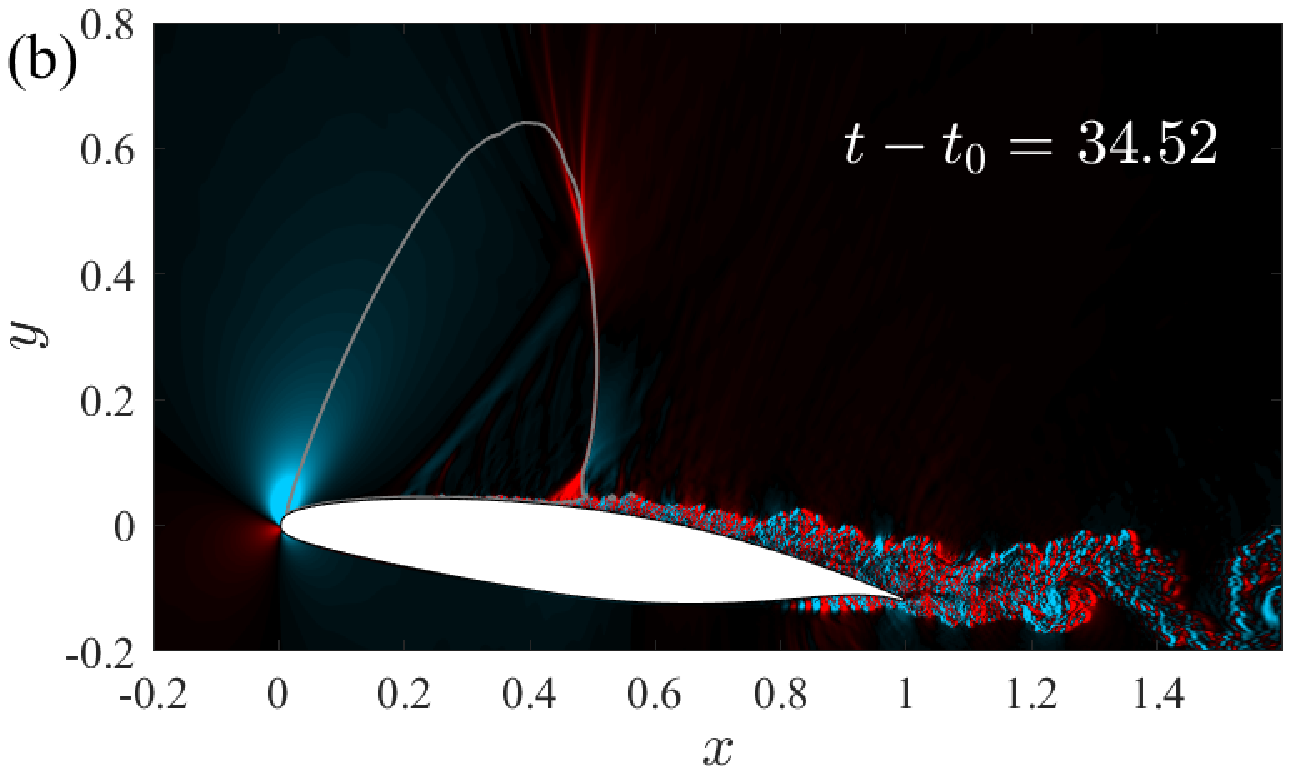}
	\includegraphics[trim={0cm 1cm 0cm 2cm},clip,width=.45\textwidth]{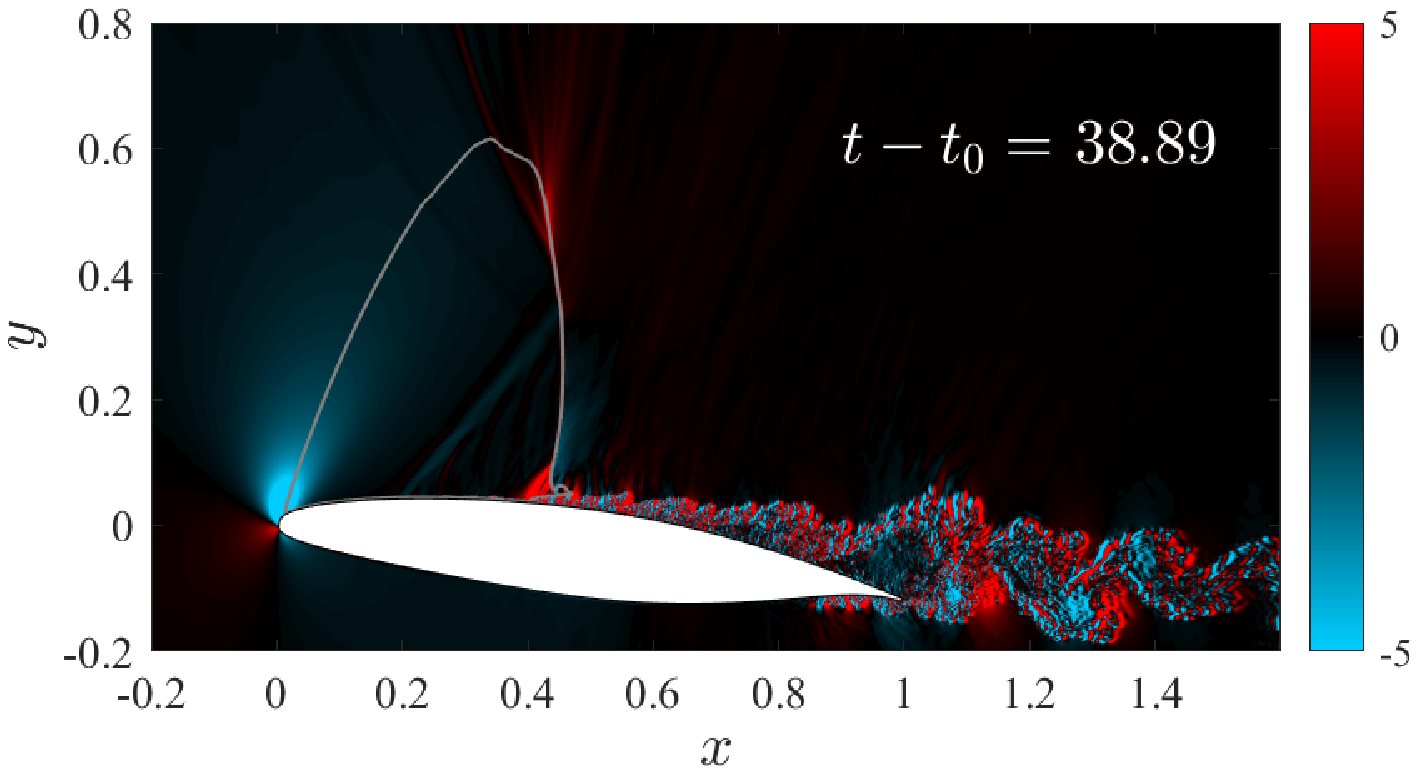}
	\caption{Streamwise density-gradient contours on the $\boldsymbol{z=0}$ plane at high- (left) and low-lift (right) phases for (a) A7M7F and  (b) A7M7S cases (LES).}
	\label{figLES_RhoContours_A7M7}
\end{figure}
\begin{figure}[t]
	\centering
	\includegraphics[trim={0cm 0cm 0cm 0cm},clip,width=.45\textwidth]{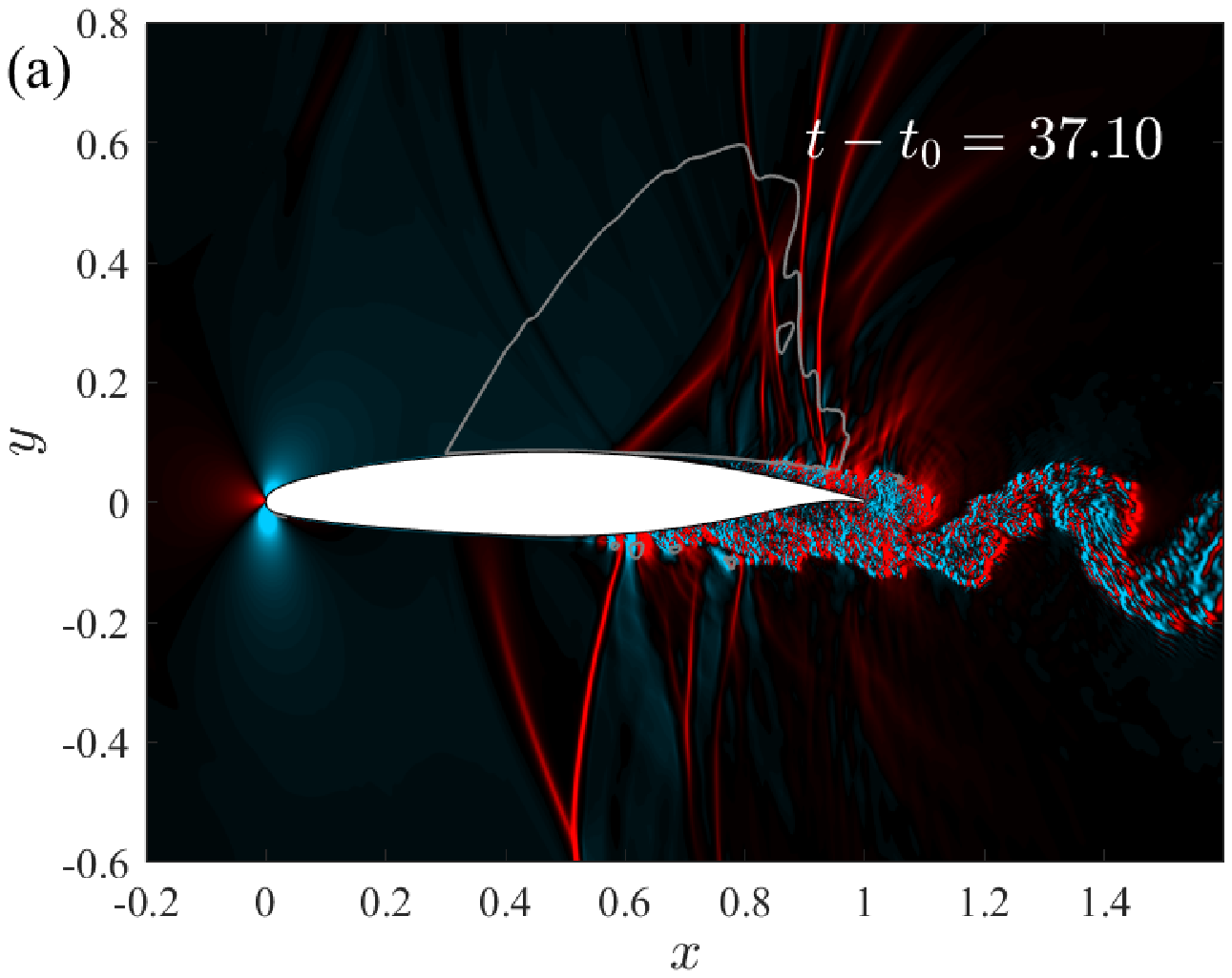}
	\includegraphics[trim={0cm 0cm 0cm 0cm},clip,width=.45\textwidth]{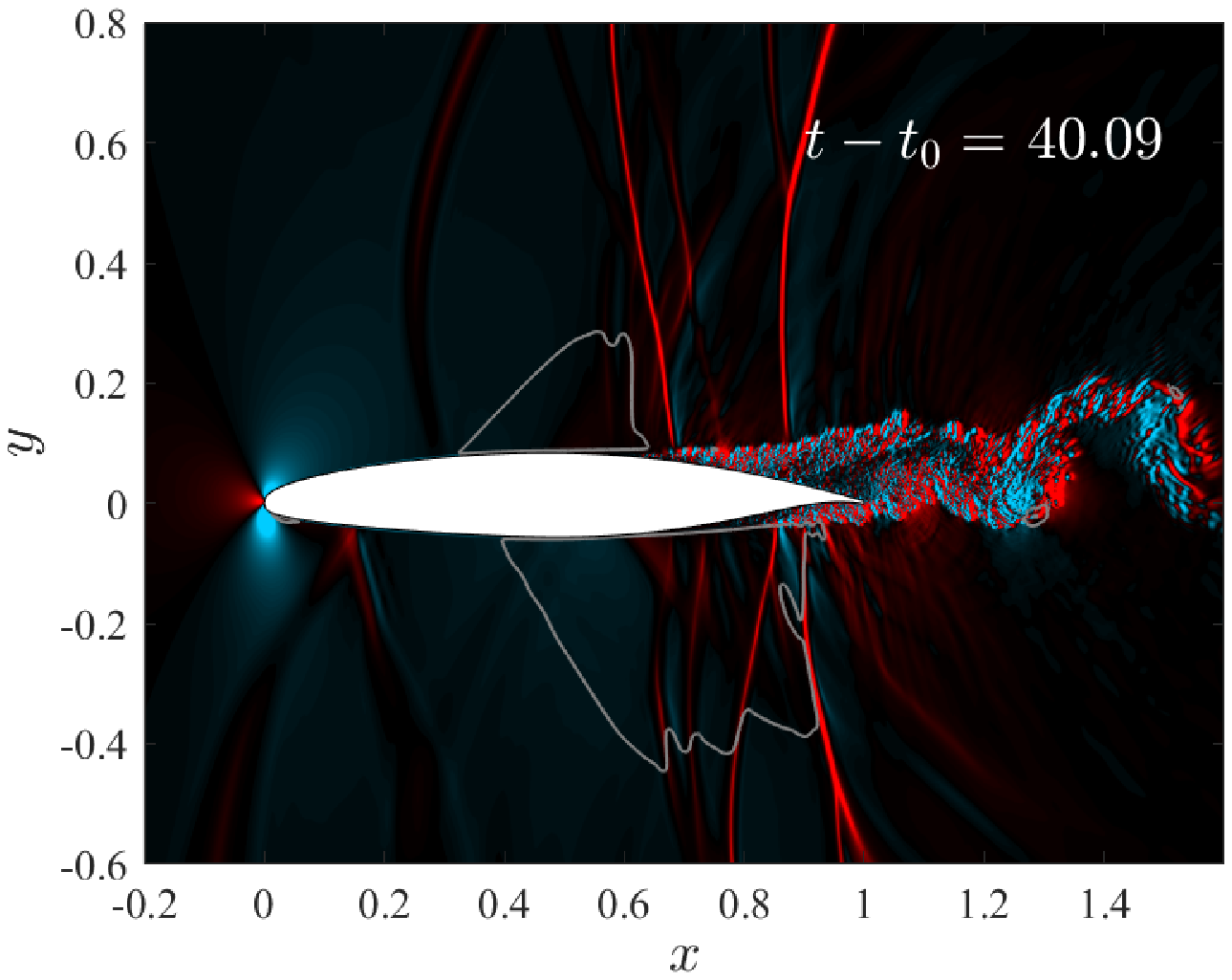}
	\includegraphics[trim={0cm 0cm 0cm 0cm},clip,width=.45\textwidth]{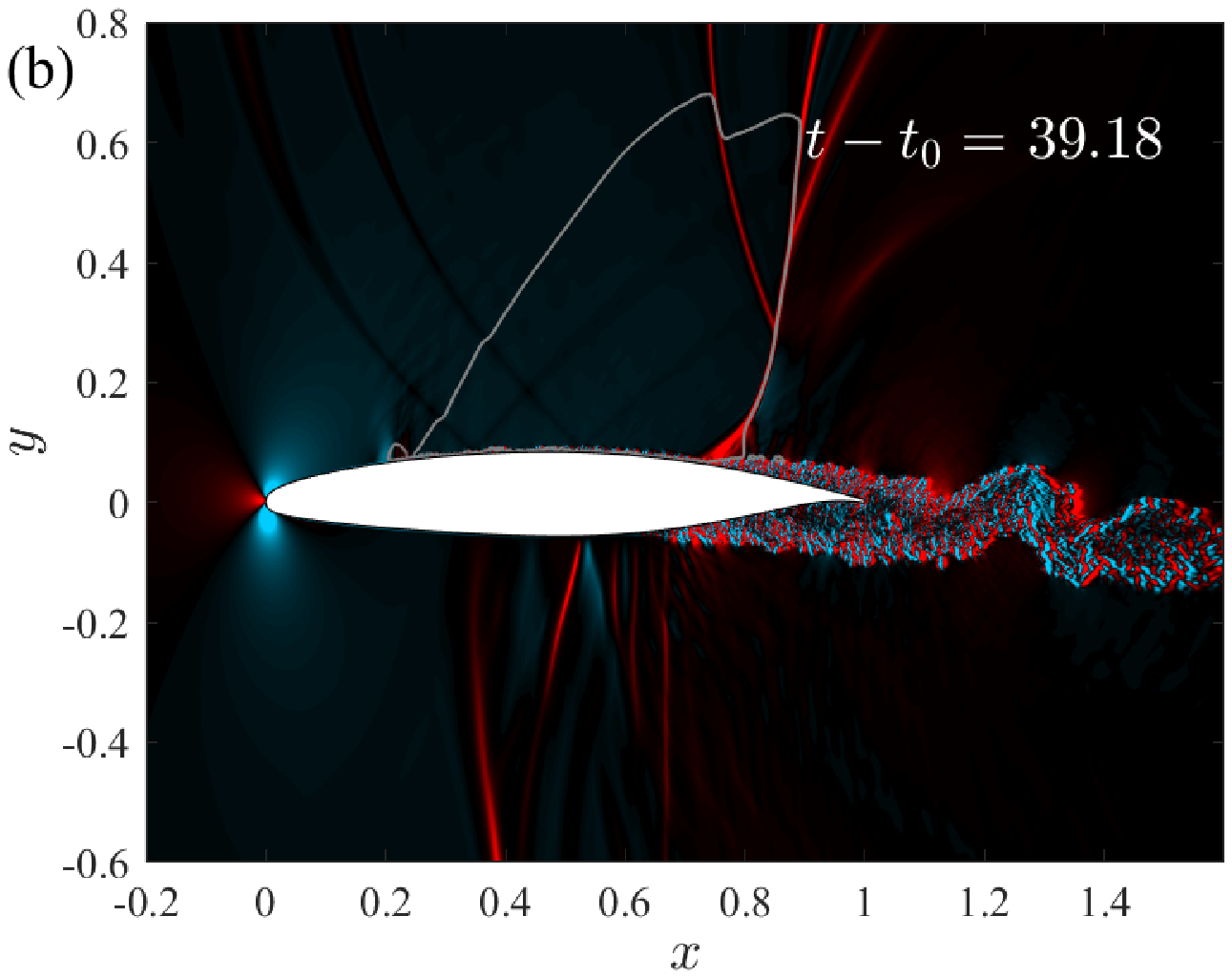}
	\includegraphics[trim={0cm 0cm 0cm 0cm},clip,width=.45\textwidth]{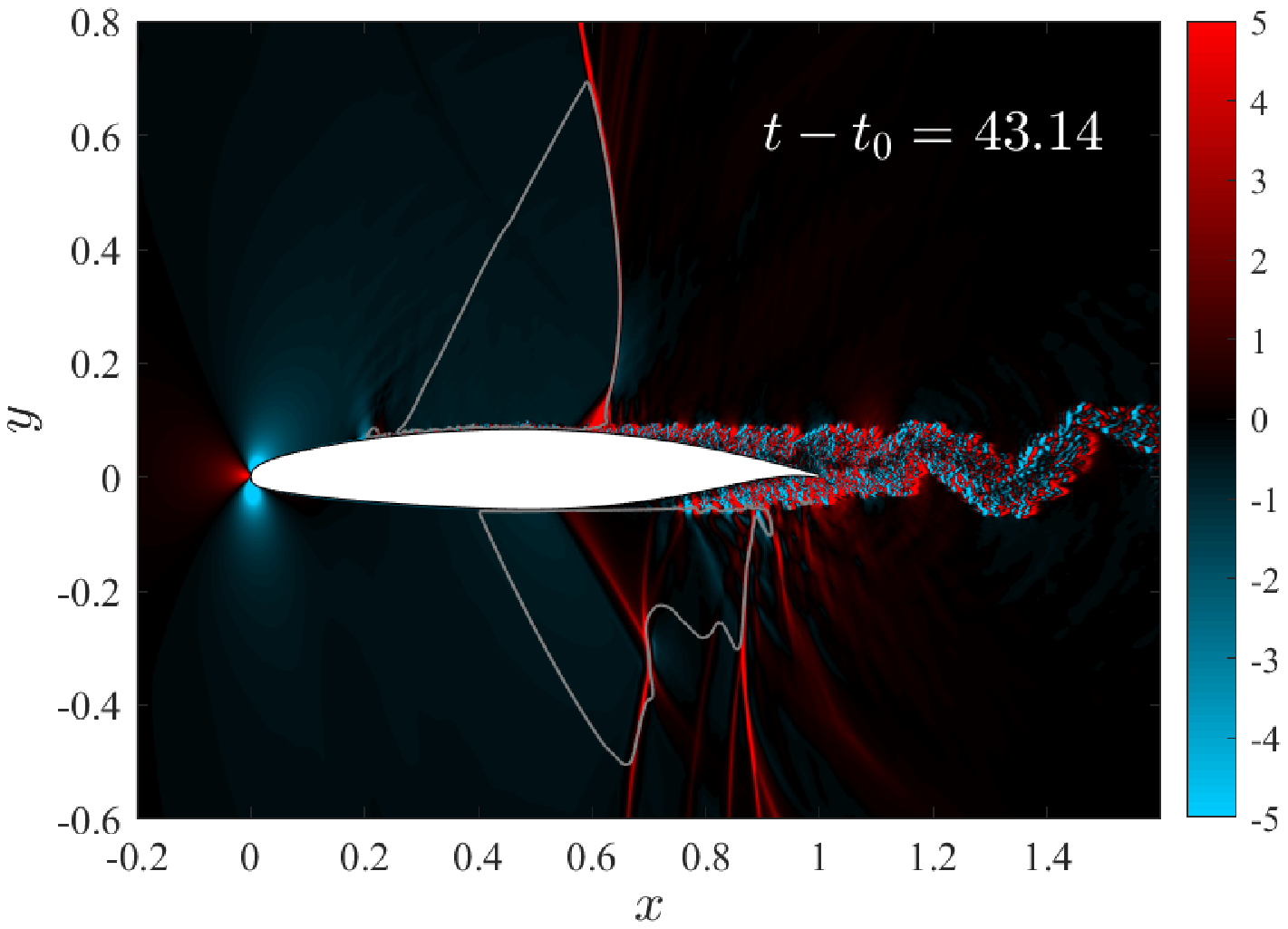}
	\caption{Streamwise density-gradient contours on the $\boldsymbol{z=0}$ plane at high- (left) and low-lift (right) phases for (a) A0M8F and  (b) A0M8S cases (LES).}
	\label{figLES_RhoContours_A0M8}
\end{figure}

Contours of streamwise density gradient on the $z = 0$ plane at high- and low-lift phases are shown in Fig.~\ref{figLES_RhoContours_A7M7} for cases A7M7F and A7M7S. The similarities and differences noted when comparing free- and forced-transition cases at $\alpha = 5^\circ$ and $M = 0.735$ (see Fig.~\ref{figLES_RhoContours_M735}) are seen to be also valid here. For cases A0M8F and A0M8S,  the contours and the sonic line (gray curve) shown in Fig.~\ref{figLES_RhoContours_A0M8} indicate the development of a supersonic region and shock waves on the pressure side of the aerofoil during the low-lift phase of the buffet cycle. This suggests that the lift oscillations observed in Fig.~\ref{figLES_ClVsT_PSD_StrongBuff} for both these cases are associated with a Type I buffet. Multiple shock waves seen for the free-transition case are replaced by a single shock wave on the suction side when transition is forced. However, there are strong similarities between the two cases at both phases shown, including in the extent of the supersonic region in each phase and the wake deflection. These results also corroborate the proposition noted previously that the occurrence of multiple shock waves or a single shock wave is dependent on the boundary layer type (laminar or turbulent, respectively) but buffet characteristics are essentially the same irrespective of the shock wave structure. This will be further corroborated using SPOD in Sec.~\ref{secSPOD}.

%%%%%%%%%%%%%%%%%%%%%%%%%%%%%%%%%%%%%%%%
\subsection{Comparison of LES and RANS results}
\label{subSecRANS}
%%%%%%%%%%%%%%%%%%%%%%%%%%%%%%%%%%%%%%%%
\begin{figure}[t]
	\centering
	\includegraphics[width=.495\columnwidth]{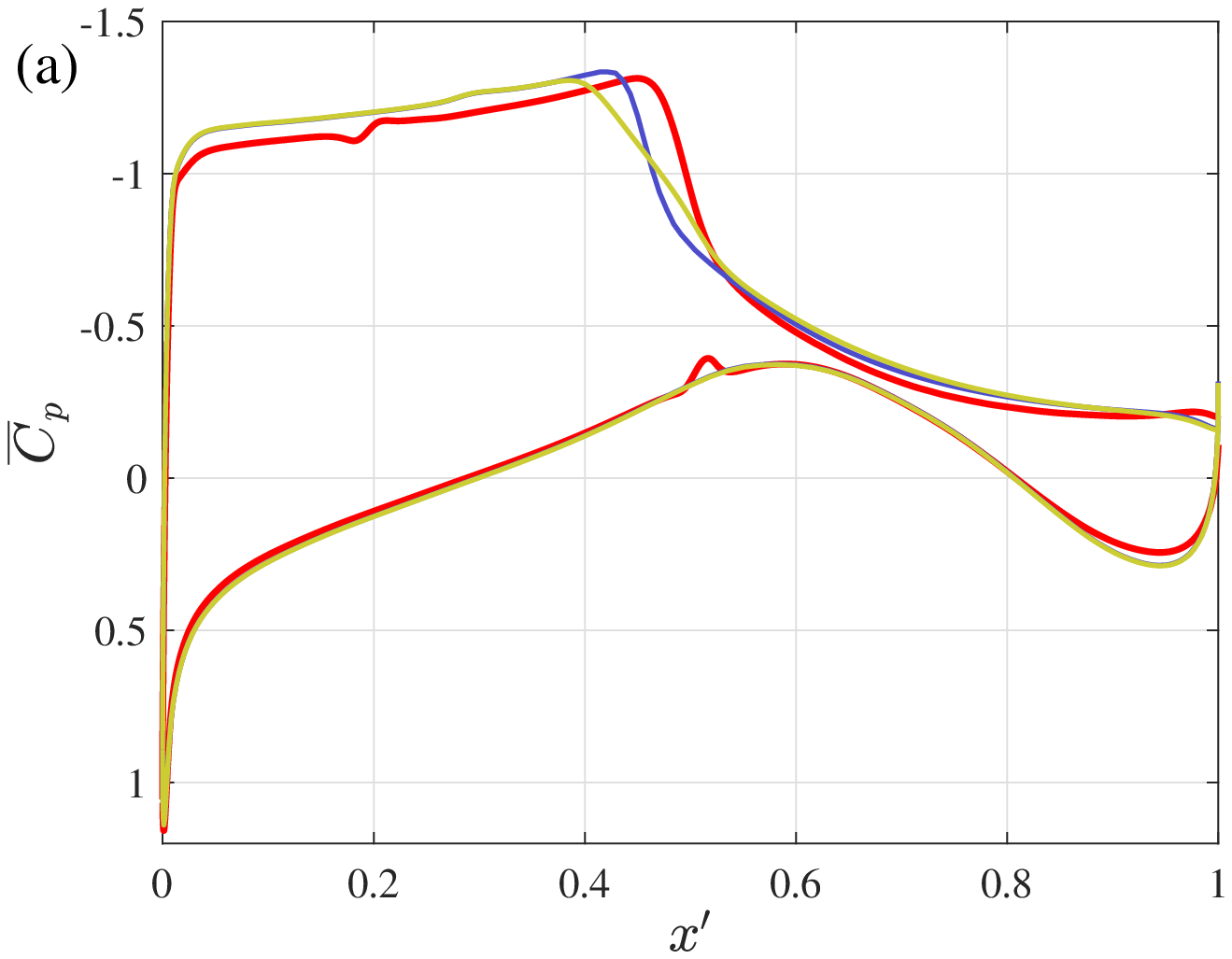}
	\includegraphics[width=.495\columnwidth]{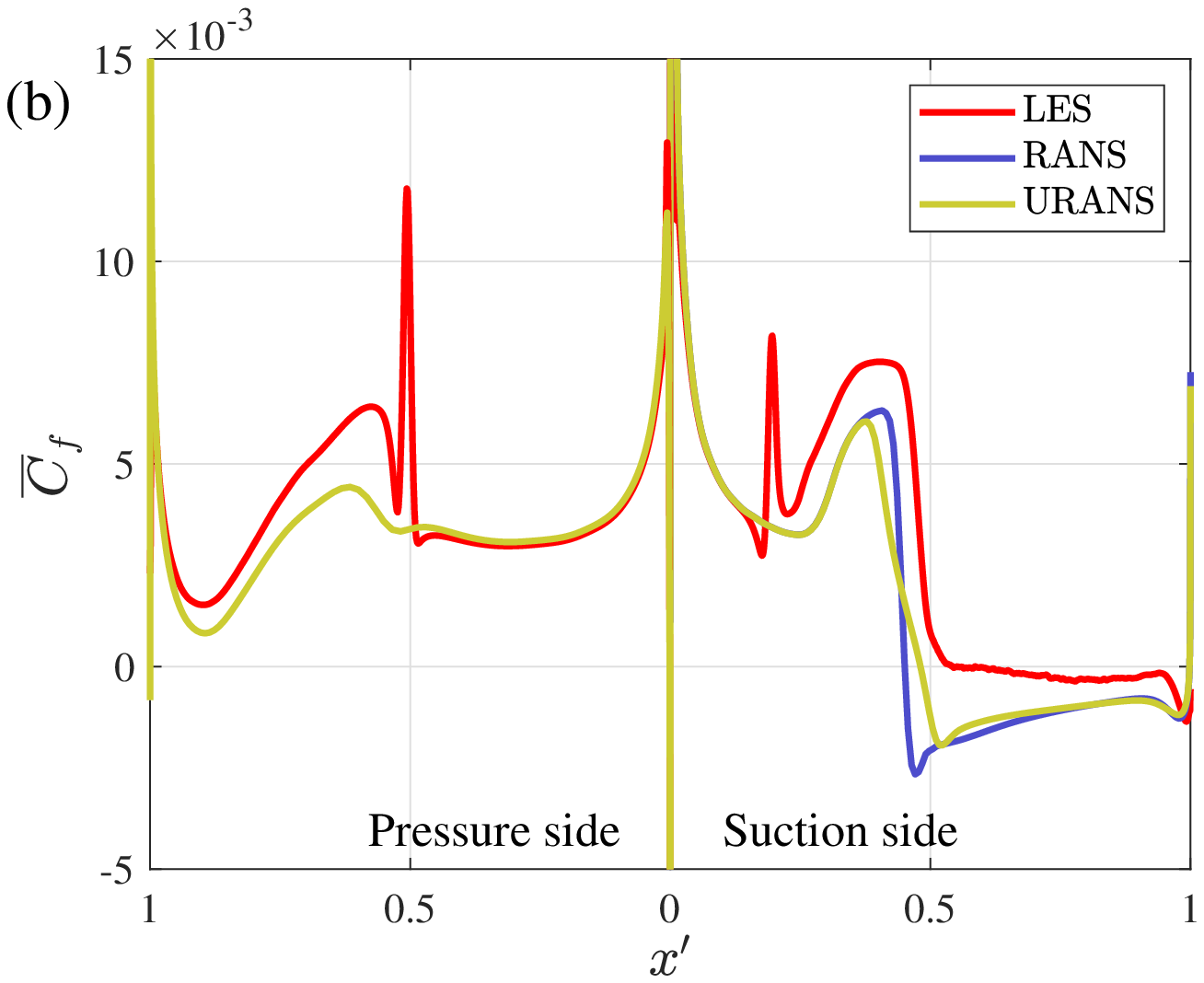}
	\caption{Comparison of (a) pressure coefficient and (b) skin-friction coefficient for LES, steady and URANS at A5M735SP.}
	\label{figLESvsRANSCp}
\end{figure}

Case A5M735SP is used for comparing steady and unsteady RANS simulation with LES. The LES results reported in Sec.~\ref{subSecRefCaseLES} show that this case is close to turbulent buffet onset (see Fig.~\ref{figLESTripDiffSidesCL}). This is useful when comparing modal features, since nonlinear effects due to interactions between the unstable global mode and the base flow are relatively weak when buffet is incipient. The mean pressure coefficient from the LES, steady and unsteady RANS simulations is shown in Fig.~\ref{figLESvsRANSCp}\textit{a}. Although there are no significant differences on the pressure side, there are some interesting differences on the suction side. Comparing LES with steady RANS simulation, we see that the mean chordwise shock location (indicated by the somewhat abrupt increase in $\overline{C}_p$ with $x'$ around mid-chord) is further downstream in the LES. There is a slightly stronger shock wave motion in the URANS simulations, as indicated by the `smearing' of $\overline{C}_p$ seen for $0.4 \leq x' \leq 0.5$. 

A comparison between the different simulation approaches based on $\overline{C}_f$, is shown in Fig.~\ref{figLESvsRANSCp}\textit{b}. For all three approaches, a steep increase in $\overline{C}_f$ immediately downstream of the trip location ($x_{ts} = 0.2$ and $x_{tp} = 0.5$) indicates the transition of the boundary layer to turbulence. Qualitative features are similar, although there are some quantitative differences, in addition to the differences in mean shock location and amplitude of shock motion identified previously. For example, the values of $\overline{C}_f$ immediately downstream of the trip locations are larger for the LES, while the separation bubble on the suction side ($\overline{C}_f < 0$) downstream of the shock has stronger reverse flow for the RANS solutions. Nevertheless, the good qualitative agreement facilitates comparisons of buffet features observed for the three approaches.

\begin{figure}[t]
	\centering
	\includegraphics[width=.45\textwidth]{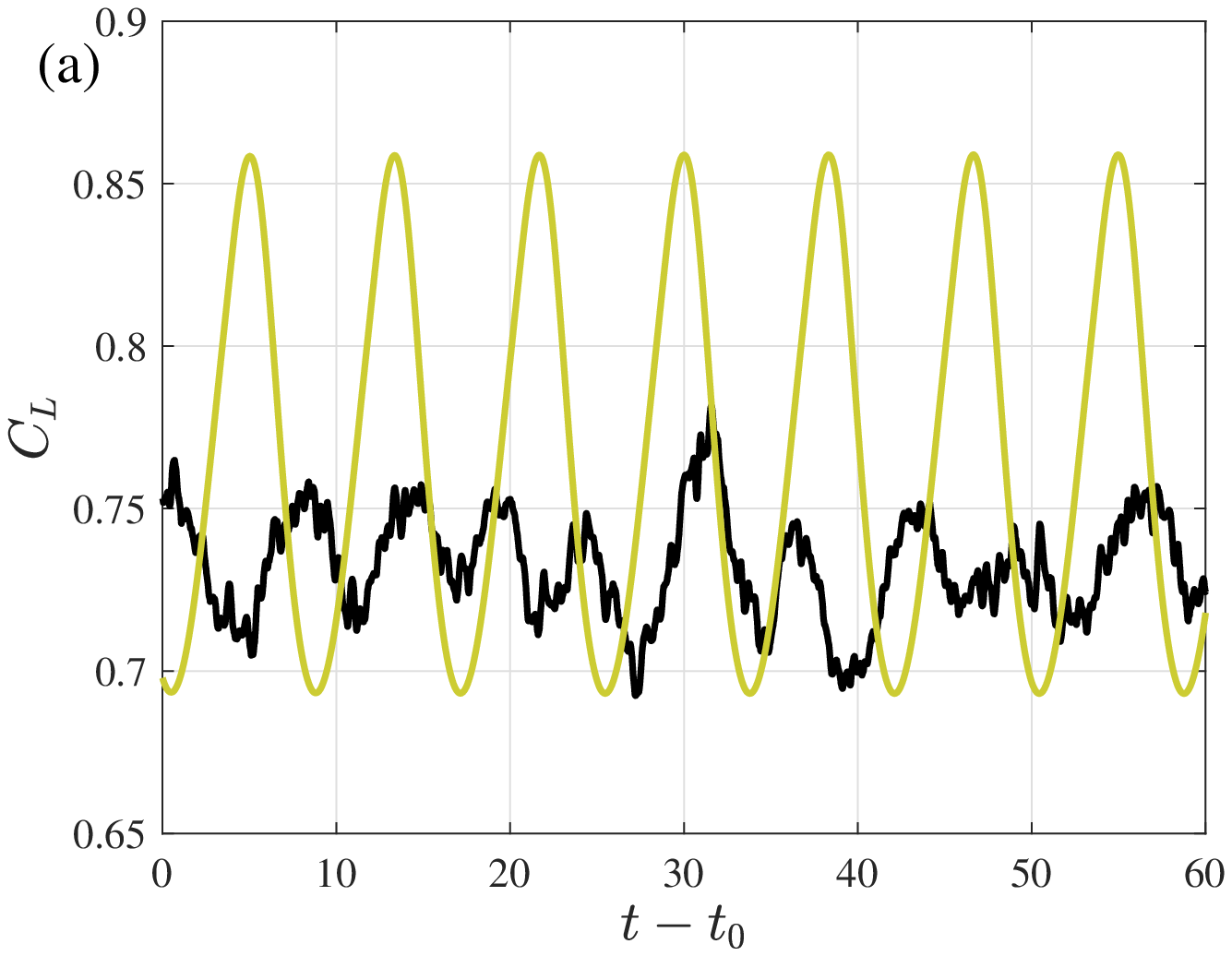}
	\includegraphics[width=.45\textwidth]{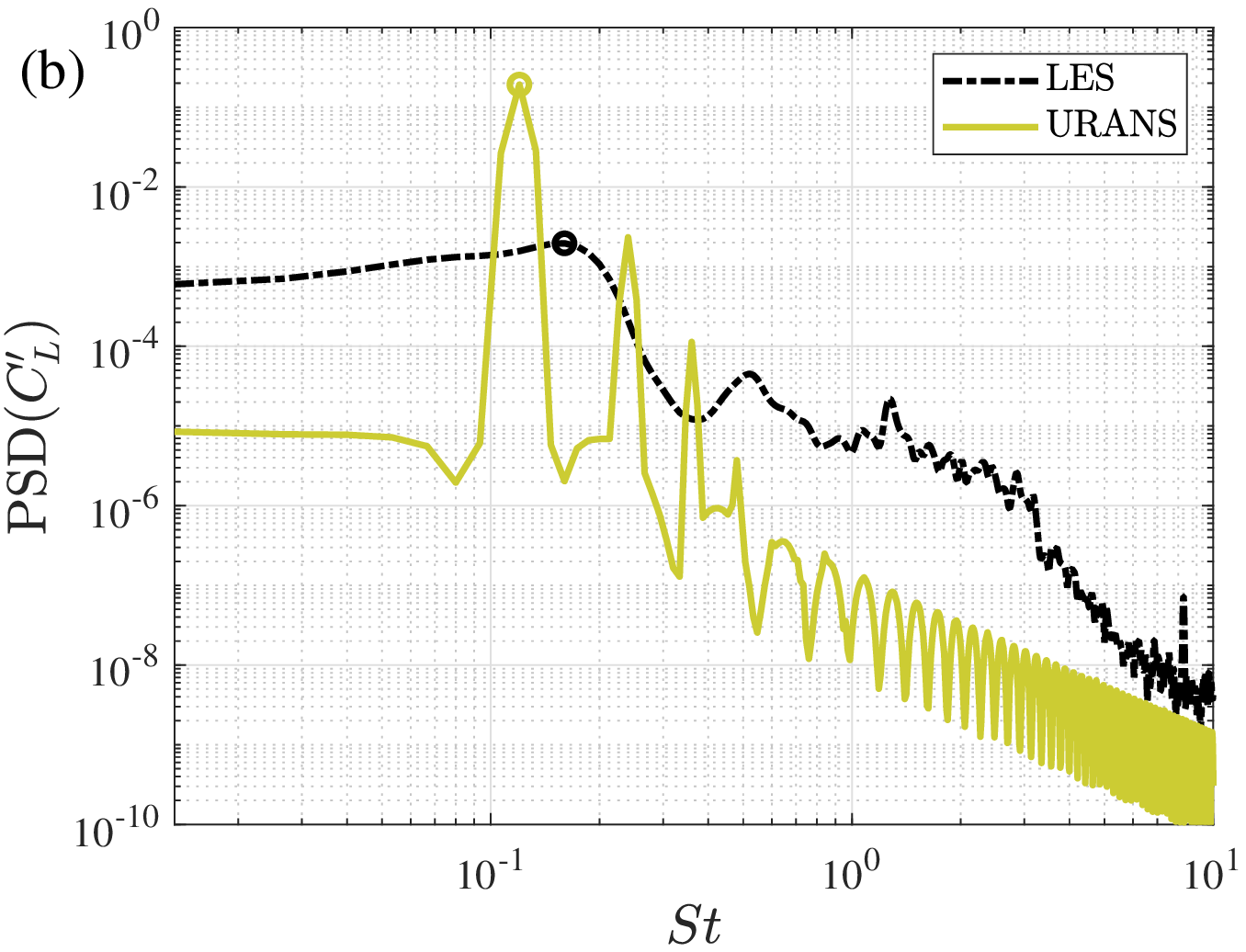}
	\caption{(a) Temporal variation of lift coefficient past transients and (b) power spectral density of its fluctuating component for LES and URANS at A5M735SP.}
	\label{figLES_ClVsT_PSD_LESvsURANS}
\end{figure}

 For the LES and URANS simulation, the temporal variation of the lift coefficient past transients and the PSD of its fluctuating component are compared in Fig.~\ref{figLES_ClVsT_PSD_LESvsURANS}. In contrast to the LES, the lift coefficient oscillations in URANS simulation are regular and periodic, with the PSD showing peaks only at the buffet (fundamental) frequency and its harmonics. Importantly, the bump seen in the spectra at $St \approx 1$ for the LES is absent in the URANS results. As will be shown in Sec.~\ref{secSPOD}, wake modes associated with vortex shedding occur at these frequencies in the LES, which is also predicted from a resolvent analysis of steady RANS solutions. However, URANS simulations appear to not capture these features. This was further confirmed by performing the SPOD of URANS results (not shown). A similar inference was made by \citet{Moise2022} based on results from previous studies including those that use detached-eddy simulations in addition to RANS-level simulations, GLSA, resolvent analysis and dynamic mode decomposition \citep{Sartor2015, Grossi2014, Crouch2009, Poplingher2019, Giannelis2018, Iovnovich2012, Szubert2016, Xiao2006a}. The reasons underlying this behaviour remain unclear but it could have important ramifications. For example, since the amplitude and frequency does not vary significantly between the URANS simulations and the LES, these controlled numerical experiments suggest that buffet features remain essentially unaffected by higher frequency vortex shedding. Since the acoustic field is predominantly associated with vortices interacting with the trailing edge, this indicates that models for buffet based on an acoustic feedback mechanism, such as those of \citet{Lee1990} and \citet{Hartmann2013}, in which acoustic waves are taken to accumulate at the shock, might be incorrect. 

\section{Discussion of Modal Features}
\label{secDecomNRecon}
%%%%%%%%%%%%%%%%%%%%%%%%%%%%%%%%%%%%%%%%

%Results from SPOD (LES), GLSA and resolvent analysis (RANS) are presented here.

%%%%%%%%%%%%%%%%%%%%%%%%%%%%%%%%%%%%%%%%
\subsection{Spectral proper orthogonal decomposition of LES results}
\label{secSPOD}
%%%%%%%%%%%%%%%%%%%%%%%%%%%%%%%%%%%%%%%%

\begin{figure}
	\centering
	\includegraphics[width=0.495\columnwidth]{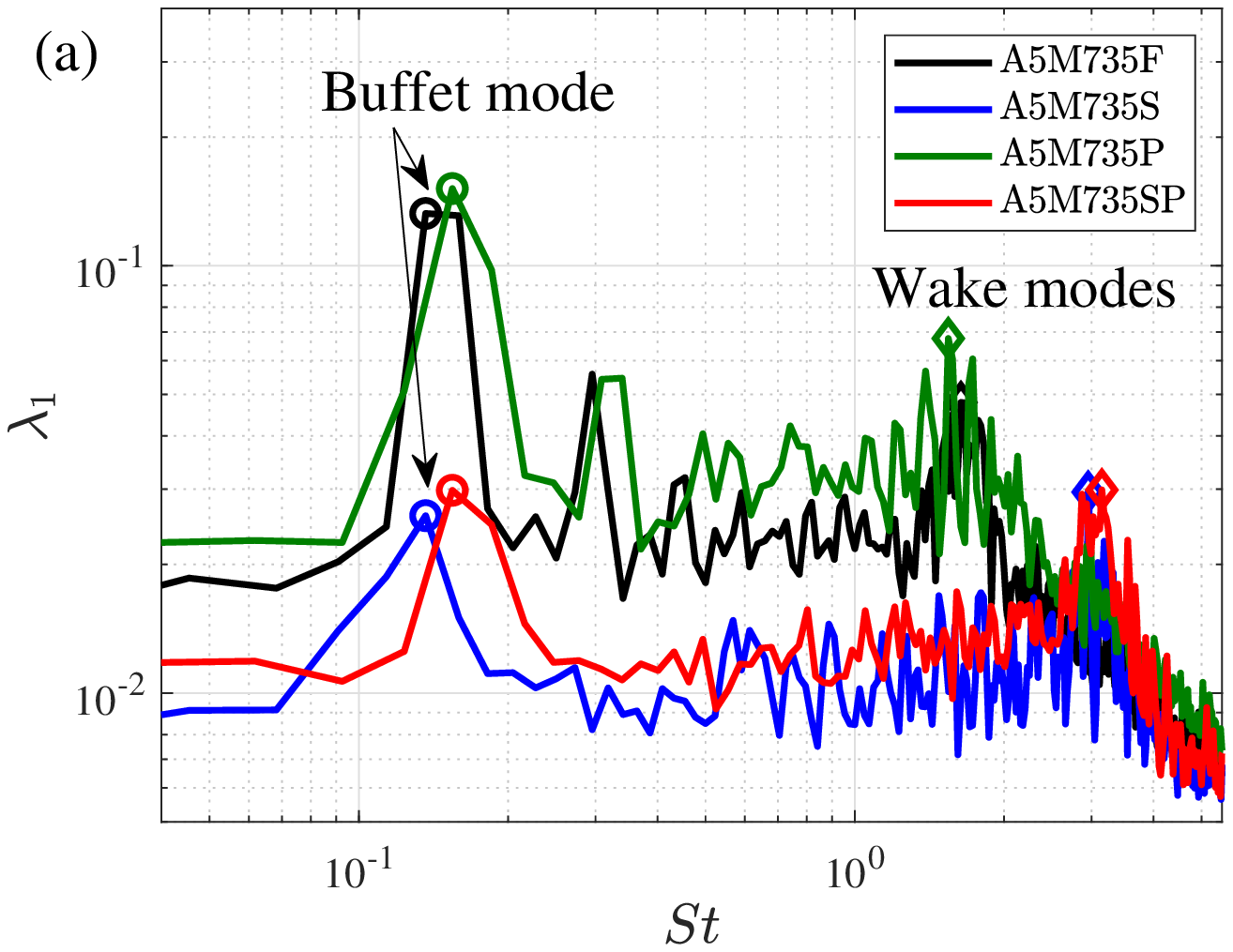}
		\includegraphics[width=0.495\columnwidth]{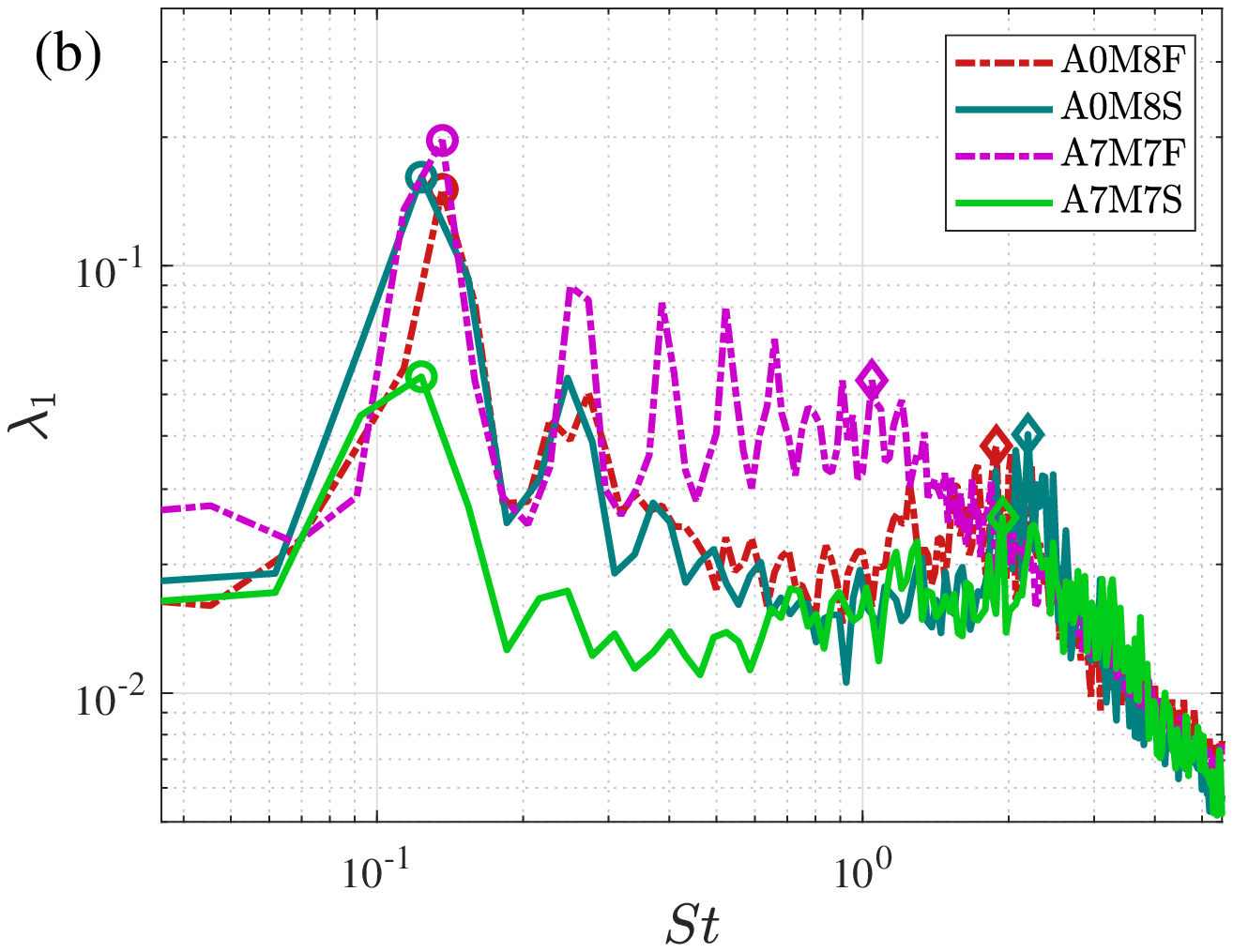}
	\caption{The eigenvalue spectra of the dominant eigenvalue from SPOD for (a) tripping on different sides for $\boldsymbol{M = 0.735}$, $\boldsymbol{\alpha = 5^\circ}$ and (b) for other parametric values.}
	\label{figLES_SPODspectrum}
\end{figure}

\begin{figure}[!ht]
	\centering
	\includegraphics[trim={2.9cm 0cm 3.9cm 0cm},clip,width=.22\columnwidth]{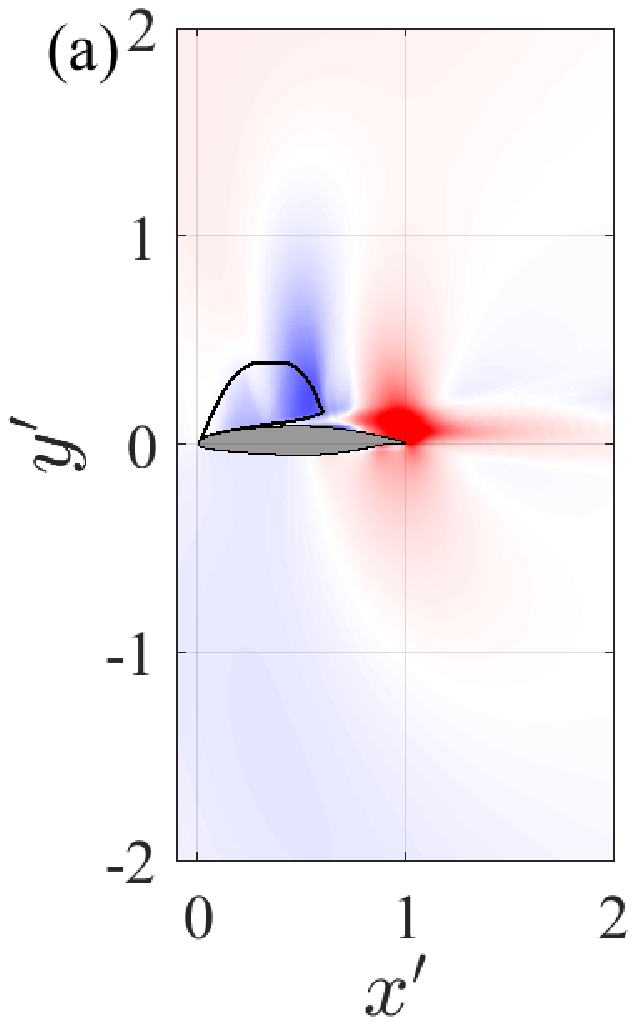}
	\includegraphics[trim={2.9cm 0cm 3.9cm 0cm},clip,width=.22\columnwidth]{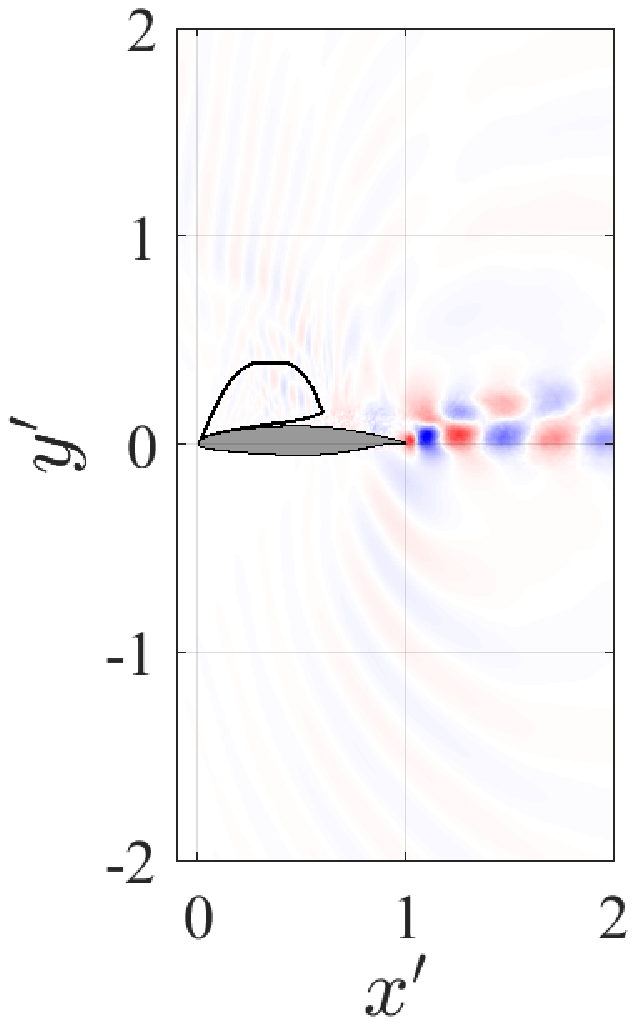}
	\includegraphics[trim={2.9cm 0cm 3.9cm 0cm},clip,width=.22\columnwidth]{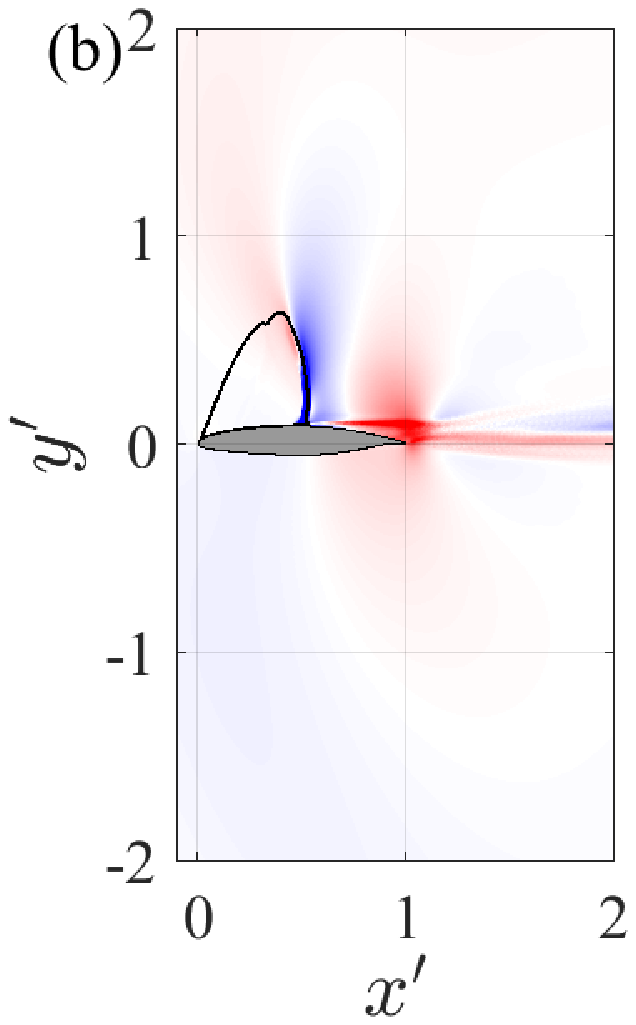}
	\includegraphics[trim={2.9cm 0cm 3.9cm 0cm},clip,width=.22\columnwidth]{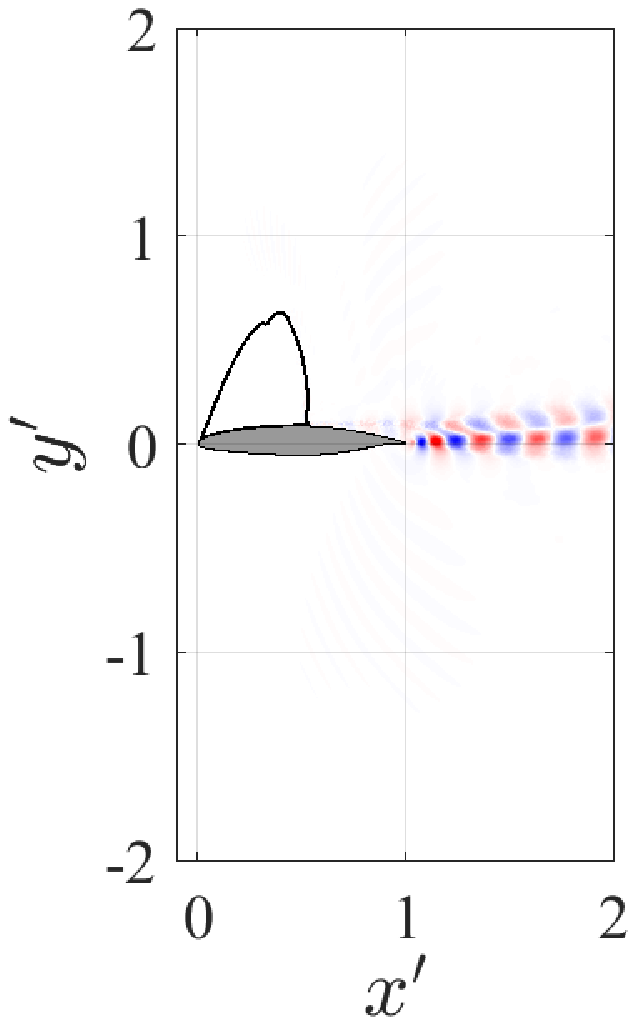}
	\includegraphics[trim={2.9cm 0cm 3.9cm 0cm},clip,width=.22\columnwidth]{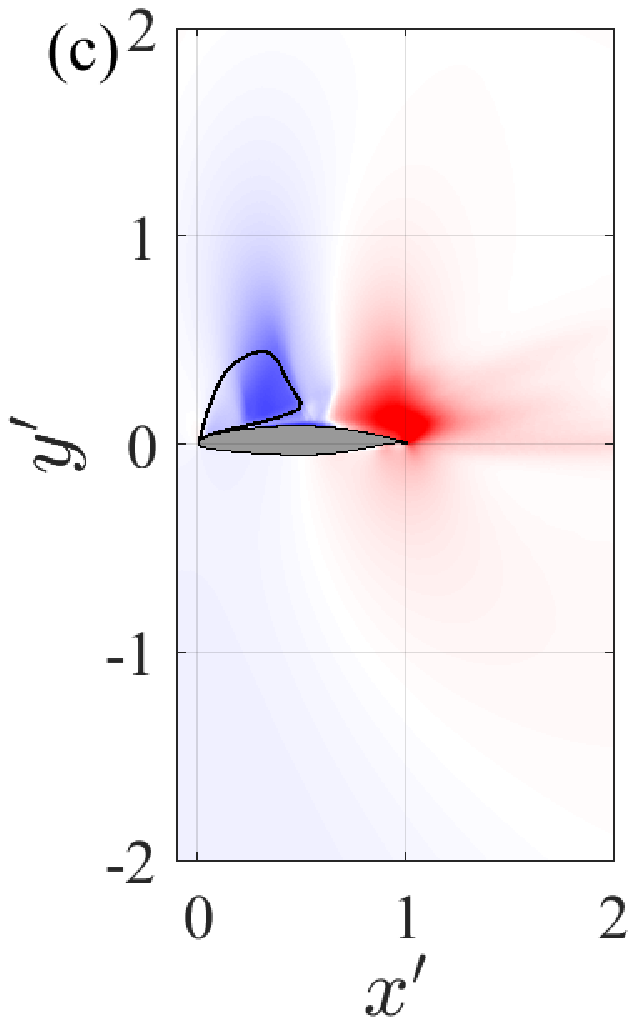}
	\includegraphics[trim={2.9cm 0cm 3.9cm 0cm},clip,width=.22\columnwidth]{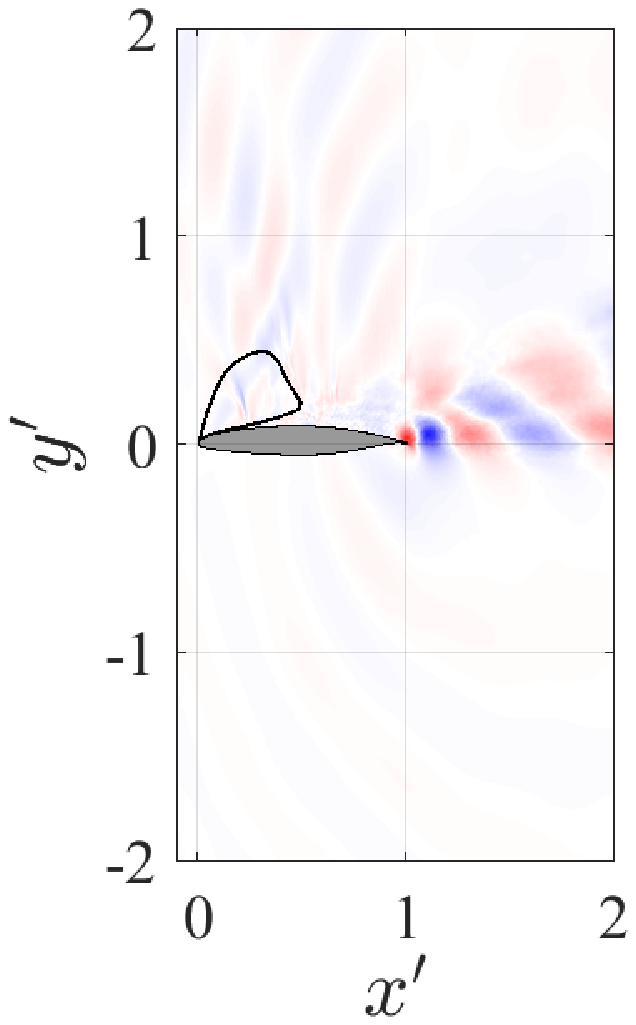}
	\includegraphics[trim={2.9cm 0cm 3.9cm 0cm},clip,width=.22\columnwidth]{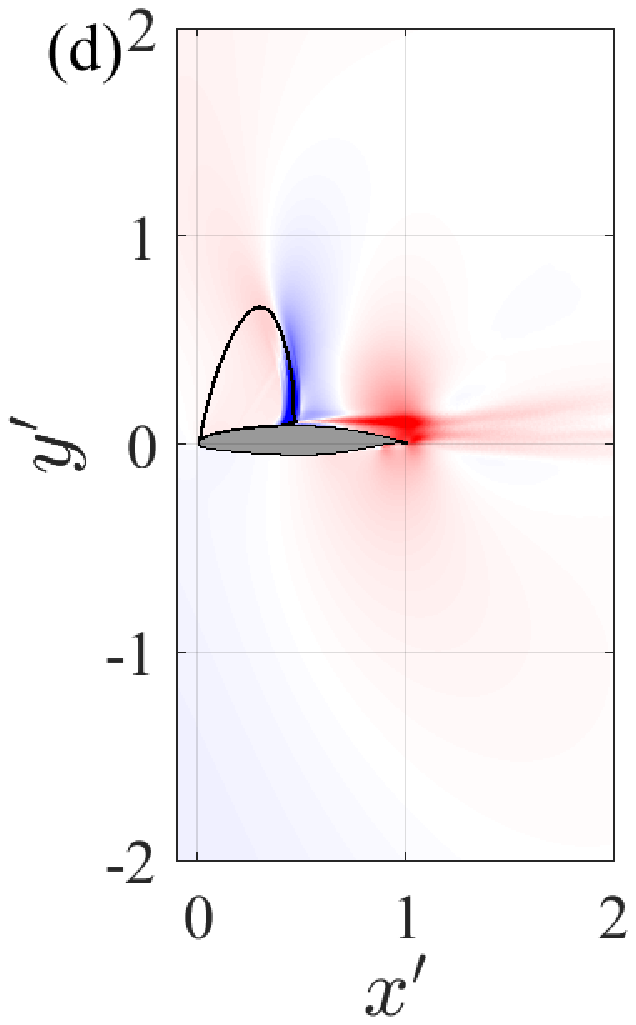}
	\includegraphics[trim={2.9cm 0cm 3.9cm 0cm},clip,width=.22\columnwidth]{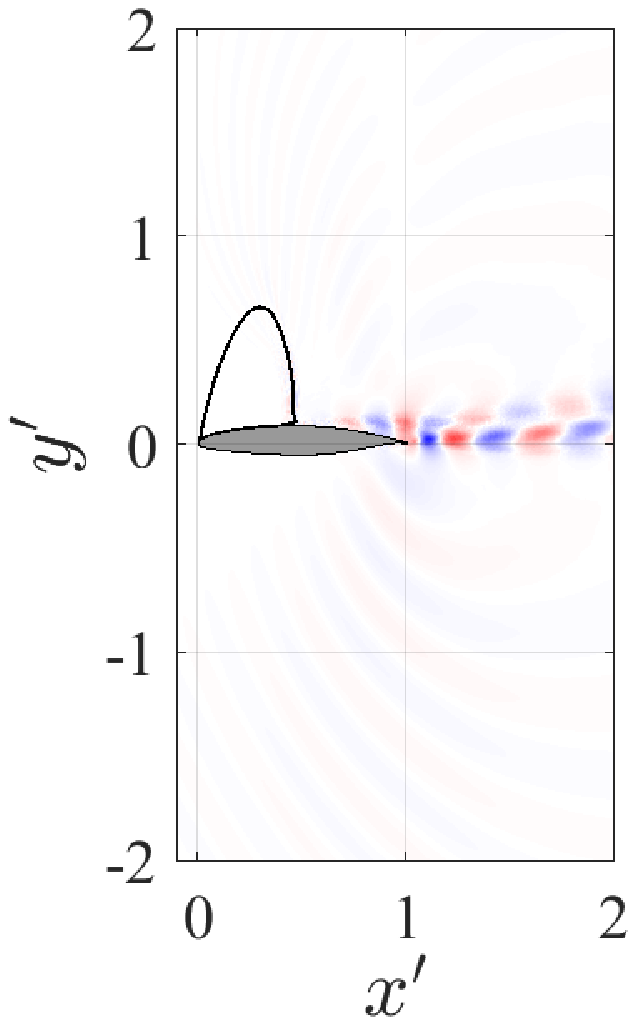}
	\includegraphics[trim={2.9cm 0cm 3.9cm 0cm},clip,width=.22\columnwidth]{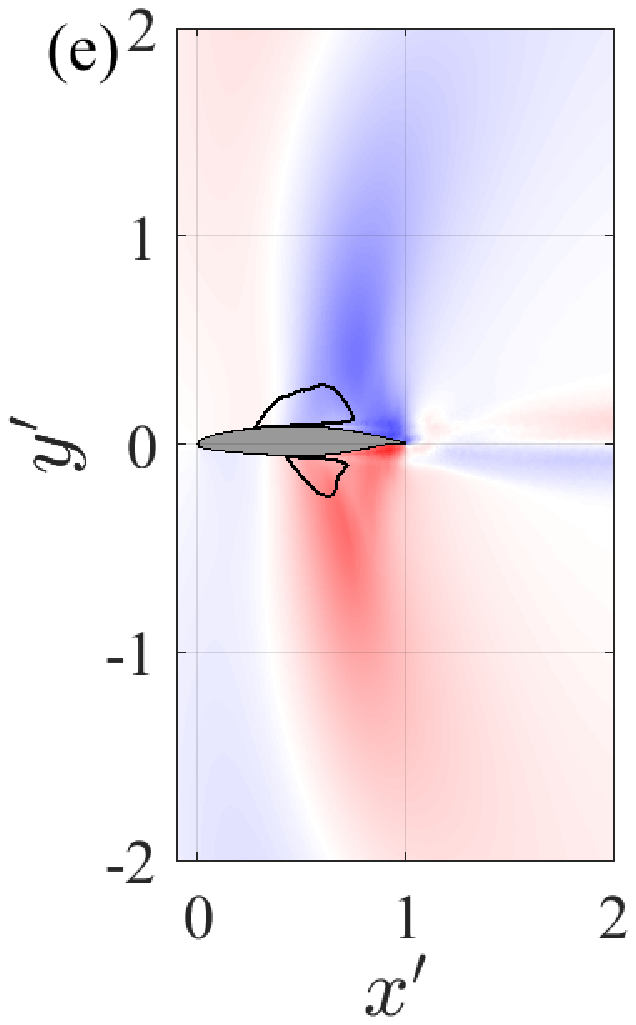}
	\includegraphics[trim={2.9cm 0cm 3.9cm 0cm},clip,width=.22\columnwidth]{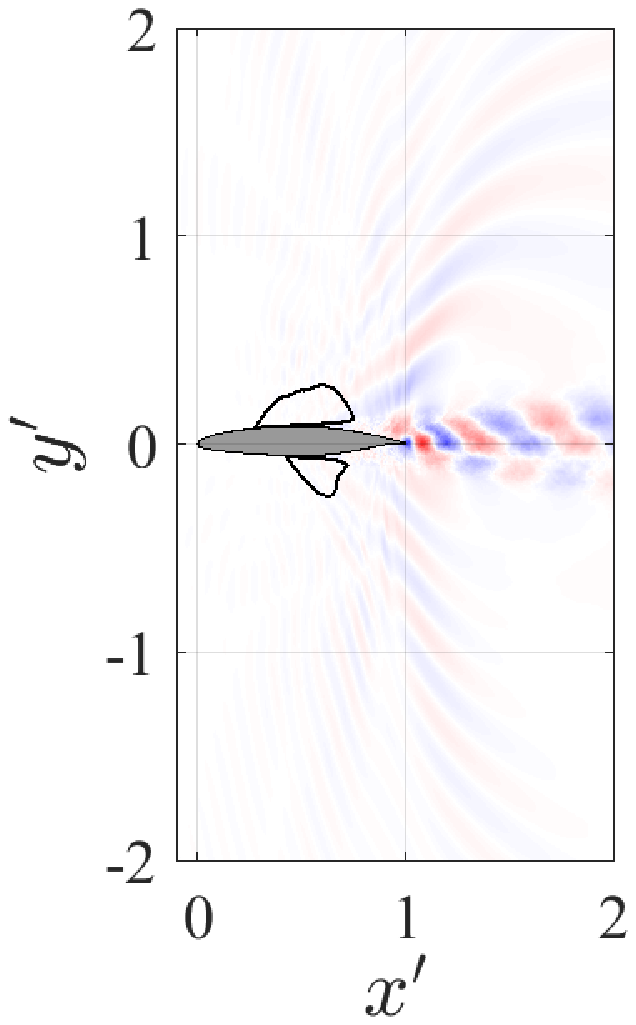}
	\includegraphics[trim={2.9cm 0cm 3.9cm 0cm},clip,width=.22\columnwidth]{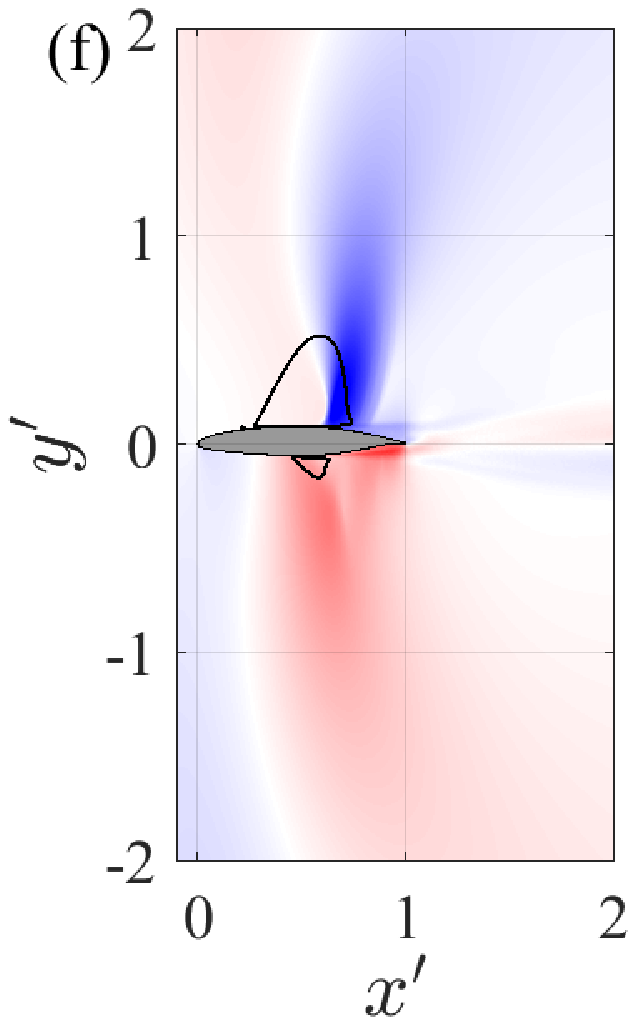}
	\includegraphics[trim={2.9cm 0cm 3.9cm 0cm},clip,width=.22\columnwidth]{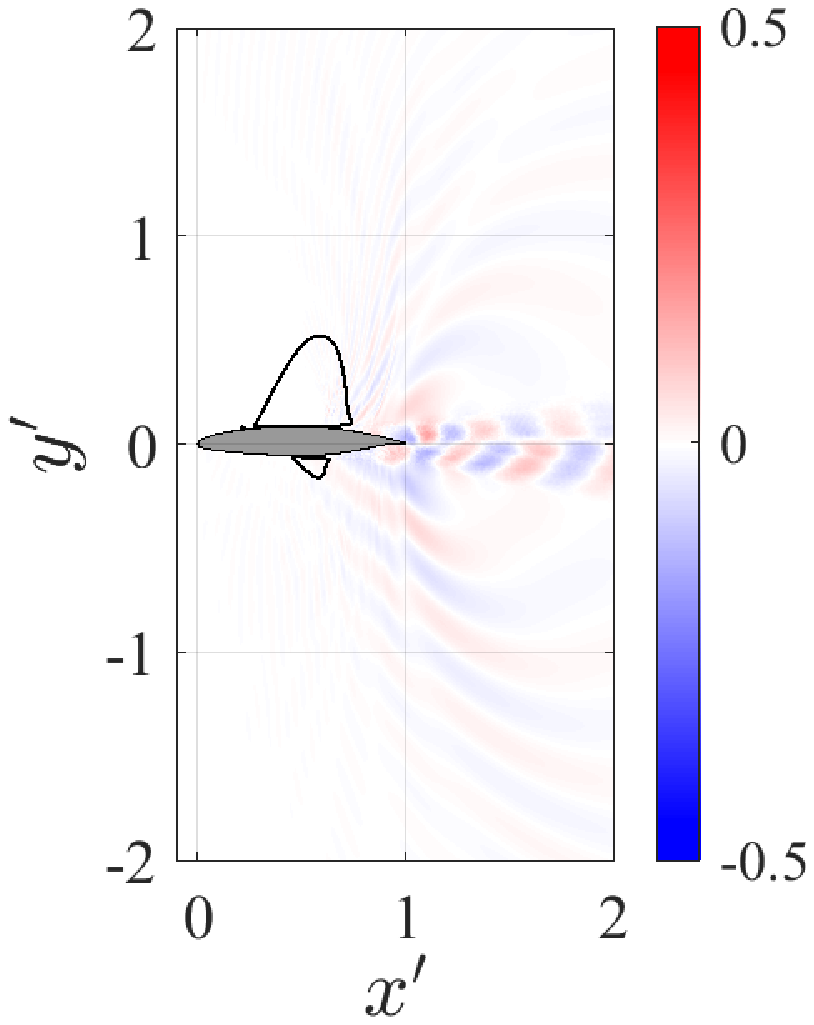}
	\caption{Contours of real part of density of the buffet (left) and wake (right) modes for (a) A5M735F, (b) A5M735SP, (c) A7M7F, (d) A7M7S, (e) A0M8F and (f) A0M8S. The sonic line based on time-averaged flow is shown using the black curve} 
	\label{figLES_SPODModes}
\end{figure}

Eigenvalue spectra, based on the dominant SPOD eigenvalue for different cases at $M = 0.735$ and $\alpha = 5^\circ$ (A5M735F, A5M735S, A5M735P and A5M735SP), are compared in Fig.~\ref{figLES_SPODspectrum}\textit{a}. The peaks at low-frequency (highlighted by circles) shown here match the buffet frequency seen for the lift coefficient in Fig.~\ref{figLESTripDiffSidesCL}\textit{b}. The SPOD modes associated with these peaks will henceforth be referred to as ``buffet modes". Since the intermediate frequency seen in  Fig.~\ref{figLESTripDiffSidesCL}\textit{b} occurs close to the harmonic of the buffet mode and since no clear peak is present in the SPOD spectra it is not explored further in this study. Bumps in the spectra in the higher frequency range of $St \sim O(1)$ (largest amplitude in the range highlighted by a diamond) are discernible and are related to a von K\'arm\'an vortex street (see Fig.~\ref{figLES_SPODModes}). Following \citet{Moise2022}, these will be referred to as ``wake modes" due to their high energy content in the wake of the aerofoil. The eigenvalue spectra for parametric conditions discussed in Sec.~\ref{subSecDeepBuffet} are shown in Fig.~\ref{figLES_SPODspectrum}\textit{b}. Peaks in the spectra associated with the buffet and wake modes are also observed here for all cases.

Contours of the real parts of the buffet modes and the most-dominant wake modes obtained using SPOD are compared for the various cases in Fig.~\ref{figLES_SPODModes}. Cases A5M735S and A5M735P are not shown for brevity because they closely resemble A5M735SP and A5M735F, respectively. For comparison purposes, the coordinate system is now rotated to chordwise direction ($x'$) and the direction normal to it ($y'$) so that the aerofoils are aligned for all cases. Furthermore, since SPOD yields orthonormal spatio-temporal modes at a given frequency, $St$, which are complex fields of the form $\boldsymbol{\psi}_1(\mathbf{x},St) \exp{(i\phi)}$ with $\phi = 2\pi St t$ (see Eq.~\ref{eqnSPOD}), reliable comparisons of their instantaneous spatial structures requires choosing the phase for each case appropriately. Here, we chose $\phi$ by requiring that the spatio-temporal density field is purely real-valued at the top edge of the blunt trailing edge for all cases (\textit{i.e.}, $\mathrm{Im}\left\{\mathbf{q}(\mathbf{x}_\mathrm{TE}) \exp{(i\phi)} \right\}) = 0$). Additionally, animations based on the buffet modes which show the variation of density over one period of buffet motion (\textit{i.e.} $\mathrm{Re}(\rho \exp(i\phi))$ at different~$\phi$) are provided as Supplementary Material. 

The wake modes for all cases show a clear von K\'arm\'an vortex street pattern, with the vortical length scale being larger for the free transition cases (Fig.~\ref{figLES_SPODModes}\textit{a,c,e}) as opposed to those for which transition is forced (Fig.~\ref{figLES_SPODModes}\textit{b,d,f}). The buffet modes for the different cases also have a similar flow structure and agree qualitatively with the global linear instability mode reported by \citet{Crouch2007} for the NACA0012 aerofoil. The similarity between the free- and forced-transition cases can be seen, for example, when considering how the mode interacts with the mean field. In particular it can be inferred that when a reduction in density from the mean value occurs on the suction surface (blue region), the density in the wake of the aerofoil increases above the mean value (red region) or vice versa, implying that these two regions are out of phase in a buffet cycle. Similarly, the maximum reduction (or increase) occurs approximately at the mean shock wave position. These variations are best seen by comparing the animations provided (Supplementary material). For the cases where Type I buffet occurs (A0M8F and A0M8S, Fig.~\ref{figLES_SPODModes}e and f), a similar relation between the density fluctuations on the pressure side (red region) and the wake (blue region) can be observed. As expected for this buffet type, the fluctuations on the suction and pressure side are out of phase with each other. Note that some minor differences exist, such as a more smeared modal field for the free transition case. However, these can probably be accounted for by the presence of multiple shock waves and larger shock excursions. 

% \begin{figure}
% 	\centering
% 	\includegraphics[width=0.5\columnwidth]{Graphics/compareSpectrumSPOD.eps}
% 	\caption{The eigenvalue spectra of the dominant eigenvalue from SPOD for free and forced transition at $M = 0.735$.}
% 	\label{figLES_SPODspectrumTripDiffSides}
% \end{figure}

% \begin{figure}
% 	\centering
% 	\includegraphics[trim={3cm 0cm 3.9cm 0cm},clip,width=.245\columnwidth]{Graphics/SPODMode_Real_Free7.eps}
% 	\includegraphics[trim={3cm 0cm 3.9cm 0cm},clip,width=.245\columnwidth]{Graphics/SPODMode_Real_Free59.eps}
% 	\includegraphics[trim={3cm 0cm 3.9cm 0cm},clip,width=.245\columnwidth]{Graphics/SPODMode_Real_Tripped7.eps}
% 	\includegraphics[trim={3cm 0cm 3.9cm 0cm},clip,width=.245\columnwidth]{Graphics/SPODMode_Real_Tripped131.eps}
% 	\caption{Contours of density of the buffet and wake modes for the free-transition (left) and forced-transition (right) cases for $M = 0.735$. The sonic line based on time-averaged flow is shown using the black curve.}
% 	\label{figLES_SPODModes_M735TripDiffSides}
% \end{figure}

%%%%%%%%%%%%%%%%%%%%%%%%%%%%%%%%%%%%%%%%
\subsection{RANS-level global linear stability and resolvent analyses}
\label{subSecGLSAResolvent}
%%%%%%%%%%%%%%%%%%%%%%%%%%%%%%%%%%%%%%%%

% \begin{figure}
%     \centering
%     \includegraphics[width=0.49\columnwidth]{Graphics/Gain_2nd.eps}
%     \caption{Gain as a function of Strouhal number based on the resolvent analysis of RANS solutions at different $Re$.}
%     \label{figGainResolvent}
% \end{figure}

\begin{figure}
	\centering
	\includegraphics[trim={0cm 0.8cm 1cm 1cm},clip,width=0.33\columnwidth]{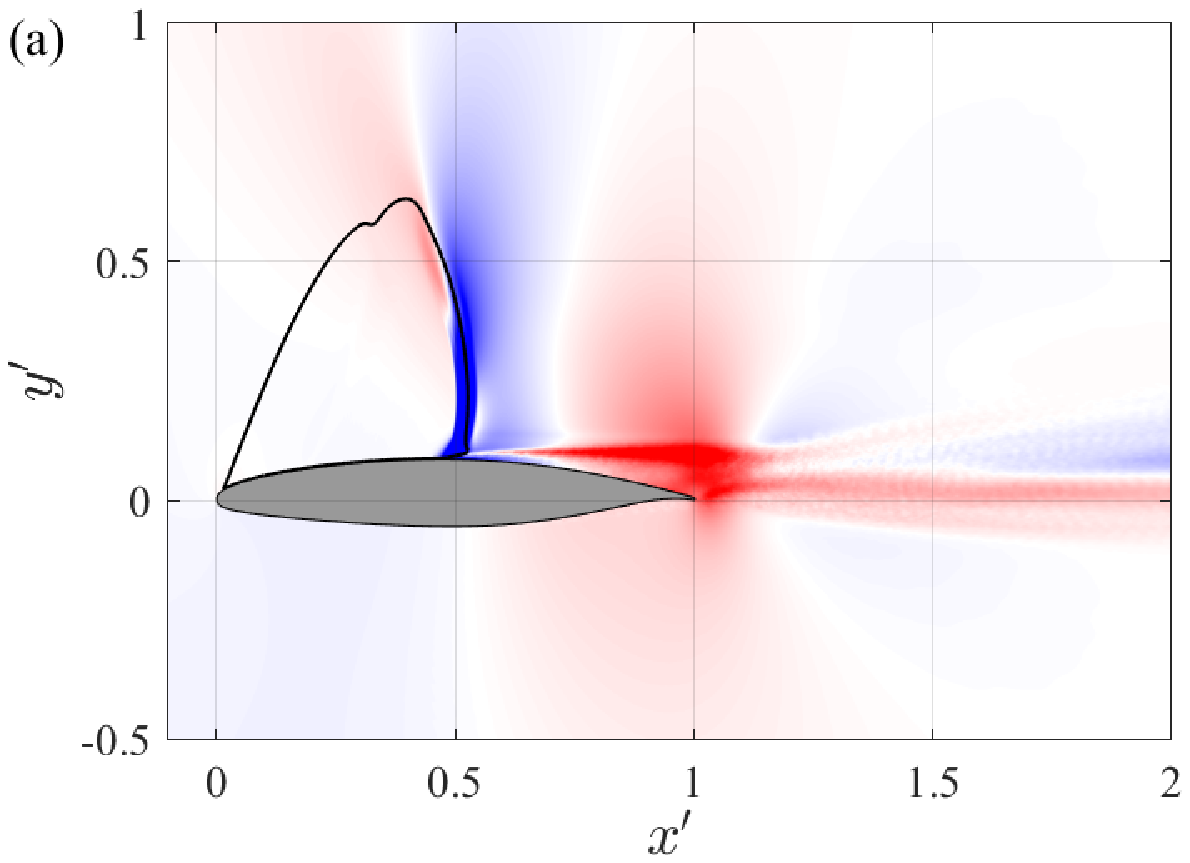}
	\includegraphics[trim={0cm 0.8cm 1cm 1cm},clip,width=0.33\columnwidth]{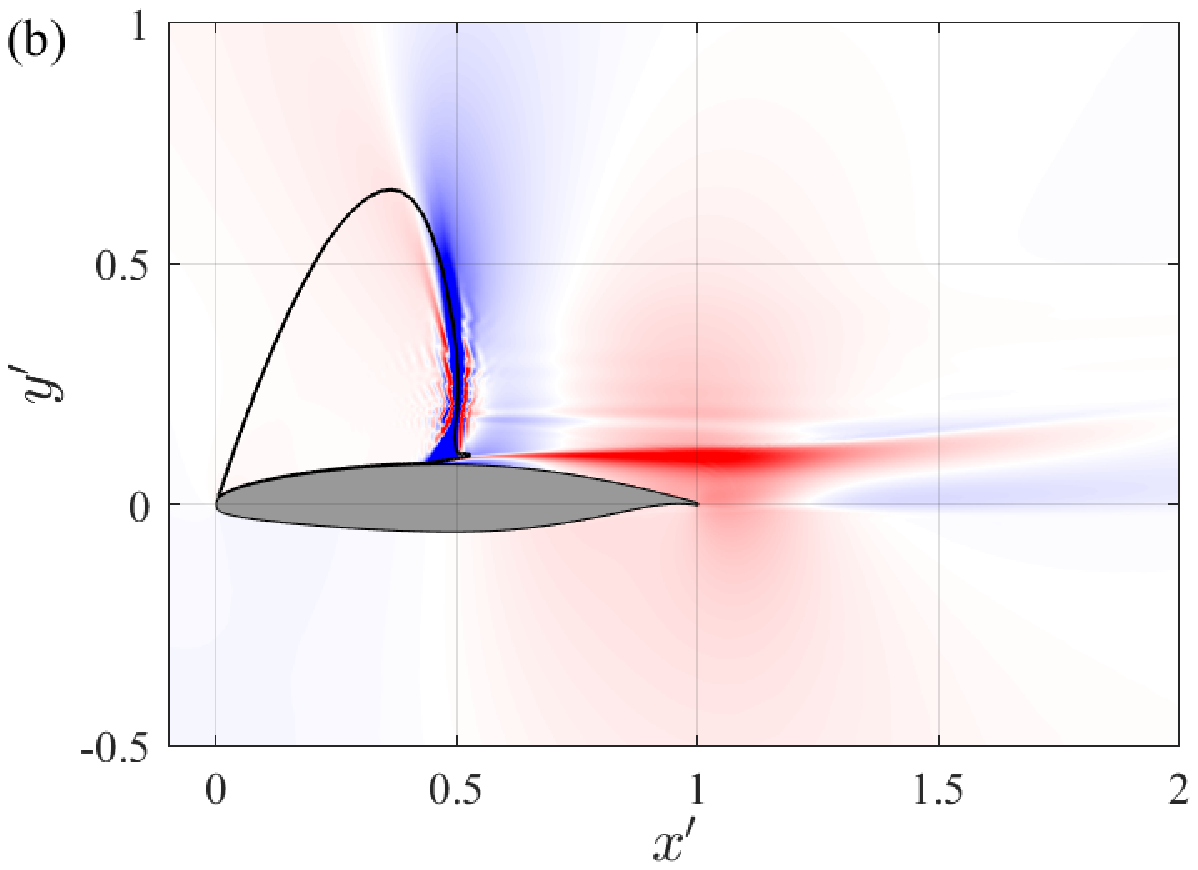}
	\includegraphics[trim={0cm 0.8cm 1cm 1cm},clip,width=0.33\columnwidth]{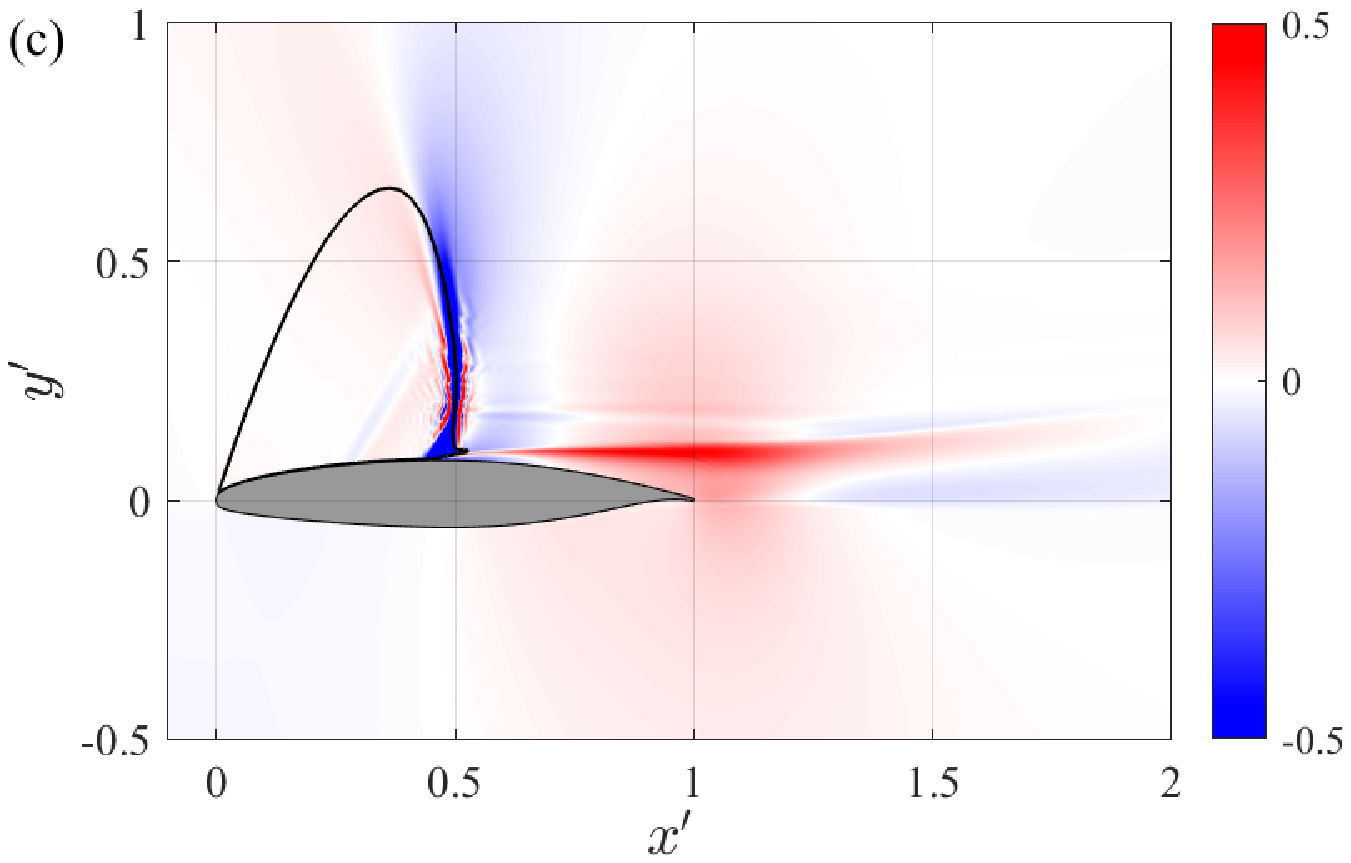}
	\caption{Contours of the real part of density of buffet mode from (a) SPOD (LES), (b) leading global mode from GLSA (RANS) and (c) resolvent mode at buffet frequency (RANS) for case A5M735SP.}
	\label{figSPODVsGLSA}
\end{figure}

GLSA of the steady RANS solution (A5M735SP) yielded a single globally unstable mode at the buffet frequency. A resolvent analysis yielded optimally amplified modes at approximately the same frequencies as those associated with the buffet and wake modes from SPOD. Note that, similar to this, two peaks were also identified using resolvent analysis by \citet{Sartor2015}. Features of the modes identified at the buffet frequency  using these approaches (\textit{i.e.}, buffet modes) are compared with those from SPOD of LES data using contours of density in Fig.~\ref{figSPODVsGLSA}. The contours shown are the real part of the density field. The sonic line based on the local Mach number is overlaid for reference. It is evident from the figure that the three approaches yield buffet modes which have a remarkably similar structure. As noted before, a decrease in density in the vicinity of the shock wave is associated with an increase in density in the wake region for all three cases. An animation based on the variation of the density contours with phase is also provided (Supplementary Material) and, comparing this with the SPOD results from LES for the same case, it is evident that the temporal variation is also accurately reproduced. There are some differences in the modal structure between the buffet mode obtained from SPOD and the others in the aerofoil's wake, with a secondary region (blue region in the wake for $x > 1.5$) which is out of phase with the primary region associated with density decrease (dominant red region in the wake). One possible reason for this difference might be that the buffet and wake modes can interact nonlinearly in the LES. These interactions would be reflected in an SPOD, whereas the linearised approaches of GLSA and resolvent analysis do not capture these. Some oscillations are present in the vicinity of the shock waves in the RANS results. These are numerical artefacts of the linearised approach when modelling a strong discontinuity. Comparing the buffet modes from GLSA and resolvent analysis, a weak wave can be observed in the latter that originates at the trip location on the suction side and extends with a positive slope along the $x-y$ plane up to the sonic line. SPOD of URANS results also yielded a peak in the eigenvalue spectrum at the buffet frequency and the mode associated with this was found to have close resemblance to the buffet modes obtained using other approaches (not shown for brevity).

\begin{figure}[!t]
	\centering
	\includegraphics[trim={0cm 0.8cm 1cm 1cm},clip,width=0.33\columnwidth]{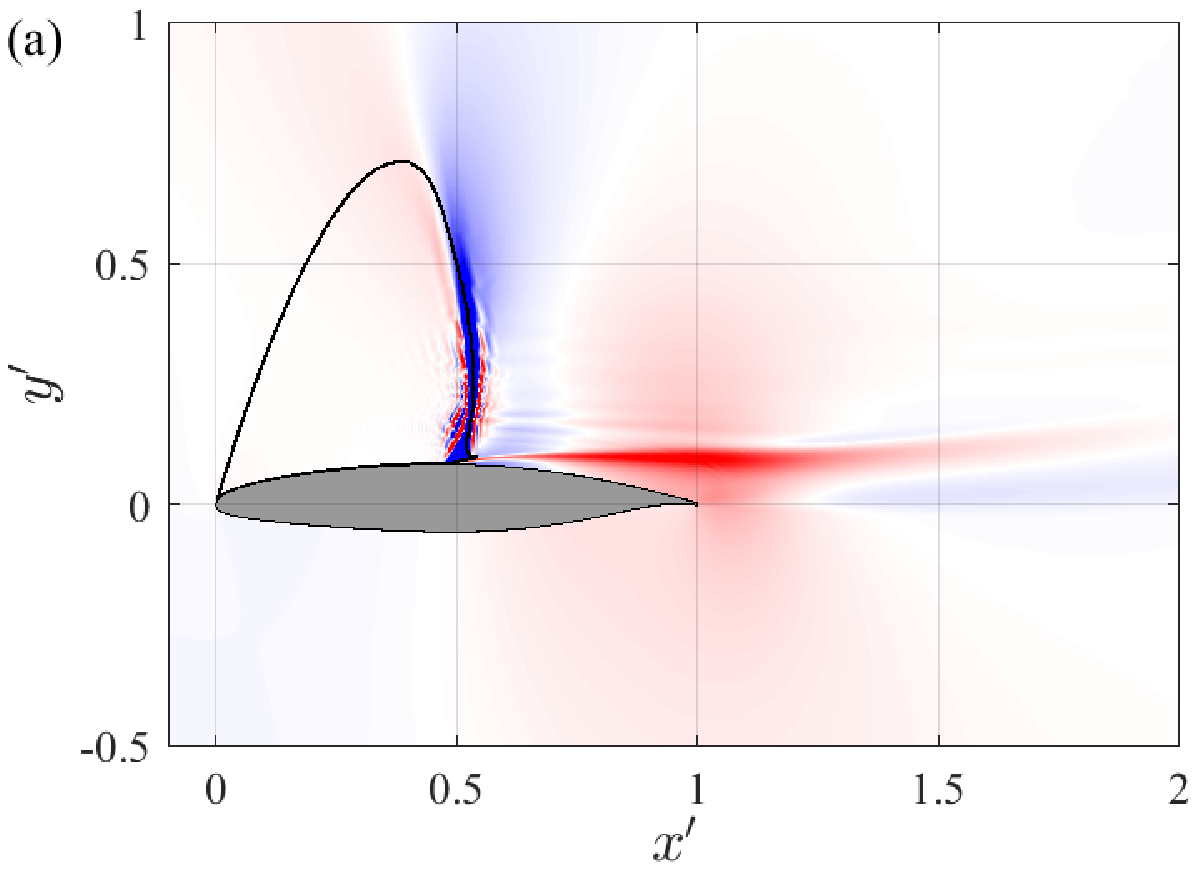}
	\includegraphics[trim={0cm 0.8cm 1cm 1cm},clip,width=0.33\columnwidth]{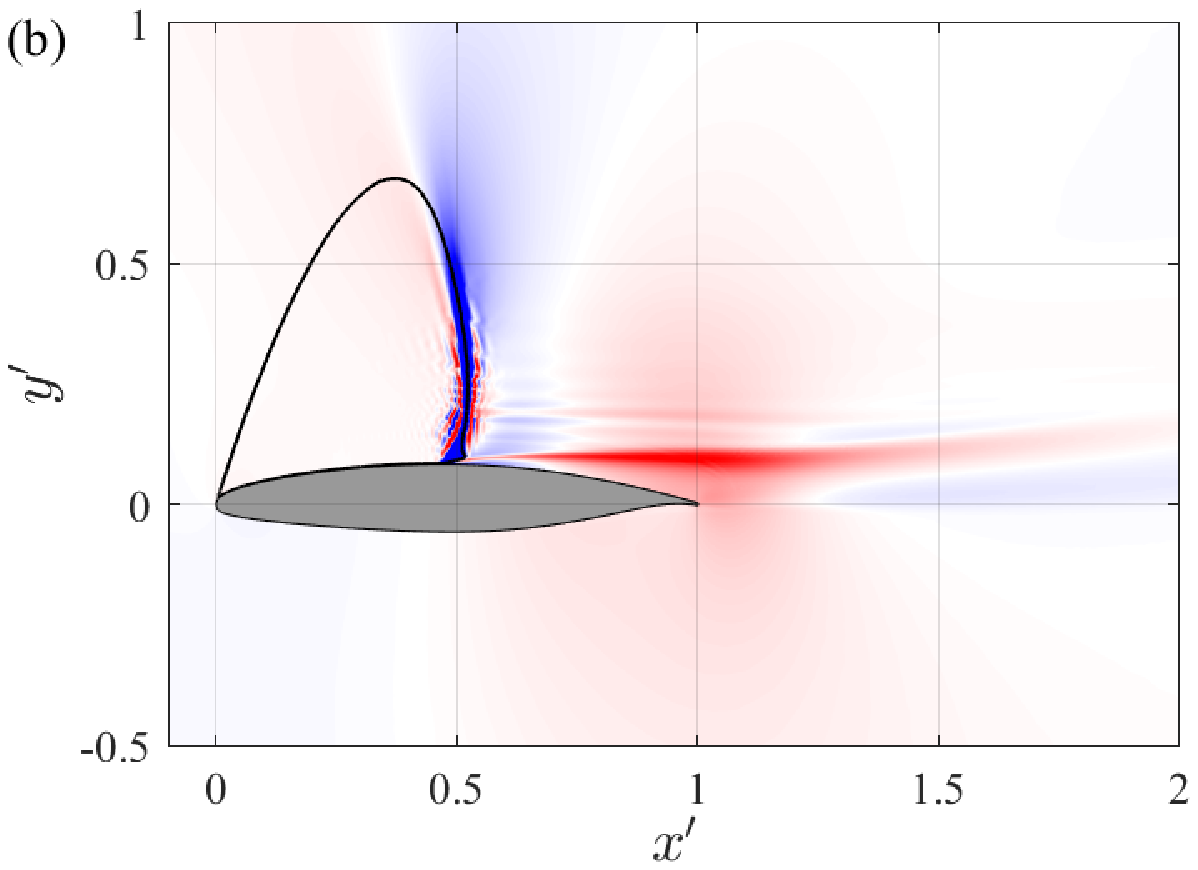}
	\includegraphics[trim={0cm 0.8cm 1cm 1cm},clip,width=0.33\columnwidth]{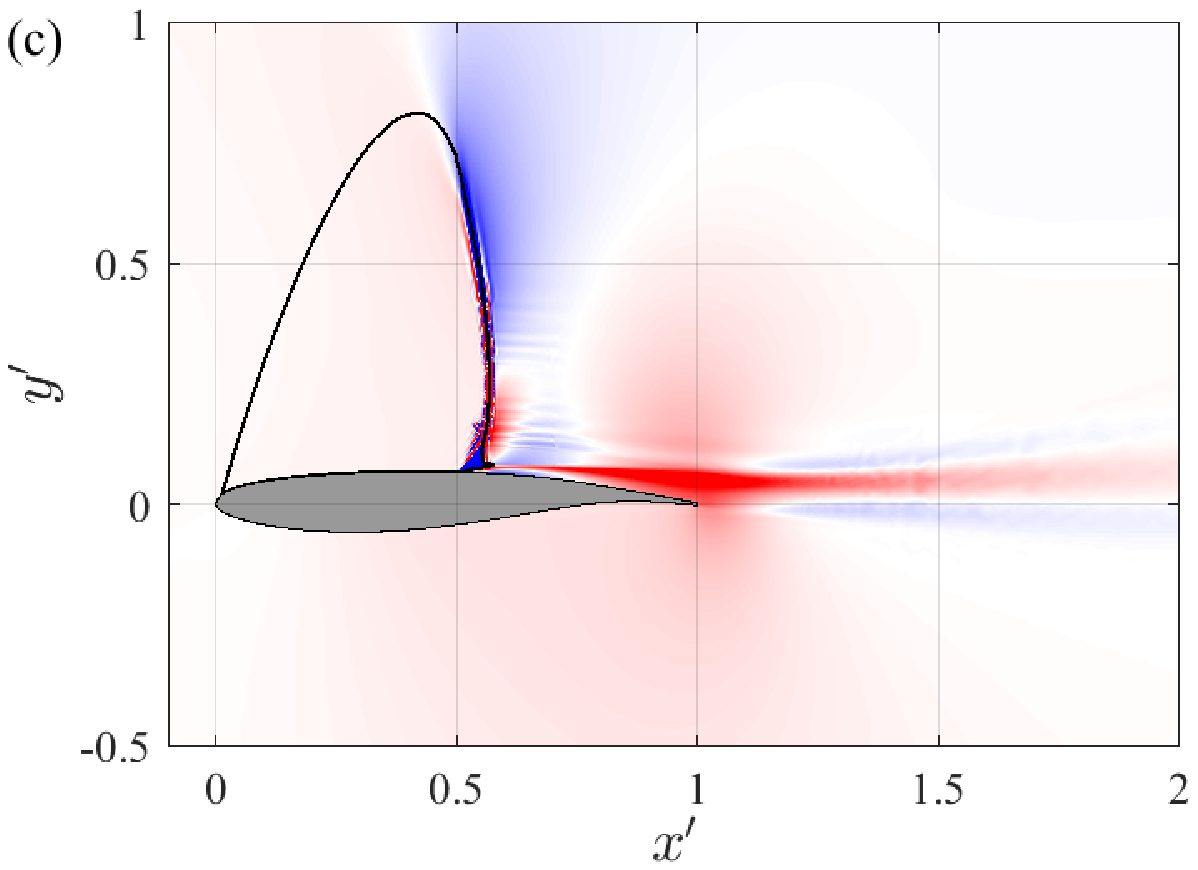}
	\caption{Contours of the real part of density of buffet mode from GLSA of RANS results at $\boldsymbol{Re = 3\times10^6}$ for (a) V2C, with tripping conditions the same as case A5M735SP, (b) V2C and (c) OAT15A, with both of the latter at fully turbulent conditions.}
	\label{figGLSAHighReOAT15a}
\end{figure}

The present and preceding section indicate that laminar and turbulent buffet have a similar flow structure and that the latter is captured well in both RANS and LES. These results have been established for the V2C aerofoil at $Re = 5\times10^5$ for free transition ($x_t \rightarrow \infty$) and specific trip locations ($x_{ts} = 0.2$ and $x_{tp} = 0.5$). To generalise these results, we also examined a higher $Re$, fully turbulent conditions and the OAT15A aerofoil. Modal features for a case with the same conditions as A5M735SP, except for a higher Reynolds number of $Re = 3\times 10^6$ are shown in Fig.~\ref{figGLSAHighReOAT15a}a. In Fig.~\ref{figGLSAHighReOAT15a}b, the conditions are the same as in Fig.~\ref{figGLSAHighReOAT15a}a, except that the transition is assumed to be at the leading edge now (\textit{i.e.}, $x_{ts} = x_{tp} = 0$) implying fully-turbulent conditions. The last panel, Fig.~\ref{figGLSAHighReOAT15a}c, is for ONERA's OAT15A aerofoil at fully turbulent conditions at $Re = 3\times10^6$, $M = 0.74$ and $\alpha = 3^\circ$. Additionally, the temporal variation in one buffet cycle for all three cases is provided as an animation (Supplementary Materials). Considering Fig.~\ref{figLES_SPODModes}, Fig.~\ref{figSPODVsGLSA} and Fig.~\ref{figGLSAHighReOAT15a}, we see striking qualitative similarities in the modal structure for all cases. % Importantly, given the almost identical mode shape obtained at a lower $Re = 500,000$ (see Sec.~\ref{secSPOD}), these results indicate that the mode shape of turbulent transonic buffet does not vary substantially with either aerofoil profile, transition conditions or $Re$.  

These similarities can be extended to other flow settings previously reported in the literature. For example, comparing Fig.~\ref{figGLSAHighReOAT15a}c with Fig.~14\textit{a} of \citet{Sartor2015} (which is for the OAT15A and for the same flow conditions except for a slightly lower $M = 0.73$ and higher $\alpha = 3.5^\circ$), there is excellent agreement in the modal structure, especially with regard to the out-of-phase relationship between the density fluctuations in the vicinity of the shock wave and the wake (same as that noted above for the V2C cases). Contours of pressure fluctuations (magnitude) based on LES of buffet shown in Fig.~7 of \citet{Fukushima2018} for the OAT15A at $Re = 3\times10^6$ and tripping at $x_{ts} = x_{tp} = 0.07$ also exhibit similar features to those seen here. The modal contours based on the streamwise velocity component for the OAT15A case (not shown for brevity) were also found to match well with Fig.~5 of \citet{Garbaruk2021} which is for the same flow conditions. Those authors also did not observe any significant change between mode shapes of turbulent buffet when the tripping location was varied. Furthermore, the temporal variations shown for the NACA0012 aerofoil in figure 5 of \citet{Crouch2009} strongly resemble the animations provided here (Supplementary Material). Thus, these results indicate that buffet features are essentially the same irrespective of boundary layer transition features and that this holds true for a range of Reynolds numbers and different aerofoil profiles.  

\begin{figure}
	\centering
	\includegraphics[trim={0cm 2cm 1cm 2cm},clip,width=0.495\columnwidth]{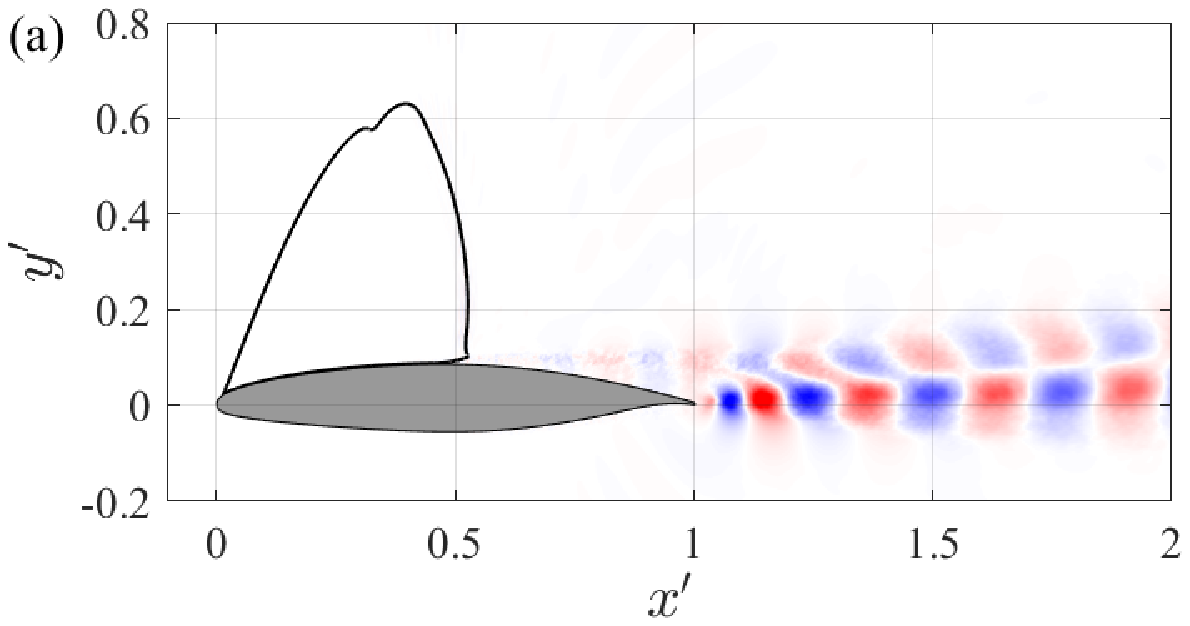}
	\includegraphics[trim={0cm 2cm 1cm 2cm},clip,width=0.495\columnwidth]{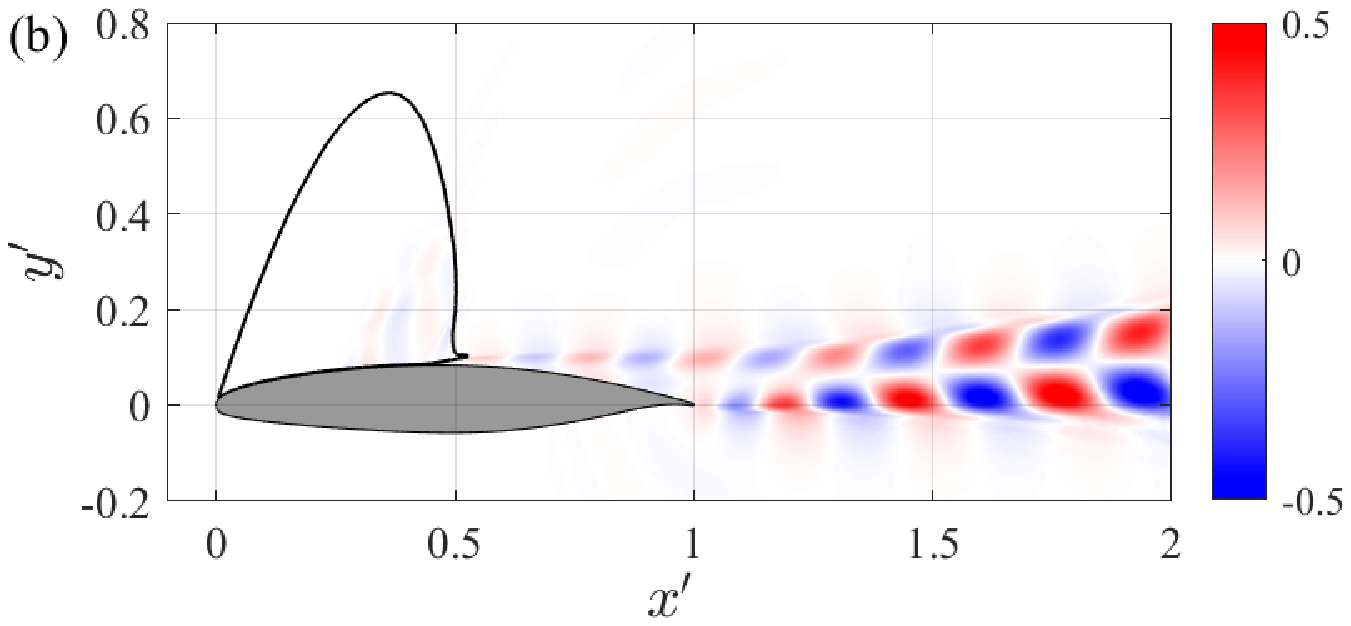}
	\caption{Contours of real part of density of wake mode from (a) SPOD of LES results and (b) resolvent mode from RANS simulation for case A5M735SP.}
	\label{figSPODVsResolvent}
\end{figure}

As previously noted in connection with the gain versus frequency spectrum from resolvent analysis, a second peak occurred at approximately the frequency associated with the wake mode observed in SPOD. The mode associated with this peak is compared with the wake mode in Fig.\ref{figSPODVsResolvent}. Although minor differences exist, it is evident that the resolvent analysis also yields a von K\'arm\'an vortex street with approximately the same width and orientation. Thus, it is clearly demonstrated here that the dominant coherent structures (buffet and wake modes) observed in the flow field can be captured in the LES and also by means of different numerical approaches applied to steady RANS solutions. However, as noted before, SPOD of URANS results did not yield any modes that resemble a vortex street, implying that URANS simulations do not capture wake mode features.

\subsection{Flow reconstruction using SPOD}
\label{secSPODModeRecon}
%%%%%%%%%%%%%%%%%%%%%%%%%%%%%%%%%%%%%%%%

\begin{figure}[!t]
	\centering
	\includegraphics[trim={1.9cm 0cm 3cm 0cm},clip,width=0.32\columnwidth]{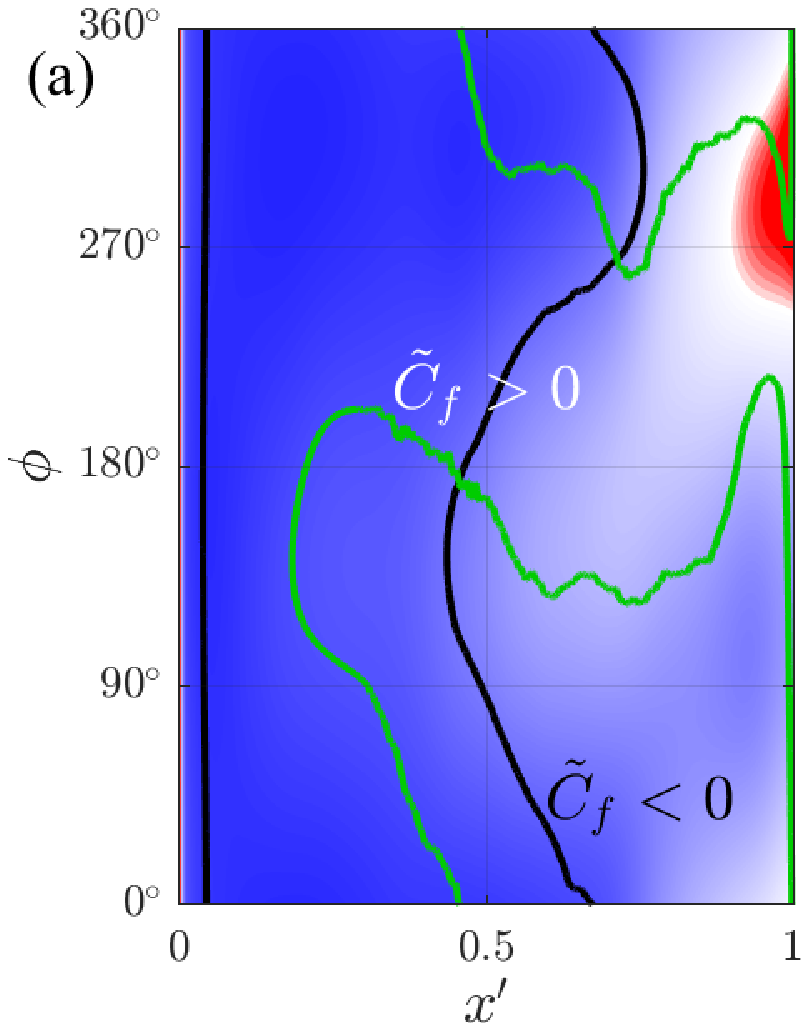}
	\includegraphics[trim={1.9cm 0cm 3cm 0cm},clip,width=0.32\columnwidth]{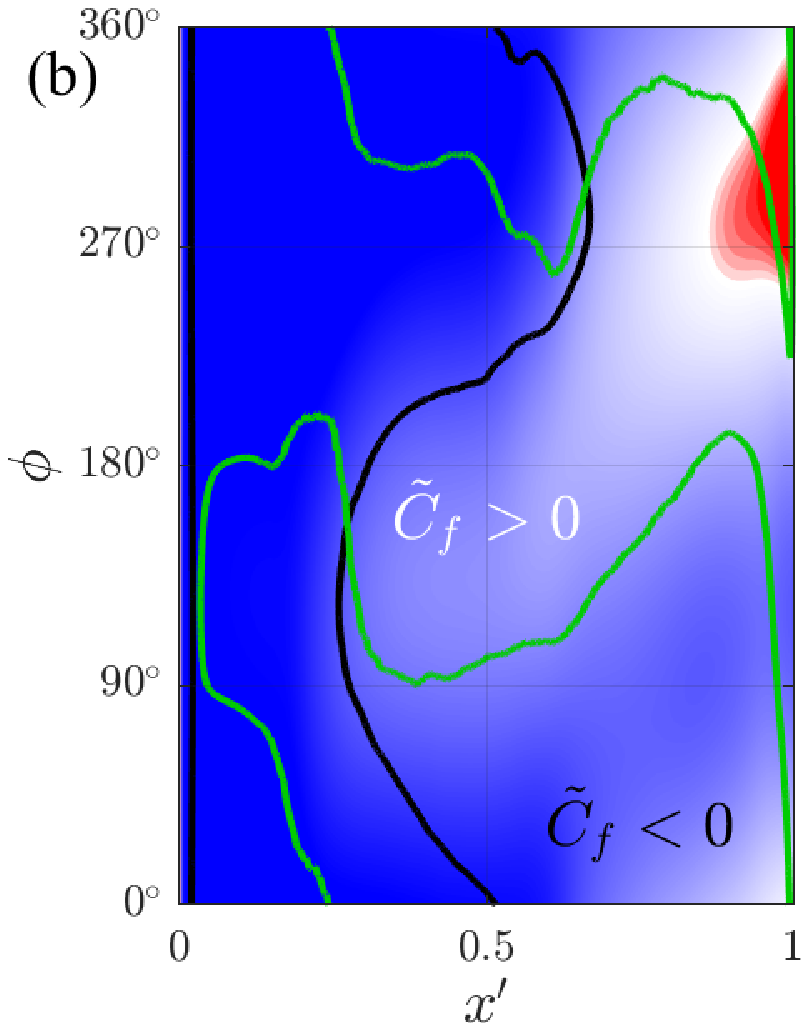}
    \includegraphics[trim={1.9cm 0cm 3cm 0cm},clip,width=0.32\columnwidth]{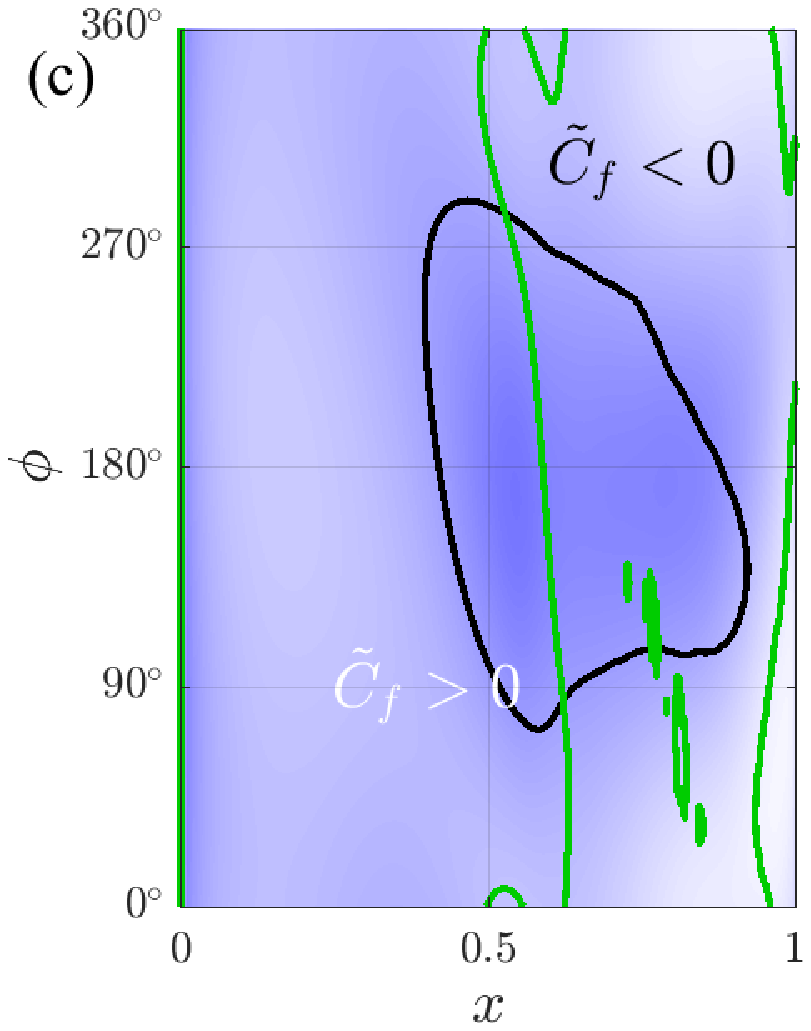}	
    \includegraphics[trim={1.9cm 0cm 3cm 0cm},clip,width=0.32\columnwidth]{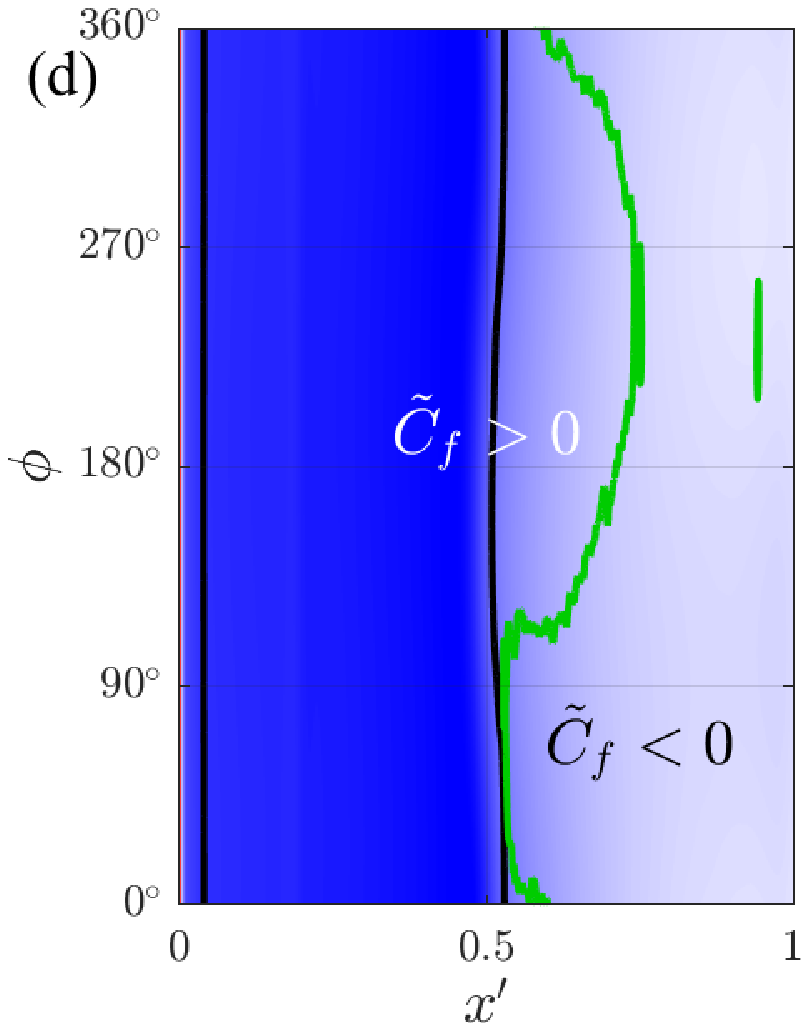}
    \includegraphics[trim={1.9cm 0cm 3cm 0cm},clip,width=0.32\columnwidth]{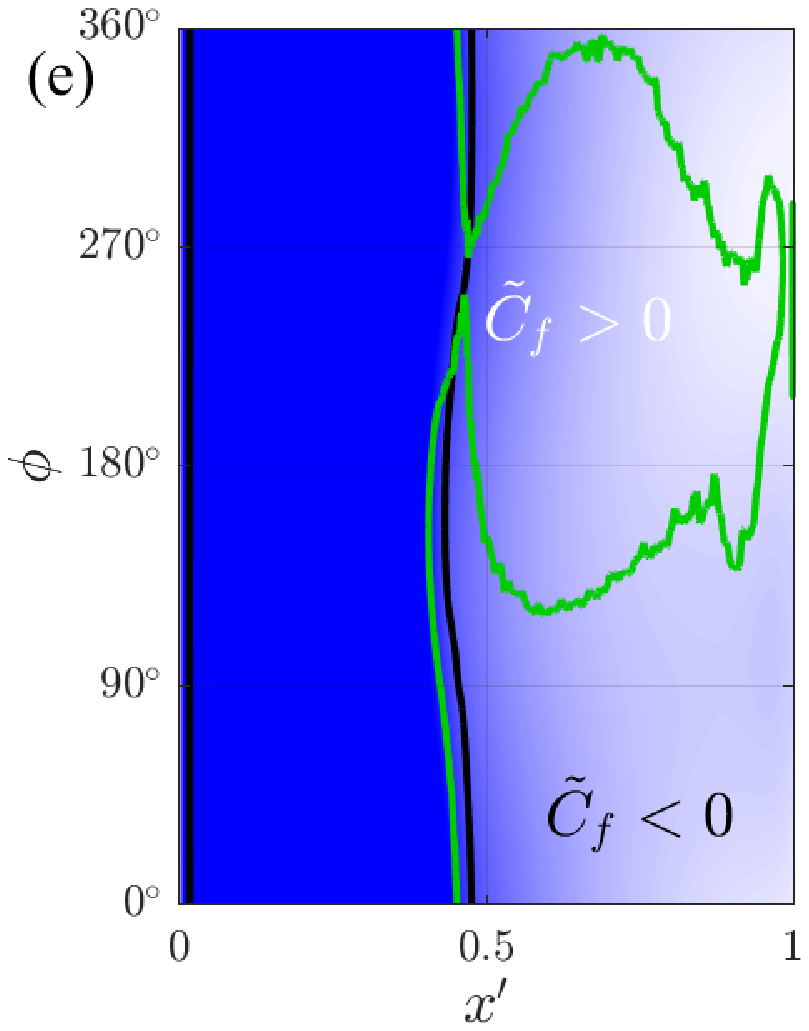}
    \includegraphics[trim={1.9cm 0cm 3cm 0cm},clip,width=0.32\columnwidth]{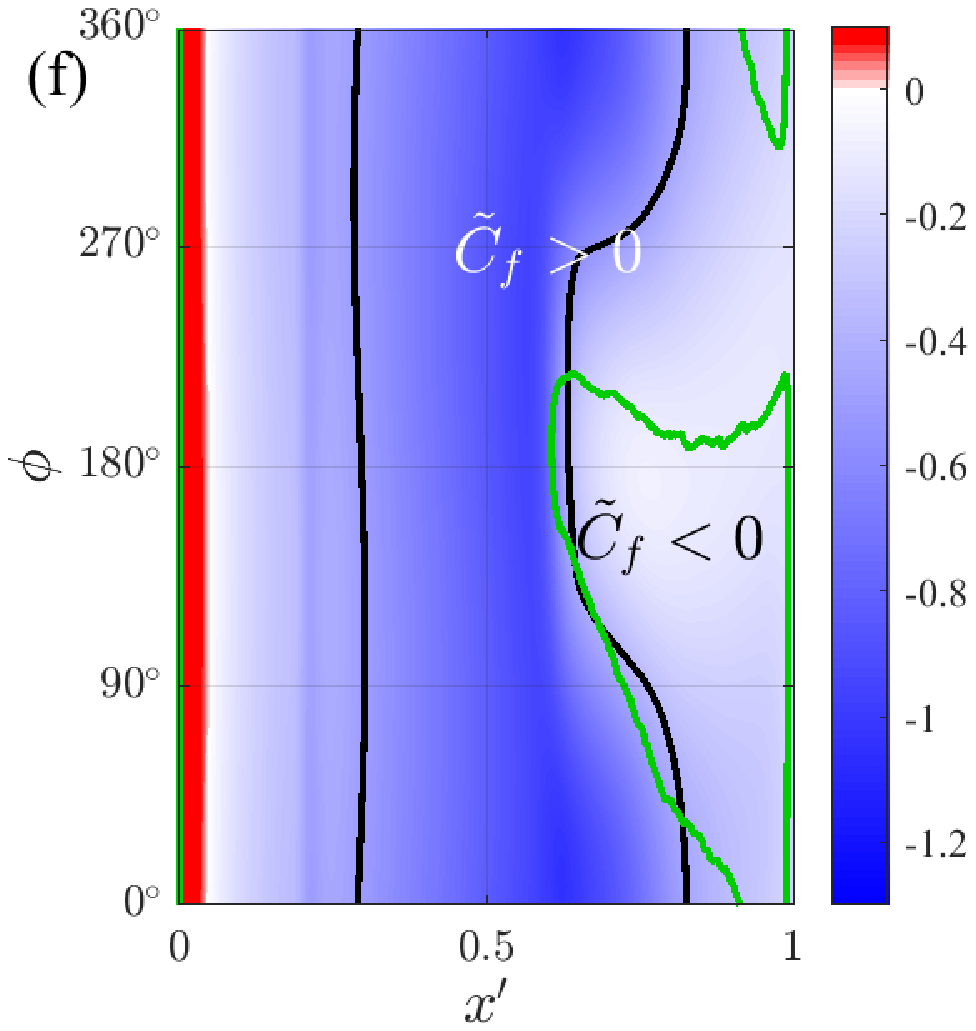}
	\caption{Spatio-temporal contours of $\tilde{C}_p$ for different cases: (a) A5M735F, (b) A7M7F, (c) A0M8S (pressure surface), (d) A5M735S, (e) A7M7S and (f) A0M8S (suction surface). The sonic line (black curve) and $\tilde{C}_f = 0$ isoline (green curve) are also overlaid.}
	\label{figLES_SPODrecon}
\end{figure}

To understand the associations between shock waves and boundary layer separation in buffet, a flow reconstruction based only on the buffet mode from SPOD of the LES data and the mean flow field (see Eq. \ref{eqnReconstr}) is examined. Spatio-temporal diagrams showing contours of the reconstructed pressure coefficient on the aerofoil surface are presented in Fig.~\ref{figLES_SPODrecon} for some of the cases studied. Curves delineating separated-flow regions (\textit{i.e.}, $\tilde{C}_f = 0$) and the sonic line (based on the local flow field at a wall-normal distance 0.05 from the aerofoil surface) are also overlaid. Note that $\phi = 0^\circ$ and $180^\circ$ denote the low-lift and high-lift phases, respectively. Only the suction-side variation is shown for all cases except A0M8S. The top set of plots (Fig.~\ref{figLES_SPODrecon}\textit{a},\textit{b},\textit{c}) are associated with free transition (since the pressure side shown in Fig.~\ref{figLES_SPODrecon} is not tripped for case A0M8S), while the bottom set (Fig.~\ref{figLES_SPODrecon}\textit{d},\textit{e},\textit{f}) correspond to forced transition. The spatial variations of $\tilde{C}_p$ are more gradual for the free-transition cases, which is expected due to the multiple unsteady shock waves present in the LES field. By contrast, the pressure jump is relatively more abrupt for the forced-transition cases. Focusing on the variations of the isolines, $\tilde{C}_f = 0$, shown using the green curves, one qualitative feature that is common to Fig. \ref{figLES_SPODrecon}a, b, d, e and f), is that the extent of separation is a minimum approximately in the phase range $180^\circ  \leq \phi \leq 270^\circ$. Note that for the pressure side of A0M8S (Fig. \ref{figLES_SPODrecon}c), the opposite trend is observed, albeit the temporal variation of the separation point location is weak. This trend is expected, given that the buffet features on this side are out of phase with the suction side by $180^\circ$. The separation behaviour further corroborates that laminar and turbulent buffet have similar characteristics. 

An interesting point to note is that there are strong dynamical variations in the separation features within the buffet cycle for the non-zero incidence cases. Indeed, for cases A5M735F, A7M7F and A7M7S (\textit{i.e.}, both laminar and turbulent buffet), the flow changes from being fully separated downstream of the shock foot at approximately $\phi \leq 90^\circ$ to almost fully reattached for $\phi \approx 250^\circ$. This behaviour has strong similarities with the ``low-frequency oscillations'' reported in the literature for incompressible flows around aerofoils at incidence angles close to stall (e.g., \citet{Zaman1989} noted that these oscillations are associated with ``a periodic switching between stalled and unstalled states"). Efforts are underway to further understand this connection.

%%%%%%%%%%%%%%%%%%%%%%%%%%%%%%%%%%%%%%%%
\section{Conclusions}
\label{secConc}
%%%%%%%%%%%%%%%%%%%%%%%%%%%%%%%%%%%%%%%%

In this study, wall-resolved large-eddy simulations of laminar and turbulent transonic buffet are performed at similar flow conditions and the results from the latter are compared with those obtained using the framework of Reynolds-averaged Navier--Stokes (RANS) equations. For the former approach, forcing transition on the pressure side does not have any significant effect, while forcing transition on the suction side causes buffet amplitude to reduce as compared to when the boundary layer transitions freely. Furthermore, a single shock wave is present when the transition is forced, while multiple shock waves are observed otherwise. However, the buffet frequency and global dynamic behaviour are similar for both cases. A spectral proper orthogonal decomposition showed that the flow is dominated by modes at two different frequencies for all cases. The lower-frequency mode was identified as due to buffet and comparisons between free- and forced-transition cases showed that, when the variations in buffet amplitude and shock structure is accounted for, the features of the modes are essentially similar. Furthermore, when considering ``Type I" buffet at zero incidence, where shock waves oscillate on both sides of the aerofoil, changing transition characteristics on one side does not have a significant effect on buffet features. These results provide strong evidence that laminar and turbulent buffet have the same origins. Furthermore, the results suggest that laminar buffet would arise as a global instability (similar to turbulent buffet), which requires confirmation using global linear stability analysis.

The framework of RANS is used to generalise the above conclusion. Global linear stability and resolvent analyses of steady solutions of the RANS equations and a spectral proper orthogonal decomposition of unsteady RANS simulations are used to show that the buffet mode obtained through these approaches has a similar structure as that seen in the large-eddy simulations for (i) the same conditions as used in the latter, (ii) at a higher freestream Reynolds number of $Re = 3\times10^6$, (iii) a different aerofoil (OAT15A) and (iv) fully turbulent conditions. Thus, this study links the turbulent buffet features at high Reynolds numbers reported in literature and laminar buffet features reported at moderate Reynolds numbers, while simultaneously making direct comparisons between the two levels of the aerodynamic modelling hierarchy. 

Another interesting result that arises from the comparison of large-eddy simulations and unsteady RANS simulations is related to the higher-frequency flow features. Whereas the wake mode, which resembles a von K\'arm\'an vortex street, occurs with significant energy content in the former approach and is also predicted by a resolvent analysis, it is not captured in the URANS simulation. Since buffet is captured in both approaches, this indicates that the mechanisms that drive it are not significantly affected by the wake mode.

%%%%%%%%%%%%%%%%%%%%%%%%%%%%%%%%%%%%%%%%
\section*{Appendix}
\label{appendix}
%%%%%%%%%%%%%%%%%%%%%%%%%%%%%%%%%%%%%%%%
%%%%%%%%%%%%%%%%%%%%%%%%%%%%%%%%%%%%%%%%%%%%%%%%%%%%%%%%%%%%%%%%%%%%%%%%%%%%%%%%
\subsection{Mesh-convergence studies for large-eddy simulations}
\label{AppLESGridConv}
%%%%%%%%%%%%%%%%%%%%%%%%%%%%%%%%%%%%%%%%%%%%%%%%%%%%%%%%%%%%%%%%%%%%%%%%%%%%%%%%
\begin{figure}
	\centering
	\includegraphics[width=.45\columnwidth]{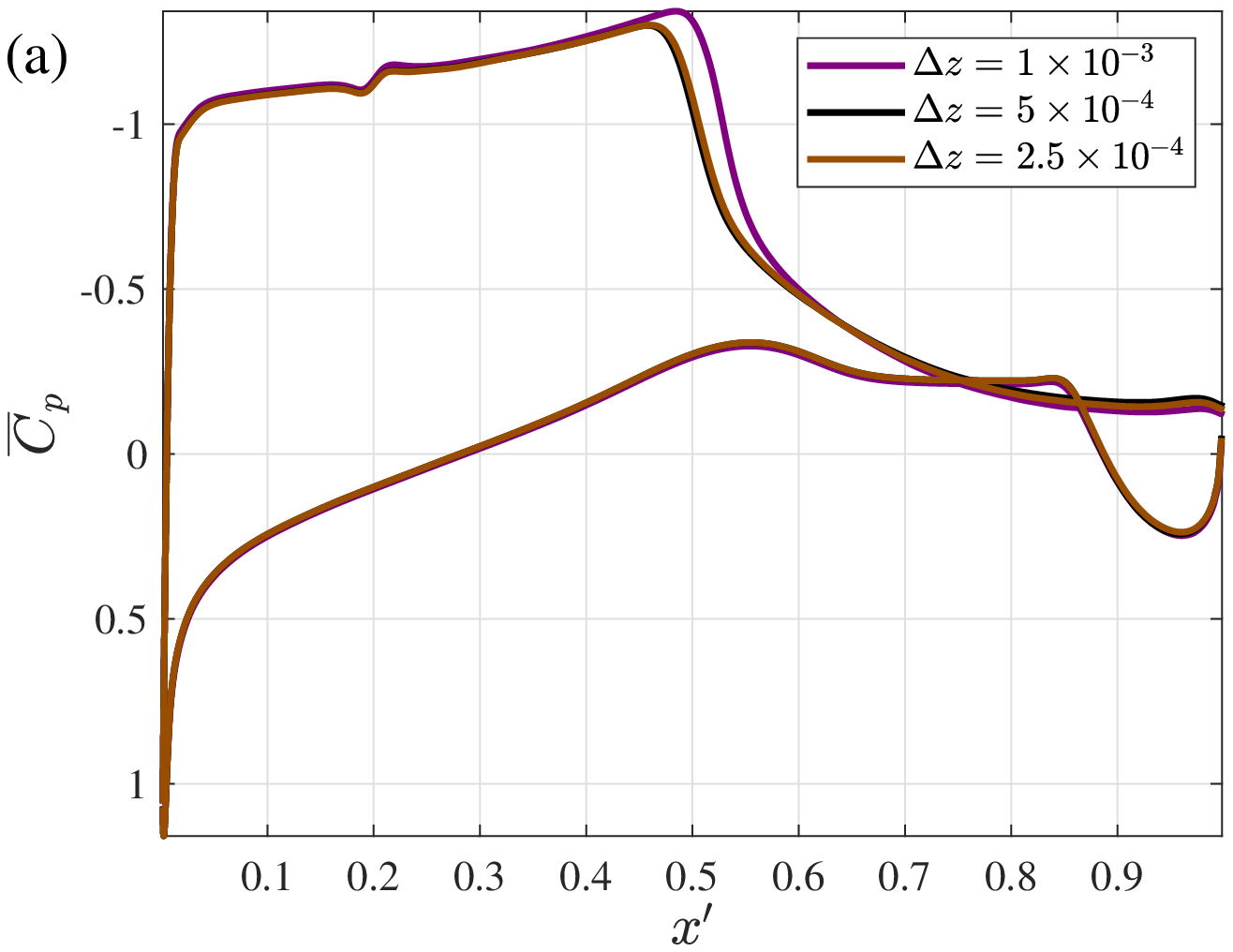}
	\includegraphics[width=.45\columnwidth]{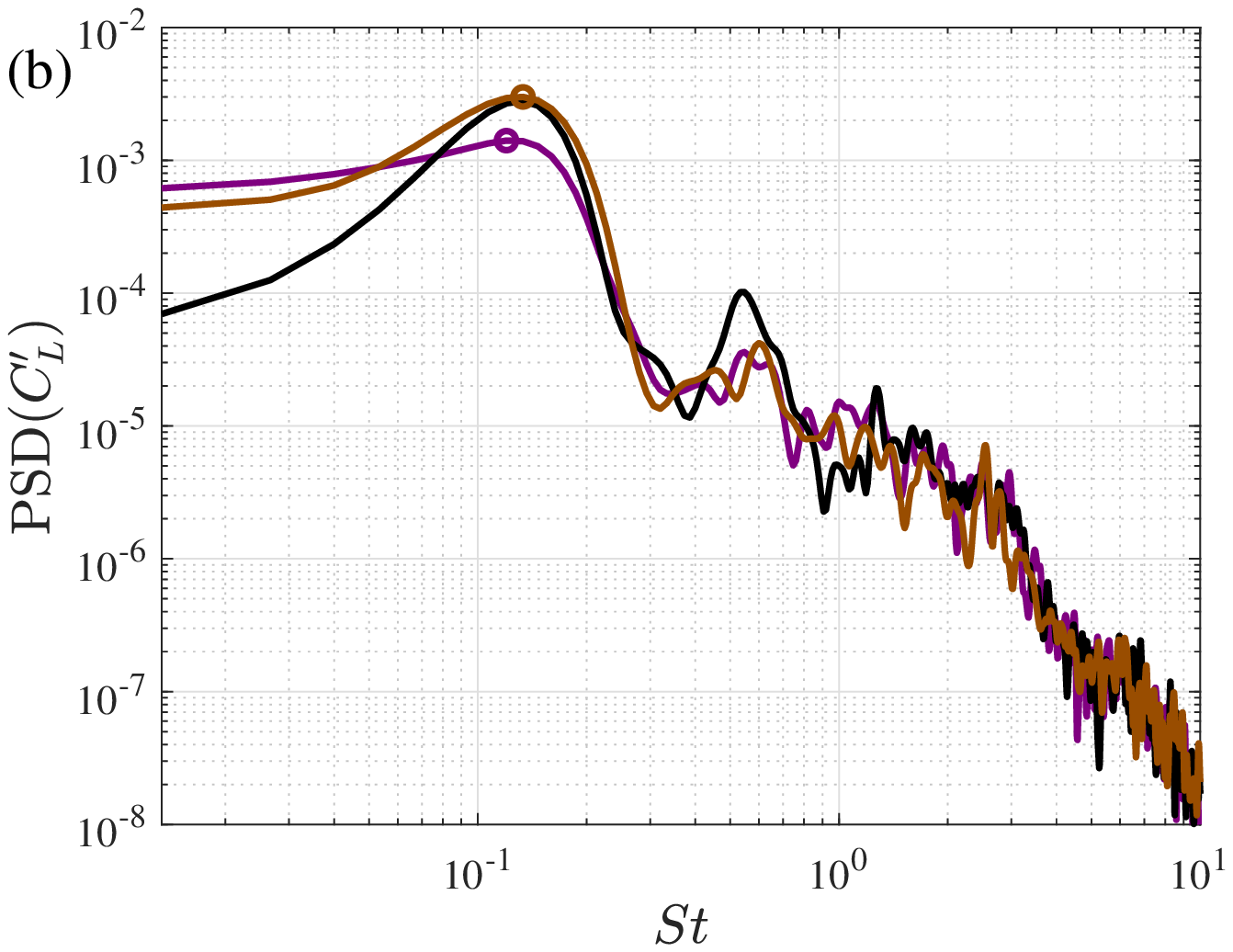}
	\caption{Comparisons of (a) $\overline{C}_p$ variation on the aerofoil surface and (b) PSD of lift fluctuations for various choices of spanwise grid spacing for case A5M735S, LES.}
	\label{figGridStudies}
\end{figure}

Studies examining the effect of the grid on the LES were performed for the forced-transition case of A5M735S (free-transition cases have been examined by \citet{Zauner2019}). The grid spacing in different directions ($x$, $y$ and $z$) was varied independently. It was observed that the solutions are most sensitive to the spatial resolution in the spanwise direction. A comparison of the mean pressure coefficient when the grid spacing in the spanwise direction is varied is shown in Fig.~\ref{figGridStudies}. The solutions for the lowest resolution is found to be inadequate, with the mean shock wave position being further downstream and buffet amplitude (circles) reduced, but for the other two cases these features and buffet frequency match adequately. Thus, $\Delta z = 5\times10^-4$ was chosen for all simulations.

%%%%%%%%%%%%%%%%%%%%%%%%%%%%%%%%%%%%%%%%
\subsection{Validation of RANS simulations}
\label{subSecRANSValidation}
%%%%%%%%%%%%%%%%%%%%%%%%%%%%%%%%%%%%%%%%

\begin{figure}[!t]
    \centering
    \includegraphics[width=0.45\textwidth]{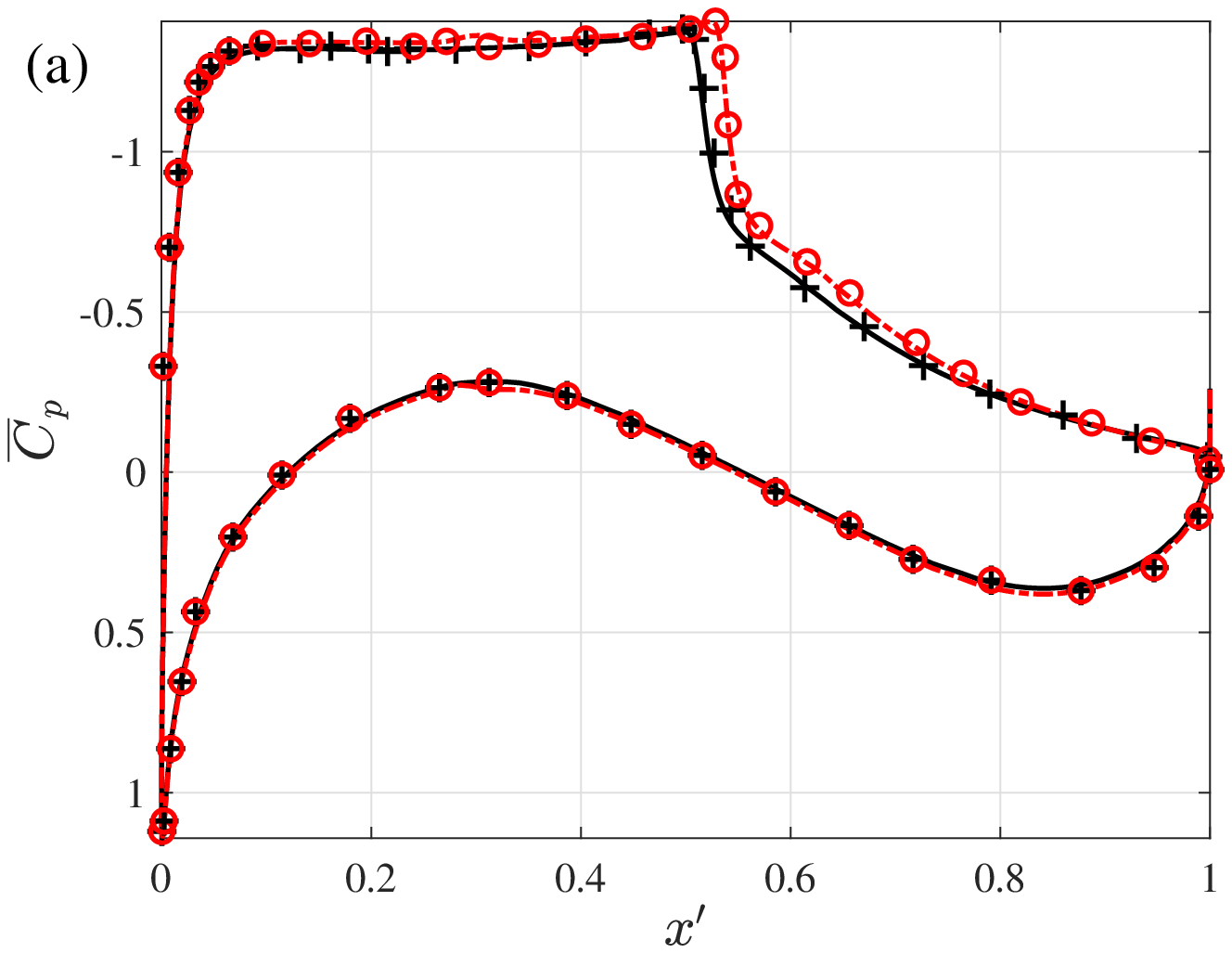} \quad
    \includegraphics[width=0.45\textwidth]{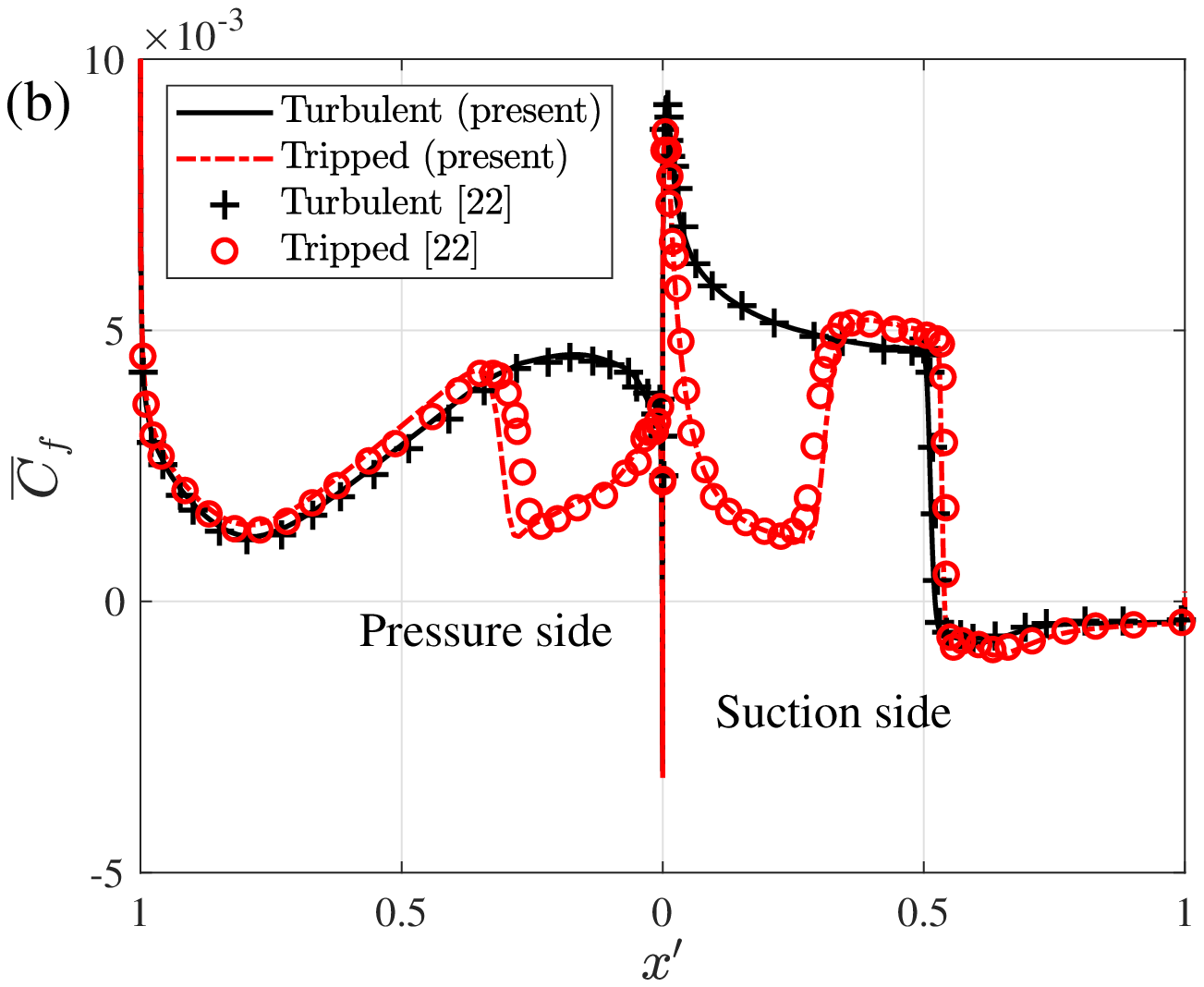}
    \caption{Validation of present RANS results with \cite{Garbaruk2021} showing chordwise variation of aerofoil coefficients at $\boldsymbol{\alpha=3.0^\circ}$ for OAT15A aerofoil for fully turbulent and tripped conditions.}
    \label{fig:comp_cpcf}
\end{figure}

The effect of the streamwise extent of the laminar zone on buffet was recently investigated by \citet{Garbaruk2021} with a RANS-based approach using the OAT15A aerofoil at $Re=3\times10^6$, $M=0.74$ and different angles of attack. We use those results to validate the RANS approach adopted here. Figure~\ref{fig:comp_cpcf} shows the mean pressure and skin-friction coefficients for (i) fully turbulent conditions (\textit{i.e.}, $x_{ts} = x_{tp} = 0$) and (ii) forced transition conditions of $x_{ts}= x_{tp} = 0.3$. While the results show excellent agreement overall, there are minor differences in the $\overline{C}_f$ distribution for the latter. Specifically, it is understood that, in contrast to our formulation of forcing transition (cf.~Sec.~\ref{subsecMethodRANS}), in \cite{Garbaruk2021} the Spalart--Allmaras turbulence was used with the original trip term, which involves an additional source term active only within a small domain of influence around the specified transition point and hence the rise in skin friction initiates slightly upstream of the transition point. The impact on the integrated coefficients of lift and drag, as well as the stability results such as the buffet onset angle of attack, was found to be minimal. 

\section*{Funding Sources}
%%%%%%%%%%%%%%%%%%%%%%%%%%%%%%%%%%%%%%%%
We acknowledge financial support by the Engineering and Physical Sciences Research Council (EPSRC) through the joint grant ``Extending the buffet envelope: step change in data quantity and quality of analysis” (EP/R037027/1 and EP/R037167/1).

\section*{Data statement}
The simulation data that support the findings of this study are available from the authors upon reasonable request. 

%%%%%%%%%%%%%%%%%%%%%%%%%%%%%%%%%%%%%%%%
\section*{Acknowledgements}
%%%%%%%%%%%%%%%%%%%%%%%%%%%%%%%%%%%%%%%%
This study used the ARCHER2 UK National Supercomputing Service (https://www.archer2.ac.uk) and we thank the assistance provided by its support staff. We would like to acknowledge the computational time on the same provided by the UK Turbulence Consortium (UKTC) through the EPSRC grant EP/R029326/1. We also acknowledge the use of the IRIDIS High Performance Computing Facility, and associated support services at the University of Southampton, in the completion of this study. The V2C aerofoil geometry was kindly provided by Dassault Aviation and we would like to acknowledge ONERA for providing the OAT15A geometry. We thank the German Aerospace Center (DLR) for access to the TAU flow solver.

\bibliography{sample}

\end{document}